\documentclass[twocolumn, twocolappendix,trackchanges]{aastex7}

\usepackage{textcomp}
\usepackage{xspace}
\usepackage{amsmath}
\usepackage{amssymb} 
\usepackage{graphicx}
\usepackage{float}
\usepackage{gensymb}
\usepackage{comment}
\usepackage{gensymb}
\usepackage{bm}
\usepackage{url}

\newcommand {\ixpe}{\text{IXPE}\xspace}

\newcommand{\maxi}{\text{MAXI}\xspace}
\newcommand{\nustar}{\text{NuSTAR}\xspace}

\begin{document}

\title{The Disk Awakens: IXPE's Sequel Observation of GS~1354$-$64 \\ Suggests Changing-Look Behavior in the Steep-Power-Law State}

\author[orcid=0000-0002-2381-4184,sname=Ravi,gname=Swati]{Swati Ravi}
\affiliation{MIT Kavli Institute for Astrophysics and Space Research, Massachusetts Institute of Technology \\ 77 Massachusetts Avenue, Cambridge, MA 02139, USA}
\email[show]{swatir@mit.edu}  

\author[orcid=0000-0002-5872-6061,gname=James F., sname='Steiner']{James F. Steiner} 
\affiliation{Center for Astrophysics \textbar\ Harvard \& Smithsonian, 60 Garden Street, Cambridge, MA 02138, USA}
\email{james.steiner@cfa.harvard.edu}

\email{hermanm@mit.edu}
\author[0000-0003-4127-0739]{Megan Masterson}
\altaffiliation{NHFP Einstein Fellow}
\affiliation{Center for Astrophysics \textbar\ Harvard \& Smithsonian,
60 Garden Street,
Cambridge, MA 02138, USA}
\email{megan.masterson@cfa.harvard.edu}

\author[orcid=0009-0001-4644-194X]{Lorenzo Marra}
\affiliation{INAF Istituto di Astrofisica e Planetologia Spaziali, Via del Fosso del Cavaliere 100, 00133 Roma, Italy}
\email{lorenzo.marra@inaf.it}

\author[orcid=0000-0002-3638-0637, sname=Kaaret, gname=Philip]{Philip Kaaret}
\affiliation{NASA Marshall Space Flight Center, Huntsville, AL 35812, USA}
\email{philip.kaaret@nasa.gov}

\author[0000-0002-5767-7253]{Alexandra Veledina}
\affiliation{Department of Physics and Astronomy, 20014 University of Turku, Finland}
\affiliation{Nordita, KTH Royal Institute of Technology and Stockholm University, Hannes Alfv\'ens v\"ag 12, SE-10691 Stockholm, Sweden}
\email{alexandra.veledina@gmail.com}

\author[orcid=0000-0003-0079-1239,gname=Michal, sname='Dov\v{c}iak']{Michal Dov\v{c}iak} 
\affiliation{Astronomical Institute of the Czech Academy of Sciences, Bo\v{c}n\'{i} II 1401, CZ-14100 Prague, Czech Republic}
\email{michal.dovciak@asu.cas.cz}

\author[orcid=0000-0002-5311-9078]{Adam Ingram} 
\affiliation{Centre for Extragalactic Astronomy, Department of Physics, Durham University, South Road, Durham DH1 3LE, UK}
\email{adam.r.ingram@durham.ac.uk}

\author[0000-0002-6384-3027]{Fiamma Capitanio}
\affiliation{INAF Istituto di Astrofisica e Planetologia Spaziali, Via del Fosso del Cavaliere 100, 00133 Roma, Italy}
\email{fiamma.capitanio@inaf.it}

\author[orcid=0000-0001-5975-1026]{Sudeb Ranjan Datta}
\affiliation{Inter-University Centre for Astronomy and Astrophysics, Post Bag 4, Ganeshkhind, Pune - 411007, India}
\email{sudeb.datta@iucaa.in}  

\author[orcid=0000-0002-1532-4142]{Elise Egron}
\affiliation{INAF-Osservatorio Astronomico di Cagliari, Via della Scienza 5, 09047 Selargius (CA), Italy}
\email{elise.egron@inaf.it} 

\author[0000-0002-5767-7253]{Javier A. Garc\'ia}
\affiliation{X-ray Astrophysics Laboratory, NASA Goddard Space Flight Center, Greenbelt, MD, USA}
\affiliation{Cahill Center for Astronomy and Astrophysics, California Institute of Technology, Pasadena, CA, USA}
\email{javier.a.garciamartinez@nasa.gov}

\author[0000-0001-8670-4575]{Ole~K\"onig}
\affiliation{Center for Astrophysics \textbar\ Harvard \& Smithsonian, 60 Garden Street, Cambridge, MA 02138, USA}
\email{ole.koenig@cfa.harvard.edu}

\author[orcid=0000-0003-2845-1009, sname=Liu,gname=Honghui]{Honghui Liu}
\affiliation{Institut f\"ur Astronomie und Astrophysik, Universit\"at T\"ubingen, Sand 1, D-72076 T\"ubingen, Germany}
\email{honghui.liu@uni-tuebingen.de}

\author[orcid=0000-0002-6492-1293, sname=Marshall,gname=Herman]{Herman L. Marshall}
\affiliation{MIT Kavli Institute for Astrophysics and Space Research, Massachusetts Institute of Technology \\ 77 Massachusetts Avenue, Cambridge, MA 02139, USA}
\email{hermanm@mit.edu}

\author[0000-0003-4216-7936]{Guglielmo Mastroserio}
\affiliation{Scuola Universitaria Superiore IUSS Pavia, Palazzo del Broletto, piazza della Vittoria 15, I-27100 Pavia, Italy}
\email{guglielmo.mastroserio@iusspavia.it}

\author[0000-0001-7374-843X]{Romana Miku\v{s}incov\'{a}}
\affiliation{INAF Istituto di Astrofisica e Planetologia Spaziali, Via del Fosso del Cavaliere 100, 00133 Roma, Italy}
\email{romana.mikusincova@inaf.it}

\author[orcid=0000-0002-8247-786X,sname='Neilsen']{Joey Neilsen}
\affiliation{Villanova University, Department of Physics, Villanova, PA 19003, USA}
\email{jneilsen@villanova.edu}

\author[orcid=0000-0002-9633-9193, sname=Nathan,gname=Edward]{Edward J. R. Nathan}
\affiliation{NASA Postdoctoral Program Fellow, X-ray Astrophysics Laboratory, NASA Goddard Space Flight Center, Greenbelt, MD 20771, USA}
\email{edward.nathan@nasa.gov}

\author[orcid=0000-0001-6061-3480,gname=Pierre-Olivier, sname=Petrucci]{P.-O. Petrucci}
\affiliation{Universit\'e Grenoble Alpes, CNRS, IPAG, 38000 Grenoble, France}
\email{pierre-olivier.petrucci@univ-grenoble-alpes.fr}

\author[0000-0001-5418-291X]{Jakub Podgorn\'y}
\affiliation{Astronomical Institute of the Czech Academy of Sciences, Bo\v{c}n\'{i} II 1401, CZ-14100 Prague, Czech Republic}
\email{jakub.podgorny@asu.cas.cz}

\author[0000-0003-2931-0742]{Ji\v{r}\'i~Svoboda}
\affiliation{Astronomical Institute of the Czech Academy of Sciences, Bo\v{c}n\'{i} II 1401, CZ-14100 Prague, Czech Republic}
\email{jiri.svoboda@asu.cas.cz}  

\author[orcid=0000-0002-2268-9318,gname=Yuexin, sname=Zhang]{Yuexin Zhang} 
\affiliation{Center for Astrophysics \textbar\  Harvard \& Smithsonian, 60 Garden Street, Cambridge, MA 02138, USA}
\email{yuexin.zhang@cfa.harvard.edu} 

\begin{abstract}
After decades of only failed transitions, the black hole X-ray binary GS~1354$-$64 has undergone a complete outburst again. We present its first X-ray polarimetric investigation in the fading steep-power-law (SPL) state, following an earlier intermediate-state \ixpe campaign during the same outburst. The 2--8~keV emission has a polarization degree (PD) of \(2.3\pm0.2\%\) and polarization angle (PA) of \(5^\circ\pm3^\circ\). Compared with the intermediate state, where \({\rm PD}=3.9^{+0.3}_{-0.2}\%\) and \({\rm PA}=-2^\circ\pm2^\circ\), the broadband PD is lower while the PA remains consistent. The intermediate-state type-C quasi-periodic oscillation state has also disappeared, marking a major reconfiguration of the accretion flow. Assuming a distance of  7~kpc, the observed the intensity drop from 2.5\% to 1.6\% of the Eddington luminosity over 10.2 days is punctuated by several broad hour-to-day-scale troughs, tracing two distinct tracks in hardness--intensity space. Simultaneous spectroscopy shows the troughs' spectra contains a cooler disk with smaller apparent normalization, while no change in the Compton component's photon index or normalization is required between tracks. The model-independent analysis identifies a steeper increase in PD with energy outside the troughs, yielding $\mathcal{G}_E=1.5\pm0.2\%~{\rm keV}^{-1}$, compared with $\mathcal{G}_E=0.7^{+0.3}_{-0.5}\%~{\rm keV}^{-1}$ during the troughs. Their duration and spectropolarimetric morphology distinguish the troughs from traditional short absorption dips seen in X-ray binaries, providing a possible stellar-mass analog of the recurrent accretion changes that drive ``changing-look'' behavior in supermassive black hole systems.
\end{abstract}


\section{Introduction} 
\setcounter{footnote}{0}
Black hole X-ray binaries (BHXBs) serve as dynamic laboratories of accretion, changing their luminosity, spectral shape, and rapid variability as they traverse a sequence of characteristic X-ray states \citep{2006ARA&A..44...49R,2007A&ARv..15....1D}. During a canonical outburst, a source generally rises into a luminous hard state dominated by Comptonized emission from a hot inner flow or corona, passes through one or more intermediate states, and enters into a thermal-dominant soft state or a steep power-law (SPL; historically, very-high) state \citep{2006csxs.book..157M,2006ARA&A..44...49R,2007A&ARv..15....1D}. The SPL state is characterized by the coexistence of substantial thermal-disk emission and a steep Comptonized tail (commonly with photon index $\Gamma\gtrsim2.4$) together with substantially weaker broadband variability than in the hard and hard-intermediate states. The SPL state can exhibit QPOs and moderate rapid variability, but its timing properties span a broader range than those of the thermal-dominant state \citep{2006csxs.book..157M,2006ARA&A..44...49R}.

The hard-to-soft transition may be accompanied by transient relativistic ejections, and once the source is in a disk-bright state, broadband X-ray variability is generally reduced and the compact, steady radio jet is quenched. Changes in the geometry of the accretion disk and corona are also inferred \citep{2004MNRAS.355.1105F,2005Ap&SS.300..107H,2006ARA&A..44...49R, 2007A&ARv..15....1D}. Although this broad phenomenology is well established, how the inner disk and corona reconfigure during state transitions and what drives substantial variability within nominally stable states remain uncertain.

The Imaging X-ray Polarimetry Explorer (\ixpe) \citep{2021AJ....162..208S,2022JATIS...8b6002W} provides a new view of this evolution by measuring the linear polarization of X-rays in the 2--8~keV band. The polarization degree (PD) and polarization angle (PA) depend on the geometry and orientation of the emitting and scattering regions, providing complementary diagnostics to X-ray spectroscopy and timing. In hard and intermediate states, the \ixpe band is often dominated by Comptonized emission whose polarization can constrain the shape and orientation of the corona \citep{2022Sci...378..650K,2023ApJ...958L..16V,2024ApJ...968...76I, 2025MNRAS.541.1774E}. 

In SPL states, \ixpe probes radiation emerging from the optically thick accretion disk alongside a substantial steep Comptonized component, making the observed polarization signal sensitive both to the intrinsic disk signal and its dilution or modification by coronal scattering. Classical models of thermal disk polarization assume a plane-parallel, pure electron-scattering atmosphere with photon sources at its base \citep{1960ratr.book.....C,Sobolev1963,Rees1975}. In this case, the PD depends primarily on inclination. However, absorption in partially ionized plasma and distributed photon sources can significantly alter this picture \citep{LightmanShapiro1975,1979Afz....15..241L,Taverna2021}. General and special relativistic effects modify the local polarization signatures by rotation of PA, resulting in PD reduction \citep{1977Natur.266..429S,1977Natur.269..128C,2008MNRAS.391...32D}. Because higher-energy disk photons preferentially originate at smaller radii, these effects are expected to produce energy-dependent polarization signatures that encode the structure of the innermost disk. Black hole spin influences this signal by setting the location of the innermost stable circular orbit (ISCO) and therefore dictating the magnitude of the PD reduction and PA rotation effects, as well as controlling the strength of returning radiation—photons that are bent back onto the disk and scatter before reaching the observer \citep{2009ApJ...701.1175S,2010ApJ...712..908S}. X-ray polarimetry in disk-bright soft and SPL states can therefore complement continuum-fitting and reflection measurements of black-hole spin while testing assumptions about disk thickness, atmospheric scattering, coronal geometry, and alignment of the innermost disk \citep{2014SSRv..183..295M,2021ARA&A..59..117R,2023MNRAS.519.6138M}.

The growing sample of disk-bright BHXBs observed by \ixpe reveals remarkable diversity. In 4U~1957$+$115, a soft-state polarization signal at the few-percent level has been interpreted in the context of relativistic disk models as evidence for a rapidly spinning black hole under the assumed system geometry \citep{2024A&A...684A..95M}. The soft-state transient 4U~1630$-$47 exhibited a much larger PD, approaching $\sim8\%$, together with a pronounced increase with energy, motivating interpretations involving a highly inclined disk, an extended scattering atmosphere, potential departures from a standard geometrically thin disk, and considerable disk-wind contributions to the observed polarimetric properties \citep{2024ApJ...964...77R}. A subsequent \ixpe observation of 4U~1630$-$47 in the SPL state likewise measured a large PD that increased strongly with energy while the PA remained approximately constant, demonstrating that pronounced energy-dependent polarization can persist when substantial thermal-disk emission coexists with a steep Comptonized component \citep{2023ApJ...958L...8R}. The low-inclination LMC~X-1 produced an upper limit of $1\%$ PD when observed by \ixpe, while the higher-inclination LMC~X-3 showed a low, few-percent PD consistent with an energy-independent PA, and consistent with relativistic disk calculations for its low spin \citep{2024ApJ...960....3S}. These observations demonstrate how the range of polarimetric morphology present among disk-bright sources can reflect differences in inclination, disk structure, returning radiation, and geometry of the residual Comptonized component.

Multi-epoch \ixpe observations have also begun to trace how polarization evolves across accretion-state transitions. In the persistent BHXB Cygnus~X-1, the hard state shows a PD of approximately $4\%$ with a PA aligned with the radio jet, favoring a corona extended in the plane of the disk \citep{2022Sci...378..650K}. In the soft state, the PD decreases while the PA remains broadly stable, consistent with an increasing contribution from thermal disk emission \citep{2024ApJ...969L..30S,2025A&A...701A.115K}. Similarly, during the 2023--2024 outburst of Swift~J1727.8$-$1613, \ixpe followed the source from the hard and intermediate states into the soft state and subsequently toward the returning hard state \citep{2023ApJ...958L..16V,2024ApJ...966L..35S,2024ApJ...968...76I,2024A&A...686L..12P}. Its PD fell from $\sim3$--$4\%$ in the hard state to below $\sim1\%$ as the thermal disk came to dominate, before the hard-state PD returned during the soft-to-hard transition. \ixpe additionally observed GX~339$-$4 during the soft-intermediate and soft states of its 2023--2024 outburst, detecting polarization during the soft-intermediate observation but obtaining only an upper limit in the soft state \citep{2025ApJ...978L..19M}. These campaigns establish that PD responds sensitively to state evolution, but often lack insight into the PA evolution across state transition, as significant disk-bright polarization has not previously been reported for a transient BHXB that was also observed polarimetrically in a harder state of the same outburst.

The Galactic transient GS~1354$-$64 (BW~Cir) offers an unusual setting in which to examine disk-bright accretion with X-ray polarimetry. Optical dynamical measurements yield a mass function of $f(M)\simeq5.7\,M_\odot$, identifying the compact companion as a black hole \citep{2004ApJ...613L.133C}. Although several studies favor a comparatively high orbital inclination, potentially near $75^\circ$, both the viewing geometry and distance remain poorly constrained \citep{2009ApJS..181..238C,2016ApJ...826L..12E,2019MNRAS.485.2642G,2023ApJ...951..145L,2024ApJ...969...40D}. In particular, recent work favors a distance of $7.1^{+4.8}_{-3.9}~\mathrm{kpc}$ \citep{2021MNRAS.503.3784P}, although published estimates have ranged from $0.7^{+3.4}_{-0.3}$~kpc to $\sim20$--$25$~kpc \citep{2009ApJS..181..238C,2019MNRAS.485.2642G}.

Much of the previous observational history of GS~1354$-$64 was defined by its reluctance to enter softer, disk-bright states. After the luminous 1987 discovery outburst \citep{1990ApJ...361..590K}, its 1997 and 2015 outbursts remained dominated by hard or hard-intermediate emission and were classified as ``failed'' or ``stalled'' transitions \citep{2001MNRAS.323..517B,2016MNRAS.459.4038S,2016MNRAS.460..942K,2017MNRAS.469..193P}. These episodes showed strong Comptonization and, at some epochs, relativistic reflection, but yielded model-dependent and disparate conclusions concerning the location of the inner disk and the black-hole spin \citep{2001MNRAS.323..517B,2016MNRAS.459.4038S,2016MNRAS.460..942K}. The eventual emergence of substantial thermal-disk emission during the 2025--2026 outburst therefore opened an accretion regime that had rarely been accessible in this source. The first \ixpe observation during the 2025--2026 outburst (taken 2026 February 07--09) sampled GS~1354$-$64 between episodes of bright X-ray flaring, while the source occupied an intermediate state \citep{2025ATel17563....1N,2026ATel17618....1N,2026arXiv260303463R}. The observation yielded a significant polarization detection, with a 2--8~keV PD of \(3.9^{+0.3}_{-0.2}\%\) at \({\rm PA}=-2^\circ\pm2^\circ\), and a PD that increased strongly with energy at \(\mathcal{G}_{E}=1.3\pm0.2\%~{\rm keV^{-1}}\), while the PA remained approximately constant \citep{2026arXiv260303463R}. It also revealed a low-frequency quasi-periodic oscillation (QPO) near 4--5.5~Hz. These properties were consistent with an accretion flow in which a structured Comptonizing region continued to contribute substantially to the 2--8~keV emission \citep{2026arXiv260303463R}. The source subsequently evolved into a substantially less variable, disk-bright configuration while retaining a steep Comptonized tail. This, together with the disappearance of the type-C QPO detected during the preceding intermediate-state observation \citep{2026arXiv260303463R}, motivates its classification as a fading SPL state and allows \ixpe to revisit the same transient after a major reconfiguration of its accretion flow. 

Here, we present the first \ixpe observation of GS~1354$-$64 in a fading SPL state and compare its polarization with that measured previously during the intermediate state. In Section~2, we describe the \ixpe and contemporaneous \nustar observations, data reduction, and supporting X-ray monitoring. In Section~3, we establish the source's position within the 2025--2026 outburst, characterize its model-independent polarization properties, and perform spectral and spectropolarimetric analyses of the two distinct intensity tracks. In Section~4, we combine these results to investigate whether the broad troughs arise from changes in the thermal disk, the reslative disk--corona contribution, or intervening material. We compare their spectral and polarimetric morphology with dipping X-ray binaries and consider whether the recurrent transitions between luminosity configurations provide a phenomenological stellar-mass analog of changing-state behavior across the black-hole mass scale.

\section{Observations and Data Reduction}
\subsection{Observations}
\subsubsection{\maxi Monitoring}
We monitored the evolution of GS~1354$-$64 throughout its 2025--2026 outburst using per-orbit \maxi light curves in the 2--4, 4--10, and 10--20~keV bands. The selected interval extends from 2025 December 29 to 2026 April 10 (MJD~61038--61141), encompassing the source's transition into the SPL state.

\subsubsection{\ixpe}
\ixpe observed GS~1354$-$64 in the SPL state during 2026 March 28--April 09 (ObsID 05250399), providing an effective exposure of approximately 450~ks for each Detector Unit (DU). We follow the reduction procedure adopted for the earlier \ixpe observation of the source \citep{2026arXiv260303463R}. Source events are extracted from the Level~2 files using a circular region with a radius of 60\arcsec, positioned with the ``centroid'' tool in \texttt{SAOImage DS9} \citep{2003ASPC..295..489J}. Given the high source count rate of 2.32~$\mathrm{s^{-1},arcmin^{-2}}$ per DU, we apply neither background rejection nor background subtraction, following \citet{2023AJ....165..143D}. We adopt the time-dependent response version 20260101 and group the spectra into 200~eV bins, corresponding to an oversampling of the instrumental energy resolution by a factor of approximately 2--3.

\subsubsection{\nustar}

GS~1354$-$64 was observed by the Nuclear Spectroscopic Telescope Array (\nustar) on 2026~March~29 (ObsID 81202310002), contemporaneously with the \ixpe observation, as indicated by the gray shaded interval in Fig.~\ref{fig:1354_lc}. The observation provides a total effective exposure of $\sim21$~ks for each focal plane module (FPM). We produce cleaned event files using \texttt{nupipeline} with CALDB version 20260903 and adopt the standard screening criteria, including \texttt{saamode=optimized} and \texttt{tentacle=yes}. Source events are extracted from circular regions with radii of 120\arcsec, centered in \texttt{SAOImage DS9} as described above \citep{2003ASPC..295..489J}. For each detector, we select a 150\arcsec-radius background region from a source-free portion of the field. Spectra and response files are generated using \texttt{nuproducts}. We optimally group the FPMA and FPMB spectra following \citet{2016A&A...587A.151K}, additionally requiring at least 30 counts per bin, and analyze them over the 3--79~keV energy range. Because this observation was obtained after the development of the NuSTAR FPMB multilayer-insulation (MLI) tear, we included the energy-dependent FPMB correction using the multiplicative table model \texttt{nuMLIv1.mod}.\footnote{\url{https://nustarsoc.caltech.edu/NuSTAR_Public/NuSTAROperationSite/mli.php}}
 We first determined the FPMB-to-FPMA cross-normalization using energies above 7~keV, where the MLI correction is negligible, and then fixed this constant while fitting the FPMB MLI covering fraction over the full 3--79~keV band. The corresponding FPMA MLI covering fraction was fixed to unity.

\subsection{Data Reduction}

\subsubsection{Model-Independent Polarimetric Analyses}

We characterize the model-independent polarization using two complementary approaches. First, we apply the weighted PCUBE algorithm implemented in \texttt{ixpeobssim} version 33.0.0, adopting the SIMPLE weighting prescription\footnote{The weighted and unweighted PCUBE measurements are consistent within $1\sigma$.} \citep{2015APh....68...45K,2022SoftX..1901194B}. We additionally perform event-level Bayesian inference with \texttt{QUEEN-BEE} version 3.3, using the constant-polarization model \texttt{scout\_const} \citep{ravi2025queenbee,2026ApJ...997...60R}.

\section{Results}
\subsection{IXPE Observation}

\begin{figure}
    \centering
    \includegraphics[width=0.45\textwidth, trim=0 0 0 5, clip]{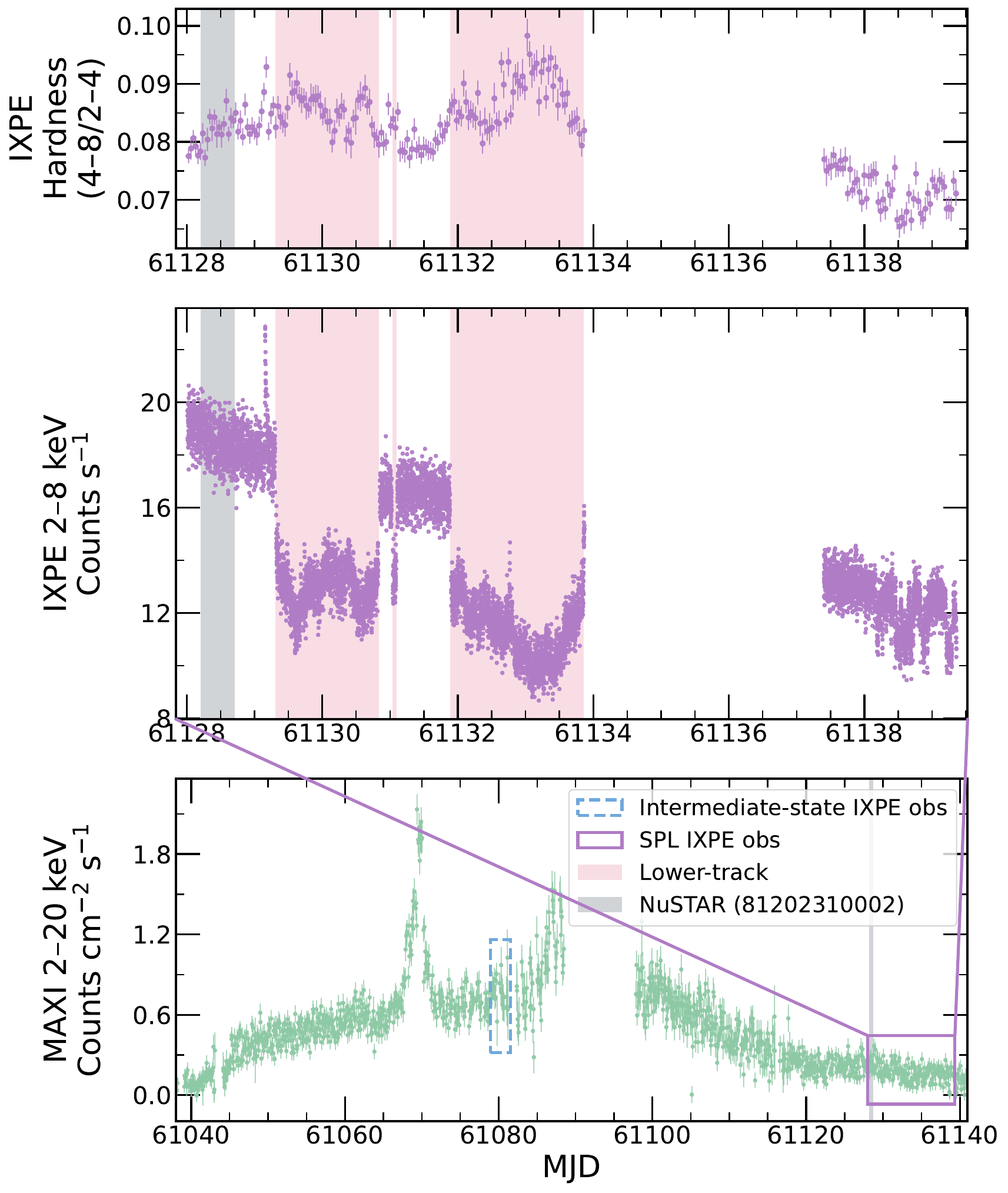}
    \vspace{-0.3em}
    \caption{\maxi/GSC light curve of GS~1354$-$64 during its 2025--2026 outburst (MJD~61038--61141). The dashed blue box in the 2--20~keV light curve depicts the intermediate-state \ixpe observation (05002101) reported in \cite{2026arXiv260303463R}, the solid purple box depicts the SPL \ixpe observation (05250399), and the gray shading denotes the contemporaneous \nustar observation (81202310002). The middle and top panels show the SPL \ixpe 2--8~keV light curve and 4--8/2--4~keV hardness ratio, respectively, calculated from the combined event lists of all three DUs and binned at 50~s and 3~ks, respectively. Pink shading marks the time intervals identified with the lower-track troughs.}
    \label{fig:1354_lc}
\end{figure}

\begin{figure}
    \centering
    \includegraphics[width=0.45\textwidth, trim=0 5 0 5, clip]{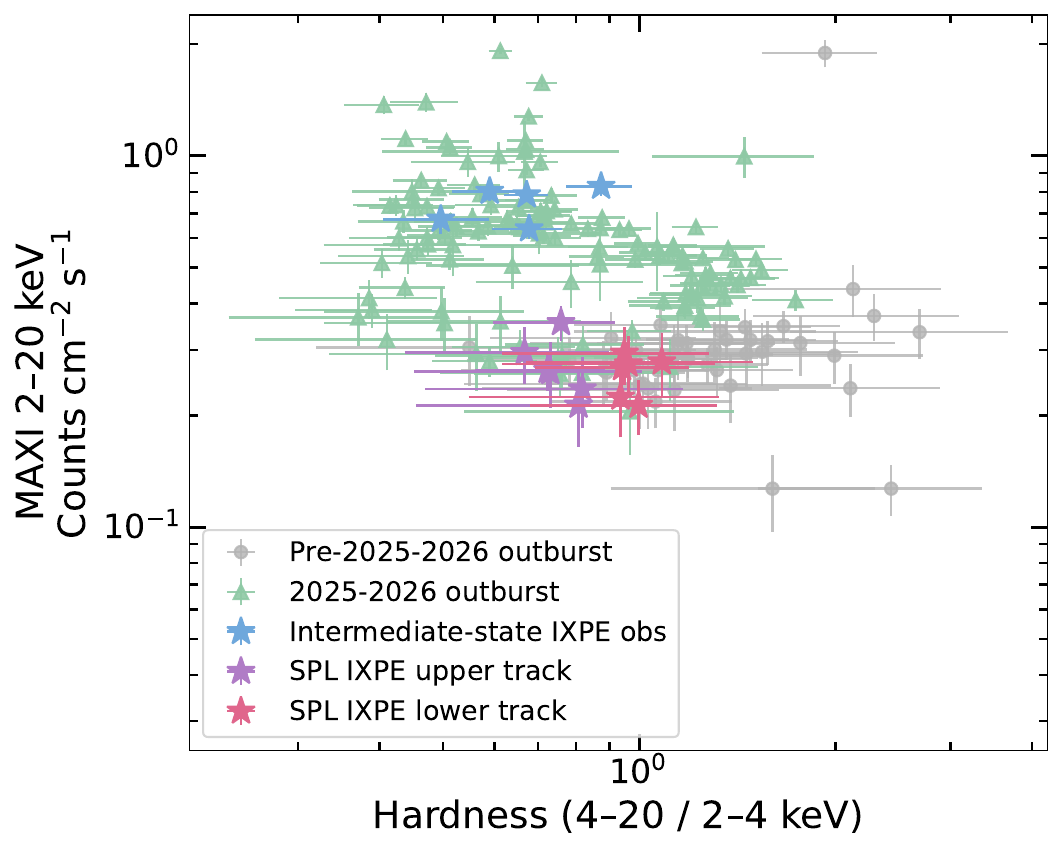}
    \vspace*{1em}
    \caption{\maxi/GSC HID of GS~1354$-$64 from MJD 55053 to 61141, using 12-hour bins. Green points illustrate the ongoing 2025--2026 outburst. Blue starred points correspond to the intermediate-state \ixpe observation (05002101) reported in \cite{2026arXiv260303463R}, while purple and pink starred points correspond to the upper and lower tracks of the SPL \ixpe observation (05250399), respectively. Data from MJD 58490--58570 are excluded to avoid contamination from the nearby transient MAXI~J1348$-$630.
    }
    \label{fig:1354_hid}
\end{figure}

\begin{figure}
    \centering
    \includegraphics[width=0.45\textwidth, trim=0 5 0 5, clip]{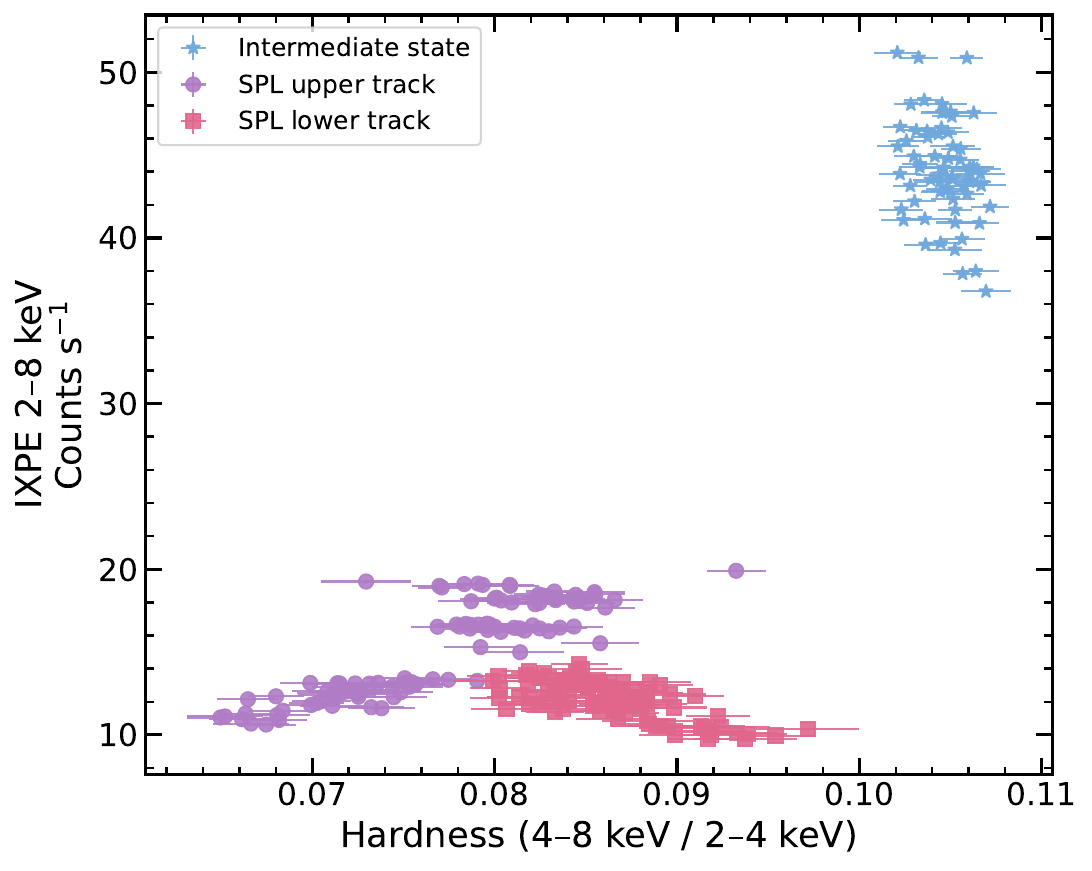}
    \vspace*{1em}
    \caption{\ixpe HID of GS~1354$-$64 during the intermediate- and SPL observations, using 3.5~ks bins. Hardness is defined as the ratio of the summed DU1--3 count rates in the 4--8 and 2--4~keV bands, while intensity is the summed DU1--3 count rate in the 2--8~keV band. Blue starred points correspond to the intermediate-state \ixpe observation (05002101) reported in \cite{2026arXiv260303463R}, while purple and pink points correspond to the upper and lower tracks of the SPL \ixpe observation (05250399), respectively.}
    \label{fig:1354_hid_ixpe}
\end{figure}

As shown in Fig.~\ref{fig:1354_lc}, the second \ixpe observation sampled the late decline of the 2025--2026 outburst, after the source had faded substantially from its earlier X-ray flares. During the observation, the \ixpe 2--8~keV count rate declined from approximately 20 to 10~counts~s$^{-1}$ while exhibiting pronounced, structured variability. Several broad troughs interrupt the otherwise comparatively smooth decline, appearing to transition between quasi-stable flux levels. These troughs are accompanied by increases in the 4--8~keV$/$2--4~keV hardness ratio. We therefore divide the \ixpe observation into two tracks: the upper track consists of ``declining-continuum'' intervals, and the lower track of ``trough'' intervals. The lower-track intervals are highlighted in pink in Fig.~\ref{fig:1354_lc}. The \maxi/GSC HID in Fig.~\ref{fig:1354_hid} is suggestive of a separation between the two track selections, with the \ixpe HID in Fig.~\ref{fig:1354_hid_ixpe} resolving them into distinct regions of hardness--intensity space separate from both each other and the previous intermediate-state \ixpe observation.

We examined each nominal trough boundary using the summed 2--8~keV light curves from the three \ixpe DUs. The light curves were constructed independently with 50- and 100-s bins, and Poisson uncertainties were calculated from the number of events in each bin. Within a \(\pm0.03\)-day interval around each boundary, we fitted a logistic transition in which the low- and high-state count rates, transition midpoint, and characteristic timescale \(\tau\) were allowed to vary. We defined the transition duration as the time required for the fitted count rate to change between 10\% and 90\% of its full amplitude, \(t_{10\text{--}90}=2\ln(9)\tau\), with uncertainties propagated from the fit covariance. Most boundaries coincide with gaps in the \ixpe coverage and are therefore unresolved or only weakly constrained. The best-sampled boundary is the egress of the second broad trough, for which the nearest bracketing bins are separated by only 50 and 100~s in the two light curves. The 50- and 100-s fits give \(t_{10\text{--}90}=0.3\pm0.1\) and \(0.6\pm0.2\)~ks, respectively, with transition midpoints agreeing to within approximately 25~s. The precise duration is therefore sensitive to the adopted binning, but both fits indicate that the egress occurred within approximately \(5\)--\(11\) minutes, more than two orders of magnitude faster than the \(\sim10^{5}\)-s trough duration.

The presence of substantial thermal-disk emission together with a steep Comptonized tail in the spectrum, combined with the marked reduction in broadband rms variability relative to the preceding intermediate-state \ixpe observation (2026 February~07--09), supports the interpretation that GS~1354$-$64 had entered a fading SPL state (see Sect.~\ref{subsec:spectroscopy},\ref{subsec:specpol}); the concurrent disappearance of the type-C QPO provides additional evidence for this transition. Formally, the Comptonized component contributes \(\sim66\%\) of the unabsorbed 2--20~keV continuum, exceeding the \(50\%\) threshold used to identify the SPL state when no QPO is detected, per \citealt{2006ARA&A..44...49R}. While the time-averaged photon index, \(\Gamma\simeq2.3\), lies slightly below the conventional \(\Gamma>2.4\) criterion for SPL, the spectrum steepens to \(\Gamma=2.8\pm0.3\) during the final 100~ks of the observation, formally crossing this threshold and motivating our interpretation of the source as evolving into a fading SPL state in this work.

\subsection{Polarimetry}

\begin{figure*}
    \centering
    
    \includegraphics[width=0.49\textwidth, trim=0 15 0 5, clip]{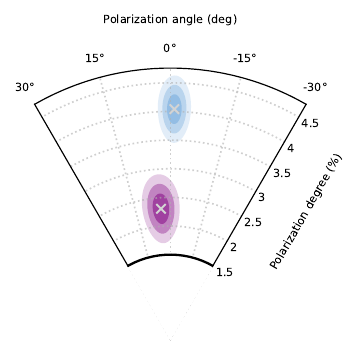}
    \hfill
    \includegraphics[width=0.49\textwidth]{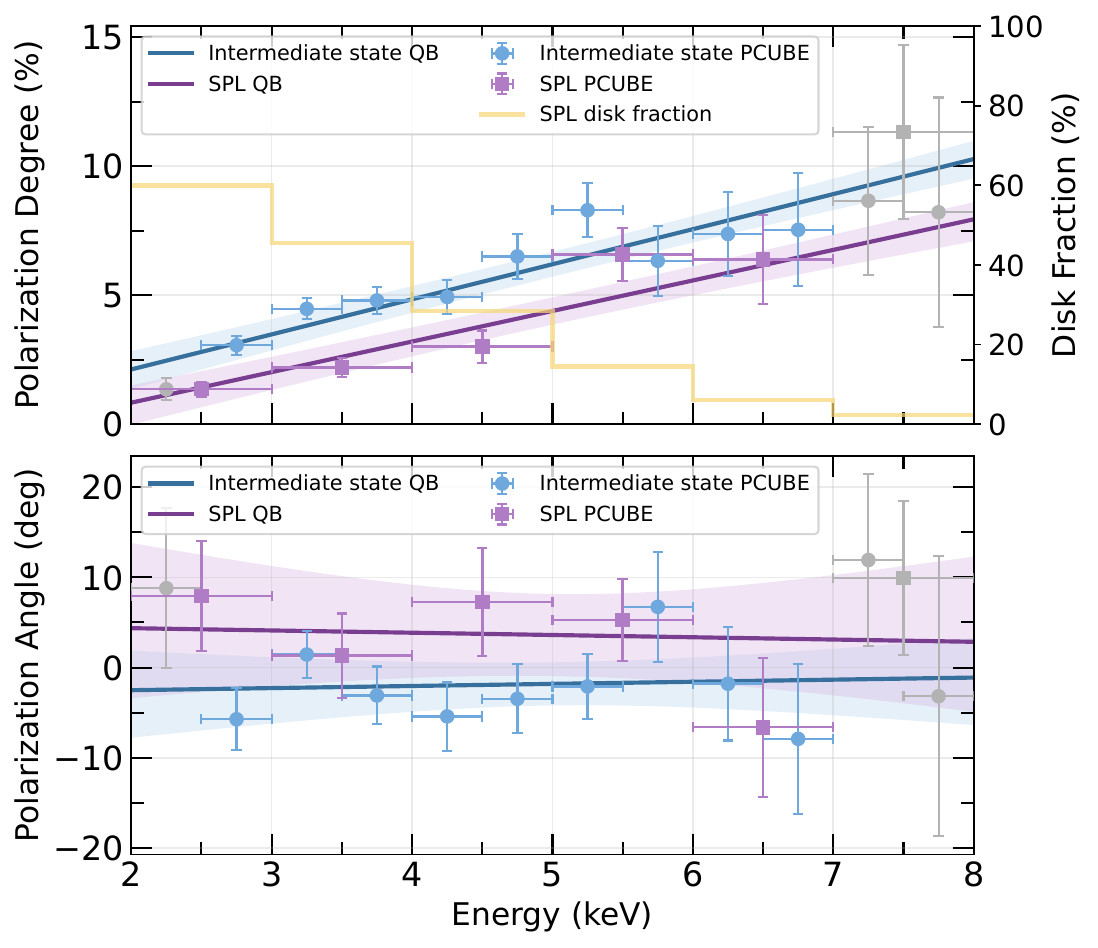}

    \caption{
    Polarimetric summary of both GS~1354$-$64 \ixpe\ observations.
    \textbf{Left:} Constant-PA (\texttt{scout\_const}) QUEEN-BEE polar plot. The intermediate-state observation (05002101) is colored blue and the full SPL observation (05250399) is colored purple. Contours represent $68\%, 95\%,~\mathrm{and}~99.7\%$ credible intervals.  
    \textbf{Right:} Energy-dependent PD and PA behavior of intermediate-state \ixpe data (05002101) in blue and full SPL \ixpe data (05250399) in purple. Binned PCUBE results are shown as individual points. Error bars denote $1\sigma$ uncertainties. Points in grey denote bins with $\textless3\sigma$ significance. Unbinned QUEEN-BEE results are shown as solid lines. Shading denotes asymmetric 90\% credible intervals. The corresponding absorbed disk fraction---defined as the ratio of disk flux to total (disk + Comptonized) flux---in each 1-keV bin across the 2--8~keV band, is overplotted in yellow.}
    \label{fig:1354_polar}
\end{figure*}

We detect significant 2--8~keV polarization from the source during the SPL \ixpe observation using both model-independent polarimetric approaches. PCUBE measures a PD of $2.3\pm0.2\%$ and a PA of $5\degr\pm3\degr$ ($1\sigma$ uncertainties), corresponding to a significance of $9.7\sigma$. The QUEEN-BEE \texttt{scout\_const} analysis  (Fig.~\ref{fig:1354_polar}) returns polarization parameters consistent with the PCUBE results, with a PD of $2.3^{+0.2}_{-0.3}\%$ and a PA of $4\degr\pm3\degr$ ($90\%$ credible intervals). Comparison of the constant-polarization model with the null unpolarized hypothesis yields a log Bayes factor of $\Delta\ln(Z)=108\pm1$.

The energy-binned PCUBE analysis (right panel of Fig.~\ref{fig:1354_polar}) reveals a significant increase in PD with energy, while the PA remains constant across the \ixpe band. Consistent with this result, the QUEEN-BEE \texttt{scout\_energy\_evol} model finds decisive evidence, per the modified Jeffreys scale \citep{Jeffreys1986}, for PD energy dependence at a rate of $\mathcal{G}_E=1.2^{+0.1}_{-0.2}~\%/\mathrm{keV}$. Comparing this model, which assumes a linear increase in PD with energy across the \ixpe bandpass, with the constant-polarization model yields $\Delta\ln(Z)=28\pm1$.

In contrast, the QUEEN-BEE \texttt{scout\_energy\_rot} analysis finds no evidence for energy-dependent PA rotation. The comparison with the constant-PA model yields $\Delta\ln(Z)=-4\pm1$. The inferred rotation rate is $\omega_E=-0.3^{+2.7}_{-2.1}{}^\circ~\mathrm{keV}^{-1}$, consistent with zero (Ravi et al., in prep.). We likewise find no significant temporal rotation of the PA using the QUEEN-BEE \texttt{scout\_rot} analysis. The time-rotating and constant-PA models yield $\Delta\ln(Z)=-3\pm1$. 

\begin{figure}
    \centering
    \includegraphics[width=0.45\textwidth, trim=0 5 0 5, clip]{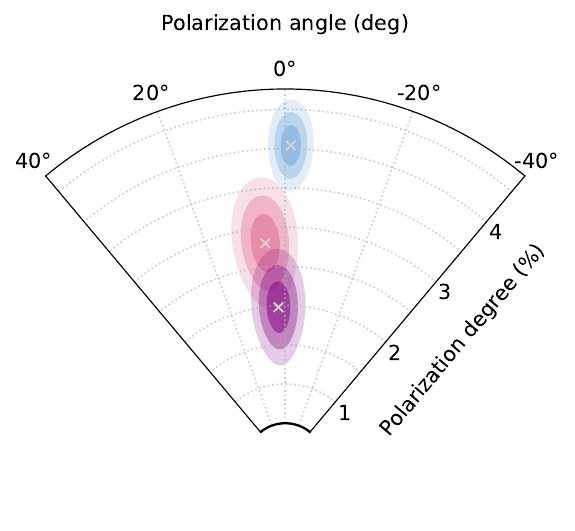}
    \vspace{-0.7em}
    \caption{Constant-PA (\texttt{scout\_const}) QUEEN-BEE polar plot. The intermediate-state observation (05002101) is colored blue. The upper-track SPL observation (05250399) is colored purple while the lower track SPL observation is colored pink. Contours represent $68\%, 95\%,~\mathrm{and}~99.7\%$ credible intervals.  }\label{fig:1354_polar_tracks}
\end{figure}

We also performed track-resolved QUEEN-BEE analyses of the upper and lower tracks (Fig.~\ref{fig:1354_polar_tracks}). For the upper track, the \texttt{scout\_const} analysis gives \({\rm PD}=2.0\pm0.3\%\) and \({\rm PA}=3^{+5}_{-4}{}^\circ\), while the lower-track analysis gives \({\rm PD}=2.8\pm0.4\%\) and \({\rm PA}=5^{+5}_{-4}{}^\circ\). The constant-polarization model is decisively favored over the unpolarized hypothesis in both intervals, with \(\Delta\ln(Z)=40.9\pm0.8\) and \(59.7\pm0.5\) for the upper and lower tracks, respectively.

\begin{figure}
    \centering
    \includegraphics[width=\columnwidth]{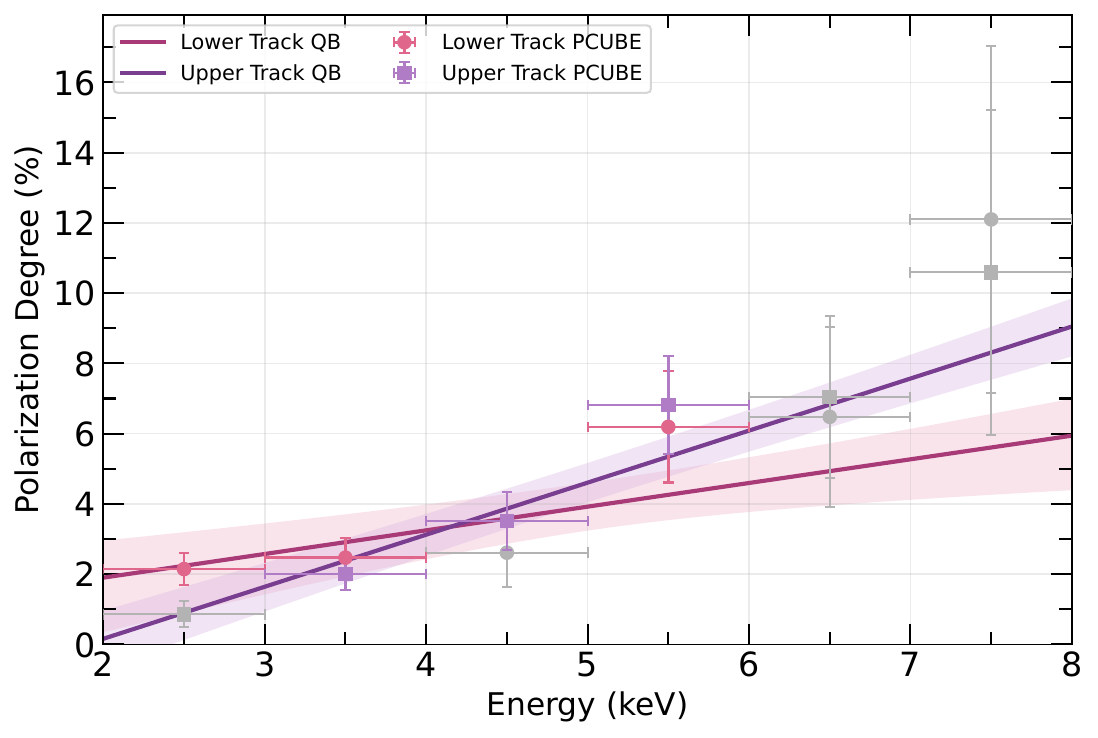}
    \caption{Energy-dependent PD for the upper and lower tracks of the SPL \ixpe observation. The purple squares and pink circles show the energy-binned PCUBE measurements for the upper and lower tracks, respectively; measurements below \(3\sigma\) significance are shown in gray. The solid curves show the corresponding QUEEN-BEE \texttt{scout\_energy\_evol} results, with shaded regions denoting the asymmetric 90\% credible intervals.}
    \label{fig:1354_track_pd_energy}
\end{figure}

The track-resolved \texttt{scout\_energy\_evol} analyses give \(\mathcal{G}_{E,\rm upper}=1.5\pm0.2\%~\mathrm{keV}^{-1}\) and \(\mathcal{G}_{E,\rm lower}=0.7^{+0.3}_{-0.5}\%~\mathrm{keV}^{-1}\), with log Bayes factors of \(\Delta\ln Z=26\pm1\) and \(1.2\pm0.9\), respectively, relative to the corresponding constant-polarization models. The \texttt{scout\_energy\_rot} analyses similarly give \(\omega_{E,\rm upper}=3\pm4{}^\circ~\mathrm{keV}^{-1}\) and \(\omega_{E,\rm lower}=-2^{+4}_{-3}{}^\circ~\mathrm{keV}^{-1}\), with \(\Delta\ln Z=-4.3\pm0.7\) and \(-4.5\pm0.6\), respectively, relative to the constant-PA models. Finally, the \texttt{scout\_time\_rot} analyses give \(\omega_{t,\rm upper}=-2\pm1{}^\circ~\mathrm{day}^{-1}\) and \(\omega_{t,\rm lower}=2\pm3{}^\circ~\mathrm{day}^{-1}\), with \(\Delta\ln Z=-1.6\pm0.6\) and \(-4.0\pm0.6\), respectively. Thus, energy-dependent PD is strongly favored only for the upper track (Fig.~\ref{fig:1354_track_pd_energy}), while neither track shows evidence for energy- or time-dependent PA rotation.

The PD of each SPL track differs from that measured during the intermediate-state observation \citep{2026arXiv260303463R} at greater than \(99\%\) posterior credibility (Fig.~\ref{fig:1354_polar_tracks}). In contrast, the upper- and lower-track polarization posteriors overlap substantially: although the lower track has a marginally higher best-fitting PD, the two tracks are not statistically distinct from one another. Their PAs are likewise consistent with both each other and the intermediate-state observation.

\subsection{Spectroscopy}
\label{subsec:spectroscopy}
\subsubsection{NuSTAR spectral fitting}

We first characterized the broadband continuum using the simultaneous \nustar observation 81202310002. The FPMA and FPMB spectra were fit in the 3--79~keV band with the model \textsc{constant*\allowbreak nuMLI*\allowbreak TBabs(diskbb+\allowbreak nthComp*\allowbreak smedge+\allowbreak laor)} using \texttt{XSPEC} version 12.14.0h \citep{1996ASPC..101...17A}. Interstellar absorption was modeled with \textsc{TBabs}, adopting the abundances of \citet{2000ApJ...542..914W} and the photoelectric cross sections of \citet{1996ApJ...465..487V}. The hydrogen column density was frozen at $N_{\rm H}=7\times10^{21}~{\rm cm^{-2}}$, consistent with the value adopted in the analysis of the preceding intermediate-state observation \citep{2026arXiv260303463R}. The \textsc{nuMLI} covering fraction was fixed to unity for FPMA and allowed to vary for FPMB. The resulting 3--79~keV spectra had net count rates of 35.90 and 33.08~counts~s$^{-1}$ for FPMA and FPMB, respectively, yielding $\sim1.43\times10^{6}$ net source counts across the two FPMs.

The \textsc{diskbb} component \citep{1984PASJ...36..741M} describes the multitemperature thermal emission from the accretion disk, while \textsc{nthComp} \citep{1999MNRAS.309..561Z} represents thermal Comptonization of the disk photons. The seed-photon temperature of \textsc{nthComp} was tied to the \textsc{diskbb} temperature, a disk seed spectrum was assumed by fixing $\textsc{inp\_type}=1$, and the redshift was fixed at zero. The electron temperature was unconstrained and reached the upper boundary of the allowed range, $kT_{\rm e}=1000$~keV, which we retained in the subsequent \ixpe analysis. This boundary value should not be interpreted as a physical measurement of the coronal temperature: at relativistic electron temperatures, the approximations underlying \textsc{nthComp} no longer provide an accurate thermal-Comptonization spectrum and can produce substantial discrepancies from more exact calculations \citep{2019MNRAS.485.2942N}. Moreover, an observed power-law-like tail in this regime may more plausibly arise from Comptonization by a nonthermal or hybrid electron distribution \citep{2009ApJ...690L..97P}. We therefore use \textsc{nthComp} only as a phenomenological description of the continuum within the observed bandpass and do not assign physical significance to the fitted electron temperature or other inferred coronal plasma properties. 

The FPMA multiplicative constant was fixed to unity. We determined the FPMB-to-FPMA cross-normalization from a fit excluding the FPMB data below 7~keV, obtaining \(C_{\rm B}=1.00163\), and subsequently froze this value while fitting the full band with the MLI correction. The resulting FPMB MLI covering fraction was \(f_{\rm MLI,B}=0.89\pm0.02\). The best-fitting parameters are reported in Table~\ref{tab:spectral_parameters}. The corrected \nustar continuum was characterized by a disk temperature of \(kT_{\rm in}=0.804^{+0.009}_{-0.010}\)~keV and a photon index of \(\Gamma=2.32\pm0.01\). All reported spectral parameter uncertainties correspond to 90\% confidence.

Following the treatment adopted for the preceding intermediate-state \ixpe-\nustar joint observation \citep{2026arXiv260303463R},  we included \textsc{laor} \citep{1991ApJ...376...90L} to describe the broad Fe-K emission-line feature and \textsc{smedge} \citep{1991PhDT........55E} to describe the associated absorption-edge-like curvature. For \textsc{laor}, we adopted the default maximal-spin Kerr geometry $a_\ast=0.998$, corresponding to the Thorne limit, fixed the line energy at 6.7~keV for Fe~\textsc{xxv}, and assumed an emissivity index of $q=3$. The inner and outer radii were fixed at $1.235\,GM/c^2$ (as expected in the maximal spin case, motivated by \citealt{2026arXiv260306883L}) and $400\,GM/c^2$, respectively.\footnote{We also tested fixing the inner radius of the phenomenological \textsc{laor} component at $6\,GM/c^2$ as expected in the low-spin case. This choice changed the NuSTAR fit statistic by only $\Delta\chi^2=2.17$ and the preferred simultaneous track fit by less than one in $\chi^2$, without affecting any of the principal spectral or polarimetric conclusions.} For \textsc{smedge}, the edge width and index were fixed at 7~keV and $-2.67$, respectively. The remaining Fe-K parameters were allowed to vary, and their best-fitting values are listed in Table~\ref{tab:spectral_parameters}. These components provide a consistent phenomenological description of the Fe-K structure but should not be interpreted as a self-consistent reflection model.

\begin{figure*}[t]
    \centering
    \includegraphics[width=0.82\textwidth]
    {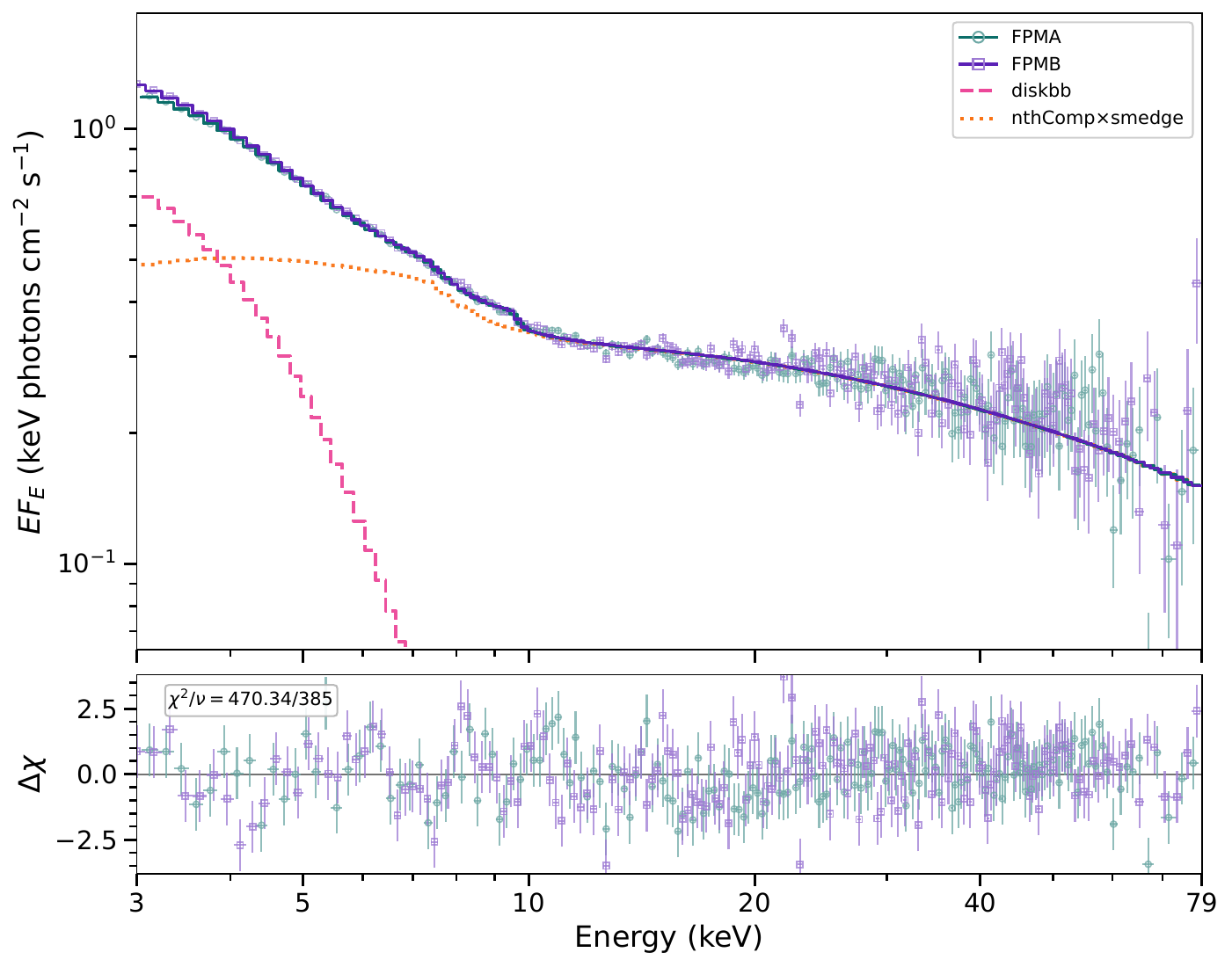}
    \caption{
    Simultaneous fit to the NuSTAR FPMA and FPMB spectra over 3--79~keV. The upper panel shows the unfolded spectra, total model, and the \textsc{diskbb} and \textsc{nthComp*smedge} continuum contributions. The weak \textsc{laor} component is included in the total model but falls below the displayed range. The lower panel
    shows the residuals in units of the measurement uncertainty. The phenomenological model describes the overall broadband continuum and Fe-K structure but leaves some statistically significant residual structure.
    }
    \label{fig:nustar_spectrum}
\end{figure*}

The best-fitting NuSTAR spectrum and residuals are shown in Fig.~\ref{fig:nustar_spectrum}. Using statistical uncertainties alone, the MLI corrected fit gave \(\chi^2/\nu=470.34/385\), indicating that some residual structure remains.\footnote{Although the MLI correction reduces the low-energy differences between FPMA and FPMB, residual discrepancies remain below approximately 5~keV. The full-band fit is formally unacceptable ($\chi^2/\mathrm{dof}=470.34/385$; $p=1.88\times10^{-3}$), whereas restricting the fit to energies above 5~keV yields an acceptable fit ($\chi^2/\mathrm{dof}=393.07/359$; $p=0.104$). We therefore attribute much of the poor full-band statistic to residual low-energy calibration and/or modeling uncertainties.} We nevertheless use this fit to establish a consistent phenomenological treatment of the disk and Compton continuum and Fe-K emission for the subsequent \ixpe analysis, rather than as a fully optimized physical description of the spectrum. Adding an energy-independent systematic uncertainty of \(1.05\%\) per spectral bin reduces the fit statistic to \(\chi^2/\nu=383.21/385\), indicating that the adopted phenomenological model reproduces the NuSTAR spectral shape to approximately the percent level.

\subsubsection{Time-averaged IXPE intensity spectrum}

We next fit the time-averaged Stokes-$I$ spectra from the three \ixpe detector units over the 2--8~keV band using the same spectral model. Starting from the \nustar solution, we thawed the \textsc{diskbb} temperature and normalization, the \textsc{nthComp} photon index, and the \textsc{nthComp} normalization. These parameters were refit because the source continuum need not remain identical between the \nustar and \ixpe spectra, particularly across the two identified light-curve tracks highlighted in Figure~\ref{fig:1354_lc}, and because the time-averaged \ixpe intensity spectrum provides the initial continuum decomposition and detector calibration for the track-resolved analysis. The hydrogen column density and electron temperature remained fixed at $N_{\rm H}=7\times10^{21}~{\rm cm^{-2}}$ and $kT_{\rm e}=1000$~keV. The \textsc{smedge} and \textsc{laor} parameters were frozen at the values obtained from the simultaneous \nustar spectral fitting.

The spectral-shape parameters were tied across the three detector units, and multiplicative constants accounted for their relative normalizations. The DU1 constant was fixed to unity, while the fitted DU2 and DU3 constants were \(1.002^{+0.006}_{-0.007}\) and \(0.979^{+0.007}_{-0.006}\), respectively.

We first performed a time-averaged calibration fit in which independent linear gain corrections were allowed for the three detector units. The resulting gain slopes were 0.9891, 0.9596, and 0.9963 for DU1, DU2, and DU3, respectively, with corresponding offsets of 0.0446, 0.1086, and 0.0406~keV. To avoid degeneracy between the absolute gain scale and the continuum parameters, DU1 was subsequently adopted as the reference detector and fixed at its calibration-fit values, while only the DU2 and DU3 gain corrections were allowed to refine. The final gain slopes were $0.960\pm0.002$ for DU2 and $0.996\pm0.003$ for DU3, with corresponding offsets of $0.109\pm0.009$ and $0.04\pm0.01$~keV, respectively. These values were frozen in the track-resolved spectral and spectropolarimetric fits. The gain corrections account for residual relative energy-scale differences among the \ixpe detector units and do not materially affect the inferred continuum.

The best-fitting time-averaged \ixpe spectral parameters are listed in Table~\ref{tab:spectral_parameters}. The spectrum is characterized by a disk temperature of \(kT_{\rm in}=0.68\pm0.01\)~keV and a photon index of \(\Gamma=2.3\pm0.1\). The fit gave \(\chi^2/\nu=107.98/77\). Because the time-averaged spectrum combines intervals with substantial intensity and spectral variability, this likely contributes to the remaining residuals and fit statistic. We nevertheless retain this phenomenological model because its primary purpose is to establish the intensity decomposition and detector calibration used in the subsequent spectropolarimetric analysis. %while avoiding a physical interpretation of small residual features.

To quantify the secular decline across the observation, we additionally fit the first and final 100~ks intervals using the same model, with the Fe-K and detector-gain parameters fixed as described above. During the first 100~ks, the spectrum had \(kT_{\rm in}=0.73\pm0.02\)~keV and \(\Gamma=2.3\pm0.3\), with \(\chi^2/\nu=61.69/81\). During the final 100~ks, the disk cooled to \(kT_{\rm in}=0.65\pm0.02\)~keV while the Comptonized continuum steepened to \(\Gamma=2.8\pm0.3\), with \(\chi^2/\nu=91.08/81\). Extrapolating the fitted models over 0.01--100~keV and removing foreground absorption gives model-dependent bolometric fluxes of \(5.37\times10^{-9}\) and \(3.42\times10^{-9}~{\rm erg~cm^{-2}~s^{-1}}\), respectively, corresponding to a decline of approximately \(36\%\).

\subsubsection{Track-resolved spectroscopy}

We divided the observation into the same upper and lower tracks identified in Fig.~\ref{fig:1354_lc} and initially fit their Stokes-$I$ spectra separately. For both tracks, the \textsc{diskbb} temperature and normalization and the \textsc{nthComp} photon index and normalization were free to vary. The absorption column, electron temperature, Fe-K parameters, and detector gain corrections were fixed as described above. The similarity between the trough separation and the binary orbital period motivates consideration of absorption as a possible contributor. We subsequently test whether allowing the neutral absorption column to differ between the upper and lower tracks, either alone or together with changes in the continuum normalizations, can explain their spectral differences.

The separate fits by track, reported in Table~\ref{tab:spectral_parameters}, show that the upper-track spectrum has a higher disk temperature than the lower track, \(0.70\pm0.01\) versus \(0.63\pm0.01\)~keV. The upper-track fit gave \(\chi^2/\nu=81.55/81\), while the lower-track fit gave \(107.59/81\), with the largest contribution in the latter arising from DU3 (\(\chi^2=50.07\) for 29 bins). Relative to the lower track, the upper track has an approximately \(12\%\) higher disk temperature and \(11\%\) higher apparent disk normalization, whereas the independently fitted photon indices differ by less than 0.03 and the \textsc{nthComp} normalizations by less than \(1\%\). We tested this interpretation more systematically through joint fitting.

\begin{deluxetable*}{llcccc}
\tabletypesize{\scriptsize}
\tablewidth{0pt}
\tablecaption{
Spectral parameters obtained from the NuSTAR, time-averaged IXPE, and independently fitted track-resolved spectra. Errors are reported at the $90\%$ confidence level.
\label{tab:spectral_parameters}
}

\tablehead{
\colhead{Component} &
\colhead{Parameter} &
\colhead{NuSTAR} &
\colhead{IXPE average} &
\colhead{Upper track} &
\colhead{Lower track}
}

\startdata
\textsc{constant}
& FPMA or DU1
& \(1^{f}\)
& \(1^{f}\)
& \(1^{f}\)
& \(1^{f}\)
\\
& FPMB/FPMA
& \(1.00163^{f}\)
& \nodata
& \nodata
& \nodata
\\
& DU2/DU1
& \nodata
& \(1.002^{+0.006}_{-0.007}\)
& \(1.004\pm0.002\)
& \(0.999\pm0.003\)
\\
& DU3/DU1
& \nodata
& \(0.979^{+0.007}_{-0.006}\)
& \(0.977\pm0.002\)
& \(0.982\pm0.003\)
\\
\hline
\textsc{nuMLI}
& FPMA covering fraction
& \(1^{f}\)
& \nodata
& \nodata
& \nodata
\\
& FPMB covering fraction
& \(0.89\pm0.02\)
& \nodata
& \nodata
& \nodata
\\
\\
\hline
\textsc{TBabs}
& \(N_{\rm H}\) (\(10^{22}\ {\rm cm^{-2}}\))
& \(0.7^{f}\)
& \(0.7^{f}\)
& \(0.7^{f}\)
& \(0.7^{f}\)
\\
\hline
\textsc{diskbb}
& \(kT_{\rm in}\) (keV)
& \(0.804^{+0.009}_{-0.010}\)
& \(0.68\pm0.01\)
& \(0.70\pm0.01\)
& \(0.63\pm0.01\)
\\
& Normalization
& \(350\pm20\)
& \(470\pm10\)
& \(252\pm5\)
& \(228\pm7\)
\\
\hline
\textsc{nthComp}
& \(\Gamma\)
& \(2.32\pm0.01\)
& \(2.3\pm0.1\)
& \(2.4\pm0.2\)
& \(2.4\pm0.1\)
\\
& \(kT_{\rm e}\) (keV)
& \(1000^{f}\)
& \(1000^{f}\)
& \(1000^{f}\)
& \(1000^{f}\)
\\
& \(kT_{\rm bb}\) (keV)
& \(=kT_{\rm in}\)
& \(=kT_{\rm in}\)
& \(=kT_{\rm in}\)
& \(=kT_{\rm in}\)
\\
& Seed type
& \(1^{f}\)
& \(1^{f}\)
& \(1^{f}\)
& \(1^{f}\)
\\
& Normalization
& \(0.267^{+0.010}_{-0.009}\)
& \(0.22\pm0.03\)
& \(0.12\pm0.02\)
& \(0.12^{+0.02}_{-0.01}\)
\\
\hline
\textsc{smedge}
& \(E_{\rm edge}\) (keV)
& \(7.35^{+0.10}_{-0.09}\)
& \(7.35^{f}\)
& \(7.35^{f}\)
& \(7.35^{f}\)
\\
& \(\tau_{\rm max}\)
& \(1.26^{+0.08}_{-0.07}\)
& \(1.26^{f}\)
& \(1.26^{f}\)
& \(1.26^{f}\)
\\
& Index
& \(-2.67^{f}\)
& \(-2.67^{f}\)
& \(-2.67^{f}\)
& \(-2.67^{f}\)
\\
& Width (keV)
& \(7^{f}\)
& \(7^{f}\)
& \(7^{f}\)
& \(7^{f}\)
\\
\hline
\textsc{laor}
& \(E_{\rm line}\) (keV)
& \(6.7^{f}\)
& \(6.7^{f}\)
& \(6.7^{f}\)
& \(6.7^{f}\)
\\
& Emissivity index
& \(3^{f}\)
& \(3^{f}\)
& \(3^{f}\)
& \(3^{f}\)
\\
& \(R_{\rm in}\) (\(GM/c^2\))
& \(1.235^{f}\)
& \(1.235^{f}\)
& \(1.235^{f}\)
& \(1.235^{f}\)
\\
& \(R_{\rm out}\) (\(GM/c^2\))
& \(400^{f}\)
& \(400^{f}\)
& \(400^{f}\)
& \(400^{f}\)
\\
& Inclination (deg)
& \(81^{+1}_{-2}\)
& \(81^{f}\)
& \(81^{f}\)
& \(81^{f}\)
\\
& Normalization ($10^{-3}$)
& \(1.0\pm0.2\)
& \(1.0^{f}\)
& \(1.0^{f}\)
& \(1.0^{f}\)
\\
\hline
Gain, DU1
& Slope
& \nodata
& \(0.989130^{f}\)
& \(0.989130^{f}\)
& \(0.989130^{f}\)
\\
& Offset (keV)
& \nodata
& \(0.0445967^{f}\)
& \(0.0445967^{f}\)
& \(0.0445967^{f}\)
\\
Gain, DU2
& Slope
& \nodata
& \(0.960\pm0.002\)
& \(0.959655^{f}\)
& \(0.959655^{f}\)
\\
& Offset (keV)
& \nodata
& \(0.109\pm0.009\)
& \(0.109062^{f}\)
& \(0.109062^{f}\)
\\
Gain, DU3
& Slope
& \nodata
& \(0.996\pm0.003\)
& \(0.996152^{f}\)
& \(0.996152^{f}\)
\\
& Offset (keV)
& \nodata
& \(0.04\pm0.01\)
& \(0.042862^{f}\)
& \(0.042862^{f}\)
\\
\hline
\textsc{diskbb}
& \(F_{2-8}\) (\({\rm erg\,cm^{-2}\,s^{-1}}\))
& \(1.006\times10^{-9}\)
& \(5.060\times10^{-10}\)
& \(3.379\times10^{-10}\)
& \(1.558\times10^{-10}\)
\\
\textsc{nthComp}
& \(F_{2-8}\) (\({\rm erg\,cm^{-2}\,s^{-1}}\))
& \(1.040\times10^{-9}\)
& \(7.109\times10^{-10}\)
& \(3.767\times10^{-10}\)
& \(3.401\times10^{-10}\)
\\
Total
& \(F_{2-8}\) (\({\rm erg\,cm^{-2}\,s^{-1}}\))
& \(2.051\times10^{-9}\)
& \(1.222\times10^{-9}\)
& \(7.201\times10^{-10}\)
& \(5.014\times10^{-10}\)
\\
Disk fraction
& \(F_{\rm disk}/(F_{\rm disk}+F_{\rm nthComp})\)
& \(0.492\)
& \(0.416\)
& \(0.473\)
& \(0.314\)
\\
\hline
\multicolumn{2}{l}{\(\chi^2/\nu\)}
& $470.34/385$
& $107.98/77$
& $81.55/81$
& $107.59/81$
\\
% \multicolumn{2}{l}{Null-hypothesis probability}
% & \(2.28\times10^{-5}\)
% & \(6.74\times10^{-3}\)
% & 0.472
% & 0.024
% \\
\enddata
\tablecomments{
The upper- and lower-track columns show the results obtained when the two tracks were fit independently. Superscripts \(f\) and \(p\) denote parameters that were fixed or pegged at a fitting boundary, respectively. The IXPE gain corrections were fit using the time-averaged spectrum and then frozen in the independently fitted track-resolved spectra. The independently fitted values are reported as a diagnostic. The spectral decomposition subsequently frozen in the spectropolarimetric analysis was taken from Case 3 of Table~\ref{tab:joint_spectral_tests}. The reported component and total fluxes are absorbed 2--8~keV model fluxes. The disk fraction is defined over 2--8\;keV from the absorbed fluxes as $F_{\rm disk}/(F_{\rm disk}+F_{\rm nthComp})$ and excludes the weak \textsc{laor} contribution. All three IXPE fits contain 87 spectral bins; the track-resolved fits have four additional degrees of freedom because the DU2 and DU3 gain slopes and offsets were fixed to the values obtained from the time-averaged spectrum.
}

\end{deluxetable*}

To determine which of these parameters were statistically required to differ, we next simultaneously fitted the six Stokes-$I$ spectra from the upper and lower tracks, comprising three DU spectra per track, using a sequence of nested linking schemes. In every test, the three detector spectra within a given track shared the same spectral parameters. The DU2 and DU3 cross-normalizations were allowed to vary but were tied between the two tracks. The detector gain corrections were frozen at the calibration values determined from the initial time-averaged \ixpe fit. For the preferred fit described below (Case~3 in Table~\ref{tab:joint_spectral_tests}), in which the disk temperature and normalization differ between tracks while the Comptonization parameters remain tied, the resulting DU2 and DU3 cross-normalizations were \(1.002\pm0.002\) and \(0.979\pm0.002\), respectively. The absorption column, electron temperature, seed type, redshift, \textsc{smedge} parameters, and \textsc{laor} parameters remained fixed at the values described above.\footnote{As a check on spectral evolution \textit{within} the upper track, we fitted its four distinct intervals jointly, allowing the disk temperature, disk normalization, and Comptonization normalization to vary. From the first to the last interval, $kT_{\rm in}$ decreased from $0.728$ to $0.677$~keV, the \textsc{diskbb} normalization increased modestly from $449$ to $495$, and the \textsc{nthComp} normalization decreased from $0.273$ to $0.157$. Allowing the photon index to vary independently improved the fit by only $\Delta\chi^{2}=3.12$ for three additional parameters ($p\simeq0.37$), providing no significant evidence for spectral-slope evolution. Thus, the gradual cooling and fading within the upper track do not drive the spectral differences between the two tracks or affect the principal track-resolved conclusions.} Before examining changes in the intrinsic continuum, we tested an absorption-only Case~0 in which all continuum parameters were tied between the two tracks. Allowing a single common column density to vary gave \(\chi^{2}/\nu=189729.7/167\). Allowing \(N_{\rm H}\) to differ between the tracks reduced the statistic to \(28403.9/166\), with nominal best-fitting values of \(N_{\rm H,upper}=2.4\times10^{22}\ {\rm cm^{-2}}\) and \(N_{\rm H,lower}=4.6\times10^{22}\ {\rm cm^{-2}}\). Despite this improvement, the fit remained grossly unacceptable and drove the tied continuum to extreme compensating values. Allowing the absorption column and both component normalizations to differ between tracks, while retaining a common disk temperature and photon index, gave $\chi^2/\nu=250.20/164$. This description remained statistically unacceptable and performed worse than the subsequently described thermal-evolution model ($209.78/166$), despite having two additional free parameters. Variable neutral absorption therefore, even when combined with component-normalization changes, cannot adequately reproduce the track-resolved spectra without spectral-shape evolution.

We subsequently considered five progressively less restrictive linking schemes, summarized in Table~\ref{tab:joint_spectral_tests}. Case 1 represents the null hypothesis that the intrinsic continuum is unchanged between the tracks. Case 2 allows only the disk temperature to differ, and Case 3 additionally allows the \textsc{diskbb} normalization to differ. Case 4 then allows the \textsc{nthComp} normalization to vary. Finally, Case 5 allows both the \textsc{nthComp} normalization and photon index to differ and is therefore the simultaneous-fit counterpart of the independently fitted spectra described above, with the detector calibration parameters treated consistently between tracks. This nested sequence determines which of the differences permitted in the independent fits are actually required by the data.

The results of the five simultaneous fitting cases are summarized in Table~\ref{tab:joint_spectral_tests}. In Case 1, all continuum parameters were tied between the upper and lower tracks. This highly restrictive model gave \(\chi^2/\nu=189730.8/168\) and was decisively rejected. Because this model forces the two tracks to have identical component normalizations as well as identical spectral shapes, its rejection primarily reflects the large intensity difference used to define the two tracks. Allowing only the disk temperature to vary in Case 2 provided a substantially improved, though still statistically unacceptable, fit with \(\chi^2/\nu=341.60/167\). Allowing the apparent disk normalization to vary in Case 3 further improved the fit by \(\Delta\chi^2=131.82\) for one additional free parameter.

Case 3 gave \(\chi^2/\nu=209.78/166\). In this model, the upper track had \(kT_{\rm in}=0.703\pm0.007\)~keV and \(N_{\rm disk}=255\pm4\), compared with \(kT_{\rm in}=0.632\pm0.009\)~keV and \(N_{\rm disk}=224\pm5\) in the lower track. The two tracks shared \(\Gamma=2.36\pm0.08\) and \(N_{\rm nthComp}=0.11\pm0.01\).

Allowing the \textsc{nthComp} normalization to differ in Case 4 improved the fit relative to Case 3 by only $\Delta\chi^2=2.10$ for one additional parameter. Additionally freeing the photon index in Case 5 changed the statistic by only $\Delta\chi^2=0.02$ for one additional parameter. Thus, even the joint-fit counterpart of the independently fitted spectra improves upon Case 3 by only $\Delta\chi^2=2.12$ for two additional parameters. The apparent agreement of the Comptonized continua in the independent fits is therefore confirmed quantitatively by the nested simultaneous comparison.

\begin{figure}
    \centering
    \includegraphics[width=\columnwidth]
    {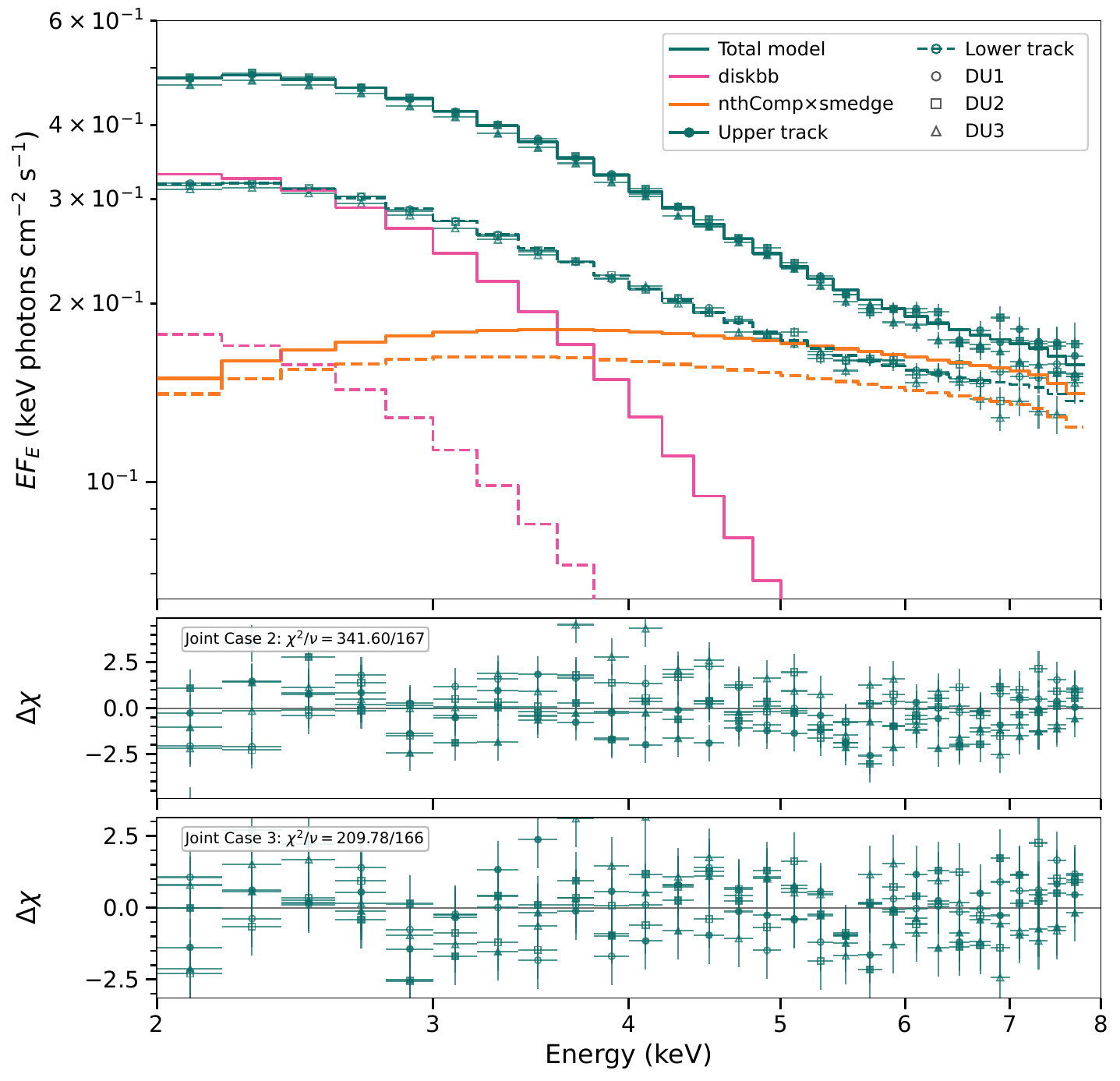}
    \caption{Simultaneous fits to the track-resolved IXPE Stokes-$I$ spectra. The upper and lower tracks are shown using solid lines with filled points and dashed lines with open points, respectively. The upper panel shows the unfolded spectra and preferred Case 3 model, in which the \textsc{diskbb} temperature and normalization were allowed to differ between the tracks, while the \textsc{nthComp} photon index and normalization were tied. The thermal disk and Comptonized continuum are shown separately, while the weak \textsc{laor} component is included in the total model but falls below the displayed range. The middle panel shows the residuals for Case 2, in which only the disk temperature was allowed to differ, while the lower panel shows the residuals for Case 3. Allowing the apparent disk normalization to differ removes the systematic residual structure and improves the fit by $\Delta\chi^2=131.82$ for one additional free parameter.}
    \label{fig:joint_track_spectra}
\end{figure}

We adopt Case 3 as the preferred description of the track-resolved spectra. It is the simplest model that captures the principal spectral differences, and the addition of track-dependent Comptonization parameters does not produce a statistically meaningful improvement. In the adopted model, the upper track has an approximately \(11\%\) higher disk temperature and an approximately \(14\%\) higher apparent disk normalization than the lower track. The spectral evolution is therefore primarily associated with the thermal component, while the data require no independent change in the \textsc{nthComp} photon index or normalization.\footnote{This is not surprising given the lack of simultaneous NuSTAR coverage across a track transition, which would provide useful observational lever above 8~keV.} The absorbed 2--8~keV flux decreased from \(7.208\times10^{-10}\) to \(5.011\times10^{-10}~{\rm erg\,cm^{-2}\,s^{-1}}\) between the upper and lower tracks. The corresponding disk flux decreased from \(3.431\times10^{-10}\) to \(1.612\times10^{-10}~{\rm erg\,cm^{-2}\,s^{-1}}\), while the \textsc{nthComp} flux decreased more modestly from \(3.722\times10^{-10}\) to \(3.344\times10^{-10}~{\rm erg\,cm^{-2}\,s^{-1}}\). The disk therefore contributes \(48.0\%\) and \(32.5\%\) of the modeled continuum flux in the upper and lower tracks, respectively. Approximately \(83\%\) of the total track-to-track flux difference is associated with the thermal component. The preferred fit and its improvement relative to Case 2 are illustrated in Figure~\ref{fig:joint_track_spectra}.

\begin{deluxetable*}{lcccccc}
\tabletypesize{\scriptsize}
\tablewidth{0pt}
\tablecaption{
Comparison of simultaneous fits to the upper-track and lower-track
Stokes-\(I\) spectra
\label{tab:joint_spectral_tests}
}
\tablehead{
\colhead{Parameter} &
\colhead{Case 0} &
\colhead{Case 1} &
\colhead{Case 2} &
\colhead{Case 3} &
\colhead{Case 4} &
\colhead{Case 5}
}
\startdata
Parameters allowed to differ
& \(N_{\rm H}\)
& None
& \(kT_{\rm in}\)
& \(kT_{\rm in},N_{\rm disk}\)
& \(kT_{\rm in},N_{\rm disk},N_{\rm nth}\)
& \(kT_{\rm in},N_{\rm disk},N_{\rm nth},\Gamma\)
\\
\hline
\\[-0.75ex]
\(N_{\rm H}\) (\(10^{22}\ {\rm cm^{-2}}\))
& \shortstack{\(2.4^{+0.3}_{-0.2}\)\\\(4.6^{+0.3}_{-0.2}\)}
& \(0.70\) (fixed)
& \(0.70\) (fixed)
& \(0.70\) (fixed)
& \(0.70\) (fixed)
& \(0.70\) (fixed)
\\
\(kT_{\rm in}\) (keV)
& \(0.36\pm0.02\)
& \(0.653\pm0.009\)
& \shortstack{\(0.703\pm0.005\)\\\(0.618\pm0.006\)}
& \shortstack{\(0.703\pm0.007\)\\\(0.632\pm0.009\)}
& \shortstack{\(0.703\pm0.007\)\\\(0.63\pm0.01\)}
& \shortstack{\(0.70\pm0.01\)\\\(0.63\pm0.01\)}
\\
\(N_{\rm disk}\)
& \(9^{+6}_{-3}\times10^{3}\)
& \(236\pm4\)
& \(254\pm4\)
& \shortstack{\(255\pm4\)\\\(224\pm5\)}
& \shortstack{\(253\pm5\)\\\(228\pm7\)}
& \shortstack{\(252\pm5\)\\\(228\pm7\)}
\\
\(\Gamma\)
& \(3.22^{+0.06}_{-0.05}\)
& \(2.48\pm0.09\)
& \(2.33\pm0.06\)
& \(2.36\pm0.08\)
& \(2.39\pm0.09\)
& \shortstack{\(2.4\pm0.2\)\\\(2.4\pm0.1\)}
\\
\(N_{\rm nthComp}\)
& \(0.74^{+0.10}_{-0.09}\)
& \(0.13\pm0.01\)
& \(0.112\pm0.008\)
& \(0.11\pm0.01\)
& \shortstack{\(0.11\pm0.01\)\\\(0.12\pm0.01\)}
& \shortstack{\(0.12\pm0.02\)\\\(0.12\pm0.02\)}
\\
\hline
\(\chi^2/\nu\)
& 28403.89/166
& 189730.8/168
& 341.60/167
& 209.78/166
& 207.68/165
& 207.66/164
\\
Null probability
& 0
& 0
& \(4.46\times10^{-14}\)
& 0.0121
& 0.0136
& 0.0119
\\
\enddata
\tablecomments{
The six cases are ordered from the absorption-only test through the most flexible model. In Case 0, the intrinsic continuum was tied while the absorption column was allowed to differ between the tracks. For parameters allowed to differ, values are listed as upper track/lower track except in the $\chi^2/\nu$ row. A single value indicates that the parameter was tied between the tracks. For Cases 1--5, the absorption column was fixed. The electron temperature, seed type, redshift, \textsc{smedge} parameters, and \textsc{laor} parameters were fixed. The detector cross-normalizations were allowed to vary but were tied between the tracks, while the detector gain corrections were frozen at their time-averaged values. Case 3 was adopted for the subsequent spectropolarimetric analysis. Uncertainties are reported at 90\% confidence; those for the strongly disfavored restrictive cases are formal confidence intervals conditional on the corresponding model.
}
\end{deluxetable*}

\subsection{Spectropolarimetry}
\label{subsec:specpol}

Following the Stokes-$\textit{I}$ analysis, we froze the spectral decomposition,
cross-normalization constants, and detector gain parameters and fit the
corresponding \ixpe Stokes-$Q$ and $U$ spectra. The Stokes spectra were
fit over 2--8~keV using the appropriate 20260101 modulation response
files. Each track was fit separately, with the polarization parameters tied among detector units.

We tested seven polarization prescriptions, summarized in
Table~\ref{tab:specpol_tracks}. These comprise a single energy-dependent \textsc{pollin} or energy-independent
\textsc{polconst} component applied to the entire continuum; separate polarization components for the disk and Comptonized emission, with their polarization angles constrained to be orthogonal; an unpolarized disk combined with either constant or linearly energy-dependent Compton polarization; and a model in which the disk polarization degree was fixed at $6\%$ and oriented orthogonally to the Comptonized component.\footnote{We adopt ${\rm PD}_{\rm disk}=0.06$ as a representative high disk polarization. In the classical electron-scattering atmosphere calculation, a PD of $6\%$ corresponds to an inclination of approximately $80^\circ$ \citep{1960ratr.book.....C}, consistent with the high inclination inferred from the \nustar fit. This remains below the theoretical edge-on maximum of approximately $12\%$ for $90^\circ$ inclination, which is unlikely given GS~1354$-$64 is not known to eclipse.} The weak \textsc{laor} component was treated as unpolarized
in all cases.

For \textsc{pollin}, the polarization degree is parameterized as
\begin{equation}
{\rm PD}(E)=A_1+(E-1~{\rm keV})A_{\rm slope},
\end{equation}
where $A_1$ is the extrapolated polarization degree at 1~keV and $A_{\rm slope}$ is its change per keV. We fixed the PA slope to zero because neither the model-independent analysis nor the spectropolarimetric fits indicated significant PA evolution. Because 1~keV lies below the fitted bandpass, \(A_1\) and \(A_{\rm slope}\) are strongly correlated. In the principal simultaneous track comparison, we therefore allowed \(A_1\) to take negative values, treating it only as an algebraic intercept, while verifying that the modeled PD remained positive throughout the observed 2--8~keV band. %% Values of $A_1$ that reach zero should therefore be interpreted as upper limits rather than as measurements of an unpolarized signal at 1~keV.

For the upper track, the common \textsc{pollin} model gave $\chi^2/\nu=178.69/171$, improving upon the common \textsc{polconst} fit by $\Delta\chi^2=14.15$ for one additional parameter. The extrapolated 1-keV polarization reached its lower boundary, with $A_1<0.6\%$ at 90\% confidence, while $A_{\rm slope}=0.8\pm0.2\%~{\rm keV}^{-1}$. The corresponding PA was $0\pm6^\circ$. The upper-track data therefore favor a polarization degree that increases with energy.

The component-resolved models provided similarly acceptable descriptions. The lowest fit statistic was obtained for orthogonally polarized disk and Comptonized components with an energy-dependent Compton polarization, yielding $\chi^2/\nu=174.53/170$. However, the disk and Compton polarization parameters were strongly covariant. Models containing an unpolarized disk and a polarized Comptonized component were also acceptable. With energy-independent Compton polarization, the inferred coronal polarization was ${\rm PD}_{\rm C}=4.2\pm1.0\%$ at ${\rm PA}_{\rm C}=1\pm7^\circ$ and gave $\chi^2/\nu=181.33/172$. Allowing the coronal PD to vary linearly with energy improved the fit to $\chi^2/\nu=178.63/171$, with $A_{1,\rm C}=2^{+3}_{-2}\%$, $A_{\rm slope,C}=0.8\pm0.8\%~{\rm keV}^{-1}$, and ${\rm PA}_{\rm C}=0\pm6^\circ$.

For the lower track, the common \textsc{pollin} and \textsc{polconst} models gave nearly identical fit statistics, $\chi^2/\nu=158.02/171$ and $158.16/172$, respectively. The common \textsc{polconst} model yielded ${\rm PD}=2.4\pm0.5\%$ at ${\rm PA}=1\pm7^\circ$, demonstrating that an energy-independent polarization adequately describes this track. The common-\textsc{pollin} fit gave $A_1=2\pm1\%$, $A_{\rm slope}=0.1\pm0.6\%~{\rm keV}^{-1}$, and ${\rm PA}=1\pm7^\circ$.

The lower-track component-resolved fits were again degenerate. In the orthogonal constant-polarization model, the disk polarization reached zero, with a 90\% upper limit of ${\rm PD}_{\rm disk}<1.4\%$. With the disk fixed as unpolarized, the Comptonized component had ${\rm PD}_{\rm C}=3.8\pm0.9\%$ at ${\rm PA}_{\rm C}=-1\pm7^\circ$, giving $\chi^2/\nu=159.36/172$. Allowing its polarization to vary linearly with energy gave $\chi^2/\nu=158.06/171$, with $A_{1,\rm C}=5\pm2\%$, $A_{\rm slope,C}=-0.6\pm0.9\%~{\rm keV}^{-1}$, and ${\rm PA}_{\rm C}=1\pm7^\circ$.

Overall, the upper-track data favor an increase in net PD with energy, whereas the lower track is adequately described by an energy-independent polarization. The component-resolved fits do not uniquely distinguish an intrinsic polarization gradient from energy-dependent dilution by the disk and Comptonized components.

\begin{table*}[t]
\centering
\caption{Track-resolved spectropolarimetric fits. Polarization degrees and \textsc{pollin} amplitudes are given in percent, and polarization slopes are given in percent per keV. The reported PA is either the common continuum PA or the PA of the Comptonized component, as appropriate. Uncertainties and upper limits are reported at the 90\% confidence level.}
\label{tab:specpol_tracks}
\scriptsize
\setlength{\tabcolsep}{4pt}
\renewcommand{\arraystretch}{1.15}
\begin{tabular}{clccccc}
\hline
Case & Polarization prescription & \({\rm PD}_{\rm disk}\) & \({\rm PD}_{\rm C}\) or \(A_{1,\rm C}\) & \(A_{\rm slope}\) or \(A_{\rm slope,C}\) & PA & \(\chi^2/\nu\) \\
& & (\%) & (\%) & (\% keV\(^{-1}\)) & (deg) & \\
\hline
\multicolumn{7}{c}{\textit{Upper Track}} \\
\hline
1 & Common \textsc{polconst} & \multicolumn{2}{c}{${\rm PD}=1.7\pm0.4$} & $\cdots$ & $1\pm8$ & 192.84/172 \\
2 & Common \textsc{pollin} & \multicolumn{2}{c}{$A_1<0.6$} & $0.8\pm0.2$ & $0\pm6$ & 178.69/171 \\
3 & Orthogonal disk \(+\) \textsc{pollin} corona & $7\pm6$ & $19\pm14$ & $-3\pm3$ & $1\pm6$ & 174.53/170 \\
4 & Orthogonal disk \(+\) \textsc{polconst} corona & $2\pm2$ & $6\pm2$ & $\cdots$ & $0\pm6$ & 176.72/171 \\
5 & Unpolarized disk \(+\) \textsc{polconst} corona & $0$ (fixed) & $4.2\pm1.0$ & $\cdots$ & $1\pm7$ & 181.33/172 \\
6 & Unpolarized disk \(+\) \textsc{pollin} corona & $0$ (fixed) & $A_{1,\rm C}=2^{+3}_{-2}$ & $0.8\pm0.8$ & $0\pm6$ & 178.63/171 \\
7 & $6\%$ disk \(+\) orthogonal \textsc{polconst} corona & $6$ (fixed) & $10.6\pm1.0$ & $\cdots$ & $-1\pm6$ & 191.85/172 \\
\hline
\multicolumn{7}{c}{\textit{Lower Track}} \\
\hline
1 & Common \textsc{polconst} & \multicolumn{2}{c}{${\rm PD}=2.4\pm0.5$} & $\cdots$ & $1\pm7$ & 158.16/172 \\
2 & Common \textsc{pollin} & \multicolumn{2}{c}{$A_1=2\pm1$} & $0.1\pm0.6$ & $1\pm7$ & 158.02/171 \\
3 & Orthogonal disk \(+\) \textsc{pollin} corona & $<7$ & $5^{+10}_{-2}$ & $-0.6^{+0.9}_{-2.0}$ & $1\pm7$ & 158.06/170 \\
4 & Orthogonal disk \(+\) \textsc{polconst} corona & $<1.4$ & $3.8^{+1.0}_{-0.9}$ & $\cdots$ & $-1^{+7}_{-6}$ & 159.36/171 \\
5 & Unpolarized disk \(+\) \textsc{polconst} corona & $0$ (fixed) & $3.8\pm0.9$ & $\cdots$ & $-1\pm7$ & 159.36/172 \\
6 & Unpolarized disk \(+\) \textsc{pollin} corona & $0$ (fixed) & $A_{1,\rm C}=5\pm2$ & $-0.6\pm0.9$ & $1\pm7$ & 158.06/171 \\
7 & $6\%$ disk \(+\) orthogonal \textsc{polconst} corona & $6$ (fixed) & $7.2\pm0.9$ & $\cdots$ & $-6\pm7$ & 180.30/172 \\
\hline
\end{tabular}
\vspace{2mm}
\begin{minipage}{0.96\textwidth}
\footnotesize
\textit{Note.} For the common-polarization models (Cases 1 and 2), a single polarization prescription is applied to the sum of the disk and Comptonized continua; the corresponding \(A_1\) or PD is therefore shown across the two component columns. For the orthogonal-component models, the disk polarization angle was constrained to satisfy \({\rm PA}_{\rm disk}={\rm PA}_{\rm C}+90^\circ\). For \textsc{pollin}, \(A_{1,\rm C}\) is the extrapolated polarization degree of the Comptonized component at 1~keV, and \(A_{\rm slope,C}\) is its linear energy slope. The PA slope was fixed at zero in all \textsc{pollin} fits. For lower-track Case 3, the confidence interval on \(A_{\rm slope,C}\) was checked using a parameter scan because the standard \texttt{error} calculation reported non-monotonicity in the fit-statistic surface. The reduced $\chi^2$ values slightly below unity for several lower-track fits ($\chi^2/\nu\simeq0.92$--$0.93$) lie within the expected statistical dispersion for approximately 170 degrees of freedom and do not, by themselves, indicate overfitting.
\end{minipage}
\end{table*}

To test the apparent difference in the energy dependence of the PD directly, we also fit the upper- and lower-track Stokes-$Q$ and $U$ spectra simultaneously with \textsc{pollin}. Because \(A_1\) is an extrapolated intercept below the fitted bandpass, we allowed it to take negative values while requiring the resulting PD to remain positive throughout the 2--8~keV band. In the null model, the two tracks were allowed to have different polarization intercepts but were constrained to share a common PD slope and PA. This model gave $\chi^2/{\rm dof}=340.01/344$. Allowing the PD slopes to differ improved the fit to $\chi^2/{\rm dof}=334.82/343$, corresponding to $\Delta\chi^2=5.19$ for one additional free parameter.

The separate-slope fit gave $A_1=-0.9\pm1.2\%$ and $A_{\rm slope}=1.2\pm0.5\%~{\rm keV^{-1}}$ for the upper track, and $A_1=2.3\pm1.5\%$ and $A_{\rm slope}=0.1\pm0.6\%~{\rm keV^{-1}}$ for the lower track. The common PA was $0^\circ\pm5^\circ$. Although the upper-track intercept is negative, the modeled PD remains positive throughout the observed band, reaching approximately $0.35\%$ at 2~keV; the negative value therefore reflects only the extrapolation of the linear model to 1~keV. Under the usual likelihood-ratio interpretation, the improvement gives a nominal $p=0.023$, while the corresponding improvement is $\Delta{\rm AIC}=3.19$. This provides modest evidence that the PD increases more steeply with energy during the upper track.

\begin{figure*}[t]
    \centering
    \includegraphics[width=\textwidth]
    {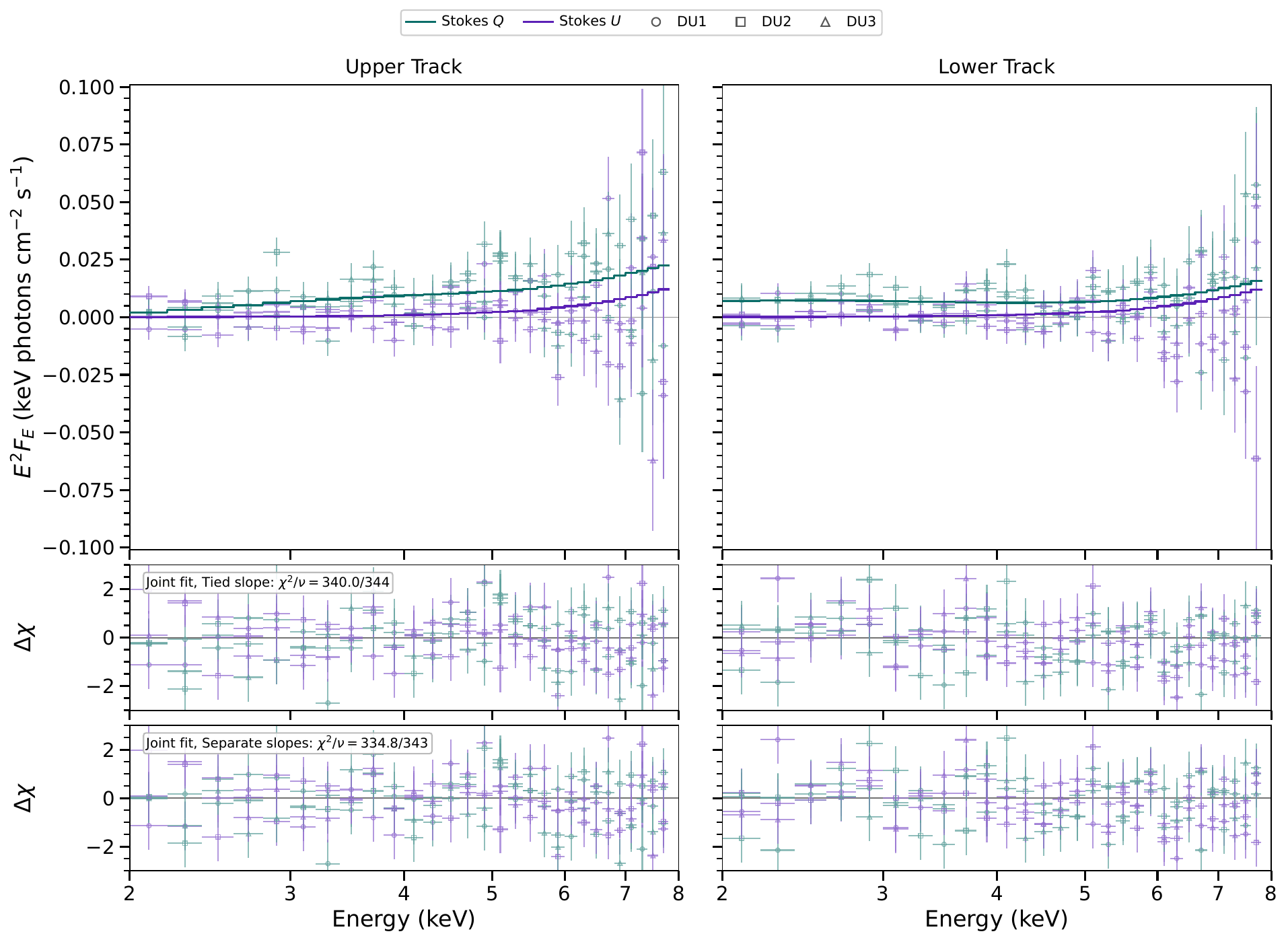}
    \caption{Simultaneous fit to the track-resolved IXPE Stokes-$Q$ and $U$ spectra. The left and right columns show the upper and lower tracks, respectively, with marker shape identifying the three detector units. The upper panels show the unfolded Stokes spectra and the model in which both $A_1$ and $A_{\rm slope}$ were allowed to differ between tracks, while the PA was tied. The middle panels show the residuals when the tracks have separate $A_1$
    values but share a common PD slope. The lower panels show the residuals when both $A_1$ and the PD slope are track dependent. Allowing the slopes to differ improves the fit by $\Delta\chi^2=5.19$ for one additional free parameter. This corresponds to marginal evidence that the PD increases more
    steeply with energy along the upper track.
    }
    \label{fig:joint_track_qu}
\end{figure*}

\begin{table*}
\centering
\caption{Simultaneous tests of track-dependent polarization using a common \textsc{pollin} prescription. Polarization parameters were tied among detector units. Values of $A_1$ and $A_{\rm slope}$ are given in percent and percent per keV, respectively. When a parameter was allowed to differ between tracks, the values are listed as upper track/lower track. The PA was tied between tracks in all four fits, and its energy dependence was fixed to zero. All uncertainties and upper limits are reported at 90\% confidence.}
\label{tab:simultaneous_specpol}
\small
\setlength{\tabcolsep}{5pt}
\begin{tabular}{llclcccc}
\hline
$A_1$ treatment &
$A_1$ (\%) &
$A_{\rm slope}$ treatment &
$A_{\rm slope}$ (\% keV$^{-1}$) &
PA (deg) &
$\chi^2/{\rm dof}$ &
% Null probability &
$\Delta\chi^2$ \\
\hline
Tied
& $0.4^{+1.0}_{-0.4}$
& Tied
& $0.8^{+0.3}_{-0.4}$
& $-1^{+5}_{-5}$
& $342.33/345$
& \nodata \\

Tied
& $0.4^{+1.0}_{-0.4}$
& Separate
& $0.7^{+0.3}_{-0.4}/0.8^{+0.3}_{-0.4}$
& $-1^{+5}_{-5}$
& $342.02/344$
& $0.31^{a}$ \\

Separate
& $0.2^{+1.0}_{-0.2}/0.8^{+1.0}_{-0.7}$
& Tied
& $0.8^{+0.2}_{-0.4}$
& $-1^{+5}_{-5}$
& $340.01/344$
& $2.32^{a}$ \\

Separate
& \({-0.9\pm1.2/2.3\pm1.5}\)
& Separate
& \({1.2\pm0.5/0.1\pm0.6}\)
& \(0\pm5\)
& 334.82/343
& 5.19 \\
\hline
\end{tabular}

\begin{flushleft}
\textit{Note.} Values separated by a slash correspond to the upper and lower tracks, respectively (except in the $\chi^2/\nu$ row). In the fit with both \(A_1\) and \(A_{\rm slope}\) separate, \(A_1\) was allowed to take negative values because it represents an extrapolated intercept below the fitted bandpass. The negative upper-track value does not imply a physical negative PD: the modeled PD remains positive throughout 2--8~keV. The PA slope was fixed at zero in all fits. 

$^{a}$ Relative to the model in which both $A_1$ and $A_{\rm slope}$ were tied.

$^{b}$ Relative to the model with separate $A_1$ values and a tied $A_{\rm slope}$.
\end{flushleft}
\end{table*}

\subsubsection{Joint comparison with the intermediate state}

We next jointly fit the Stokes-$Q$ and $U$ spectra from the intermediate-state observation and both SPL tracks to test directly whether the polarization evolved across the state transition. The spectral decomposition, detector gain corrections, and cross-normalization constants for each track were fixed at their corresponding Stokes-$I$ solutions. A common \textsc{pollin} prescription was applied to the full continuum, with the PA slope fixed at zero.

The preferred model allows the upper- and lower-track polarization amplitudes and slopes to differ, while tying the intermediate-state amplitude to that of the upper track and allowing the intermediate-state slope to vary independently. A single PA was shared by all three intervals. This model gives $\chi^2/\nu=532.40/522$ with a null-hypothesis probability of $0.367$. The shared upper-track and intermediate-state amplitude is $A_1=0.2^{+0.7}_{-0.2}\%$, with its 90\% lower confidence bound reaching the hard limit at zero. The upper-track slope is $0.8\pm0.3\%~{\rm keV^{-1}}$, while the lower-track amplitude and slope are $2\pm1\%$ and $0.1\pm0.6\%~{\rm keV^{-1}}$, respectively. The intermediate-state slope is $1.7^{+0.2}_{-0.3}\%~{\rm keV^{-1}}$, and the common PA is $-1^\circ\pm2^\circ$. All uncertainties are reported at 90\% confidence.

Tying the intermediate-state slope to the upper-track slope worsens the fit to $\chi^2/\nu=570.60/523$, corresponding to $\Delta\chi^2=38.20$ for one additional constraint and a nominal $p\simeq6.4\times10^{-10}$. The change in the PD slope between the intermediate and fading-SPL states is therefore strongly required. In contrast, allowing the intermediate-state PA to differ improves the fit by only $\Delta\chi^2=0.49$ for one additional parameter, while allowing its amplitude to differ improves the fit by only $\Delta\chi^2=1.39$. Allowing both quantities to differ gives $\chi^2/\nu=530.52/520$, an improvement of only $\Delta\chi^2=1.88$ for two additional parameters. The preferred model therefore retains a common PA and ties the intermediate-state amplitude to the upper-track value while allowing an independent intermediate-state slope.

\section{Discussion}

\subsection{Polarimetric evolution from the intermediate to the SPL state}

The two \ixpe observations provide a direct view of the polarization evolution of GS~1354$-$64 across a major accretion-state transition, evidenced by the disappearance of the Type-C QPO, strong suppression of rapid variability, increased prominence of the thermal disk, reduction in the broadband PD, and flattening of the PD energy dependence. During the intermediate state, the source had a 2--8~keV polarization degree of \(3.9^{+0.3}_{-0.2}\%\) at \({\rm PA}=-2^\circ\pm2^\circ\), whereas the subsequent fading-SPL observation gives \({\rm PD}=2.3^{+0.2}_{-0.3}\%\) at \({\rm PA}=4^\circ\pm3^\circ\). Thus, the broadband PD decreased in both observed intensity tracks as the source became more disk bright, while the PA remained statistically consistent across the transition.

The state-dependent change is particularly clear in the energy dependence of the PD. The preferred joint Stokes-\(Q\) and \(U\) fit gives an intermediate-state slope of \(1.7^{+0.2}_{-0.3}\%~{\rm keV^{-1}}\), compared with \(0.8\pm0.3\%~{\rm keV^{-1}}\) for the upper SPL track and \(0.1\pm0.6\%~{\rm keV^{-1}}\) for the lower track. Tying the intermediate-state and upper-track slopes worsens a joint fit by \(\Delta\chi^2=38.20\) for one additional constraint, corresponding to a nominal \(p\simeq6.4\times10^{-10}\). By contrast, allowing the intermediate-state PA to differ improves the fit by only \(\Delta\chi^2=0.49\). The dominant polarimetric evolution is therefore a reduction and flattening of the PD spectrum, rather than a rotation of its PA. The consistent PAs indicate that the disk--corona system retained the same projected orientation across the transition, despite changes in their relative fluxes and the disappearance of the Type-C QPO.

Part of this evolution can plausibly arise from the changing relative fluxes of the thermal and Comptonized components. If the Comptonized emission is more strongly polarized than the directly observed disk, increasing the disk contribution lowers the band-integrated PD through dilution. The energy dependence then reflects the changing component mixture across the \ixpe band, because the disk contributes most strongly at lower energies while the Comptonized component becomes relatively more important toward higher energies. This naturally produces an increasing net PD with energy. However, dilution alone is not uniquely established: changes in the intrinsic disk polarization, coronal optical depth or geometry, and scattering in the disk atmosphere could produce similar behavior. The observations therefore show that the polarization-weighted disk--corona balance changed substantially, but do not uniquely determine whether this resulted only from changing flux fractions or also from evolution of the intrinsic component polarizations.

\subsection{The origin of the two intensity tracks}

The SPL observation of GS~1354$-$64 reveals two distinct intensity tracks: a declining-continuum interrupted by two broad troughs, each lasting approximately \(10^{5}\)~s and a shorter-interval dip. The spectral differences between these tracks are dominated by the thermal component. In the preferred simultaneous fit, both the \textsc{diskbb} temperature and normalization differ between the upper and lower tracks, whereas the \textsc{nthComp} photon index and normalization can remain tied. Allowing the disk normalization to vary in addition to the temperature improves the fit by \(\Delta\chi^{2}=131.82\) for one additional free parameter, while subsequently allowing either the Compton normalization or the full Compton continuum to vary produces improvements of only \(\Delta\chi^{2}\simeq2\). The lower track spectrum is therefore not simply a fainter realization of the same continuum: it has a cooler and fainter disk with no independent change in the \textsc{nthComp} photon index or normalization required.

The inferred luminosity places this behavior near the regime in which BHXBs commonly begin the transition out of disk-bright states and complete the transition back toward the hard state. Assuming a distance of 7~kpc \citep{2021MNRAS.503.3784P}, the model-derived bolometric flux indicates that the source declined from approximately \(2.5\%\) to \(1.6\%\) of the Eddington luminosity during the observation. This overlaps the \(\sim1\)--\(2\%\) Eddington range often associated with the soft-to-hard back-transition \citep{2003A&A...409..697M,2010MNRAS.403...61D}. The repeated movement between the upper and lower tracks may therefore represent brief excursions along the intermediate-state branch as the source approached the transition threshold. This interpretation is supported qualitatively by both the source's motion in the \ixpe HID and its behavior in the PD--PA plane, both indicating a possible sputtering near the cusp of a state transition, as the source repeatedly moves between neighboring configurations before settling into the hard state again.

This behavior differs from that expected for a conventional cold-absorption dip. In wind-fed systems such as Cygnus~X-1, dips preferentially suppress the soft X-ray emission, produce spectral hardening, and are commonly described by changes in neutral or partially covering absorption. While GS~1354$-$64 is putatively a Roche-lobe-overflow low-mass X-ray binary rather than a wind-fed system \citep{2004ApJ...613L.133C}, we tested this possibility directly by tying the intrinsic continuum between the two GS~1354$-$64 tracks while allowing \(N_{\rm H}\) to differ. Although the trough favored a larger nominal column density, the resulting fit remained grossly unacceptable, with \(\chi^{2}/\nu=28403.9/166\), and drove the tied continuum to extreme compensating values. Variable neutral absorption alone therefore cannot reproduce the track differences. Recent \ixpe analysis of dipping in Cygnus~X-1 found that the broadband PD and PA remained approximately unchanged between dip and non-dip intervals, consistent with compact absorbing structures crossing the line of sight without substantially changing the geometry of the extended polarized source \citep{2026arXiv260726144L}. The GS~1354$-$64 troughs are broader and smoother and are instead described successfully by changes in the thermal emission.

More complex obscuration cannot yet be excluded completely. Although the absorption-only Case~0 rules out a change in fully covering neutral absorption as the sole origin of the two tracks, the limited spectral resolution and 2--8~keV bandpass of \ixpe provide weaker constraints on partially covering or highly ionized material. A highly ionized or nearly gray scattering screen could reduce the observed intensity without producing strong hardness changes though would not explain the suggestive temperature change across tracks. Nevertheless, maintaining the source in a low state for \(\sim10^{5}\)~s would require a large and coherent line-of-sight structure. As a simple order-of-magnitude estimate, following the crossing-time argument commonly applied to X-ray occultation events \citep[e.g.,][]{2007ApJ...659L.111R}, consider an obscuring structure moving across the line of sight with an assumed transverse velocity \(v_{\perp}\). To maintain the source in the trough for \(t_{\rm trough}\), its projected extent along the direction of motion must satisfy
\begin{equation}
D_{\perp}\gtrsim v_{\perp}t_{\rm trough}
\simeq 5\times10^{12}
\left(\frac{v_{\perp}}{500~{\rm km~s^{-1}}}\right)
\left(\frac{t_{\rm trough}}{10^{5}~{\rm s}}\right)
{\rm cm}.
\end{equation}

\noindent Here, \(500~{\rm km~s^{-1}}\) is an illustrative velocity rather than a measured value for GS~1354$-$64. This is many orders of magnitude larger than the characteristic X-ray-emitting region and may approach or exceed the scale of the binary. Since the ingress and egress are substantially shorter than the trough duration ($\sim5$--11 minutes, compared with the $\sim10^{5}$-s or $\sim28$-hr trough duration), the obscurer would additionally require a sharp leading edge and a large transverse aspect ratio. For plausible binary-scale velocities, this would require a substantially extended and coherent obscuring structure, unlike the compact clouds inferred from rapid X-ray occultation events in Active Galactic Nuclei (AGN) such as NGC~1365 \citep{2007ApJ...659L.111R}. The measured rapid egress strengthens this geometric constraint, while broader-band fits using partial-covering or ionized-absorption models would be required to test these more complex obscuration scenarios.

The fitted \textsc{diskbb} normalization should not, however, be interpreted directly as a change in the physical inner radius. In addition to its dependence on the assumed distance, inclination, color-correction factor, inner-boundary condition, and any partial occultation, the normalization may be biased by the treatment of Comptonization. In the additive \textsc{diskbb}+\textsc{nthComp} model used here, the seed-photon temperature is tied to the disk temperature, but the Comptonized flux is not coupled self-consistently to the number of disk photons removed by scattering. The fitted \textsc{diskbb} component therefore represents only the directly observed disk emission; any disk photons redistributed into the Comptonized component appear as ``missing'' disk flux, biasing the inferred radius low. A photon-conserving convolution treatment such as \textsc{simpl} would be required for a more physical radius estimate \citep{2009PASP..121.1279S}; however, at the high electron temperatures inferred here, a Comptonization model employing an accurate Klein--Nishina cross-section, such as the plane-parallel \textsc{compps} or \textsc{compps\_conv} model, would be more appropriate \citep{1996ApJ...470..249P}. Likewise, the inclination inferred from the phenomenological \textsc{laor} component should not be treated as a robust measurement of the binary or inner-disk inclination. It is sensitive to the assumed line energy, emissivity profile, disk radii, continuum model, and instrumental calibration, several of which were fixed in these fits. The combination of a lower temperature and smaller apparent normalization during the lower track could therefore represent a genuine change in the visible thermal-emitting area, a change in the disk atmosphere or spectral hardening, or obscuration of part of an extended disk. The current data establish that the phenomenological change is associated primarily with the thermal component, but do not uniquely distinguish among these physical explanations.

\subsection{Polarimetric constraints on the trough geometry}

The upper track favors a PD that increases with energy, whereas the lower track is adequately described by energy-independent polarization. The direct simultaneous comparison provides only modest evidence for different slopes, with the inferred difference sensitive to the covariance between the extrapolated \textsc{pollin} intercept and slope. The track dependence should therefore be regarded as suggestive rather than a firm detection of polarimetric evolution during the troughs.

The component-resolved fits illustrate an important degeneracy. The rising net PD along the upper track can be reproduced either by intrinsic energy dependence or by a weakly polarized disk diluting a more strongly polarized Comptonized component. Models with an unpolarized disk yield similar Compton polarization degrees in both tracks, while models with orthogonally polarized components are also acceptable but poorly constrain their individual contributions. The data therefore do not uniquely distinguish changes in intrinsic polarization or the scattering environment from changes in component weighting.

The stable PA provides a complementary constraint: neither track requires a change in the net polarization orientation. This disfavors explanations requiring a large rotation of the observed polarization during the troughs, but does not imply that the accretion geometry is otherwise unchanged. Changes in the visible disk area, temperature, or scattering properties remain compatible with the measurements.

The possible track dependence is weaker than the polarization evolution across the intermediate-to-SPL transition, where the PD spectrum becomes significantly shallower without a measurable PA change. Importantly, a shallower PD spectrum need not indicate a weaker corona. If the energy dependence arises partly from disk dilution, the reduced thermal contribution during the troughs can allow polarized Comptonized emission to dominate over more of the \ixpe band, flattening the net PD spectrum without requiring a change in its intrinsic polarization.

\subsection{A changing-state analogy across the black-hole mass scale}

The repeated movement of GS~1354$-$64 between two long-lived intensity tracks invites comparison with changing-state accretion behavior. Changing-state AGN (CSAGN) such as Mrk~590 and Mrk~1018 undergo large continuum and broad-line changes that are increasingly interpreted as changes in the accretion flow rather than solely as variable obscuration \citep{2023NatAs...7.1282R}. Mrk~1018, for example, evolved between bright and faint states over years, accompanied by substantial UV and X-ray changes and evidence for a reconfiguration of the inner accretion flow \citep{2024A&A...683A.131V}. Mrk~590 has likewise displayed fading, rebrightening, and repeated X-ray/UV activity at relatively low Eddington ratio \citep{2023MNRAS.519.3903L}. These sources demonstrate that accretion flows can occupy and revisit distinct luminosity configurations without a simple monotonic evolution. 

An additional relevant comparison is the ``flip-flop'' behavior observed in some BHXBs near global state transitions \citep{2025arXiv250208718B,2026PASA...43...44M}. These events involve rapid transitions between two distinct intensity and spectral-timing configurations, each of which can persist substantially longer than the transition itself. This morphology resembles the repeated upper- and lower-track behavior of GS~1354$-$64: the best-sampled transition, the egress from the second broad trough, has a measured 10--90\% duration of approximately \(5\)--\(11\) minutes, whereas the trough itself persists for \(\sim10^{5}\)~s.

\citet{2026PASA...43...44M} showed that the ingress, egress, and state-duration timescales of XRB flip-flops and CSAGN are broadly consistent with scaling linearly with black-hole mass. Under this scaling,
\begin{equation}
t_{\rm AGN}\simeq 1.9
\left(\frac{t_{\rm XRB}}{10~{\rm min}}\right)
\left(\frac{M_{\rm AGN}/M_{\rm XRB}}{10^{5}}\right)
{\rm yr},
\end{equation}
so the measured \(5\)--\(11\) minute egress would correspond to approximately \(1.0\)--\(2.1\)~yr for a \(10^{6}\,M_{\odot}\) AGN and a \(10\,M_{\odot}\) XRB. This scaled timescale lies within the months-to-years range observed in CSAGN such as 1ES~1927$+$654 \citep{2019ApJ...883...94T} and SDSS~J162829.17$+$432948.5 \citep{2022ApJ...939L..16Z}, although this agreement remains phenomenological and does not establish a common physical mechanism.

The luminosity at which the transitions occur provides a complementary comparison. Both XRB flip-flops and changing-state transitions preferentially occur at luminosities of a few percent of Eddington \citep{2018MNRAS.480.3898N,2026PASA...43...44M}. Extrapolating the best-fitting time-averaged \ixpe spectral model over 0.01--100~keV and removing the foreground absorption gives an unabsorbed model flux of \(F_{\rm bol}=4.19\times10^{-9}~{\rm erg~cm^{-2}~s^{-1}}\), although this bolometric correction is model dependent. For a \(10\,M_{\odot}\) black hole,
\begin{equation}
\begin{aligned}
\frac{L_{\rm bol}}{L_{\rm Edd}}
&\simeq 1.95\times10^{-2}
\left(\frac{F_{\rm bol}}
{4.19\times10^{-9}~{\rm erg~cm^{-2}~s^{-1}}}\right) \\
&\quad\times
\left(\frac{d}{7~{\rm kpc}}\right)^2
\left(\frac{M}{10\,M_{\odot}}\right)^{-1}.
\end{aligned}
\end{equation}
The large uncertainty in the distance prevents this estimate from locating GS~1354$-$64 securely relative to the putative \(1\%\) transition scale. At the nominal distance of 7~kpc, the inferred Eddington ratio is approximately \(2\%\), consistent with a source in the SPL state approaching a return to the hard state, since BHXBs commonly complete the soft-to-hard transition at a few percent of Eddington \citep{2003A&A...409..697M,2010MNRAS.403...61D}. By contrast, distances of 0.7 and 25~kpc would imply \(L_{\rm bol}/L_{\rm Edd}\simeq2\times10^{-4}\) and \(0.25\), respectively.

Despite these similarities in morphology, transition timescale, and inferred luminosity, GS~1354$-$64 does not reproduce every defining property of previously reported flip-flops. Classical examples often show marked transitions between QPO-dominated and broadband-noise-dominated variability, whereas both tracks in GS~1354$-$64 retain weak variability without a detected QPO. The present observations may therefore represent a less complete excursion between neighboring accretion configurations near the soft-to-hard transition, rather than a direct example of the classical flip-flop phenomenon.

\section{Conclusion}

The \ixpe observations of GS~1354$-$64 reveal evolution on two distinct levels. Between the preceding intermediate-state observation and the fading SPL observation, the disappearance of the Type-C QPO, strong suppression of rapid variability, increased prominence of the thermal disk, and flattening of the PD energy dependence indicate a substantial reconfiguration of the emitting and scattering flow. The persistence of a steep Comptonized tail argues against identifying the later observation as a purely thermal soft state, while the stable PA suggests that the evolution occurred without a measurable reorientation of the projected symmetry axis. 

Within the fading SPL state, the recurrent tracks are distinguished robustly by changes in the thermal component, while their polarization behavior is suggestive of a pivoting accretion power dissipation in the disk-vs-corona or changes in the optical depth or geometry of the material scattering disk photons. This interpretation nevertheless remains dependent on the adopted polarization decomposition. The marginal difference between the upper-track and lower-track PD slopes, together with the significantly steeper intermediate-state slope, indicates that the disk--corona weighting may change among the different luminosity configurations. In this sense, GS~1354$-$64 may represent a stellar-mass example of an accretion flow moving between recurrent luminosity configurations, with a phenomenological connection to changing-state behavior across the black-hole mass scale.

\appendix
\section{\ixpe Timing Analysis}

We analyze the \ixpe timing data with version 1.1.2.4 of the Python package Stingray \citep{2019ApJ...881...39H,matteo_bachetti_2023_7970570}, following the procedure outlined by \citet{2026arXiv260303463R}. After filtering the good-time intervals from the \ixpe data, we combine events from the three DUs into light curves with 1~ms time resolution, corresponding to a Nyquist frequency of 500~Hz. We compute the power density spectrum by averaging 100~s intervals and adopt the fractional-rms normalization. We correct for deadtime using the Fourier Amplitude Difference method \citep{2018ApJ...853L..21B}, then logarithmically rebin the spectrum using bins with 5\% fractional width. The Poisson-noise level is estimated from the average power above 100~Hz and then subtracted.

\begin{figure}
    \centering

    \includegraphics[
        width=0.45\textwidth,
        trim=0 5 0 5,
        clip
    ]{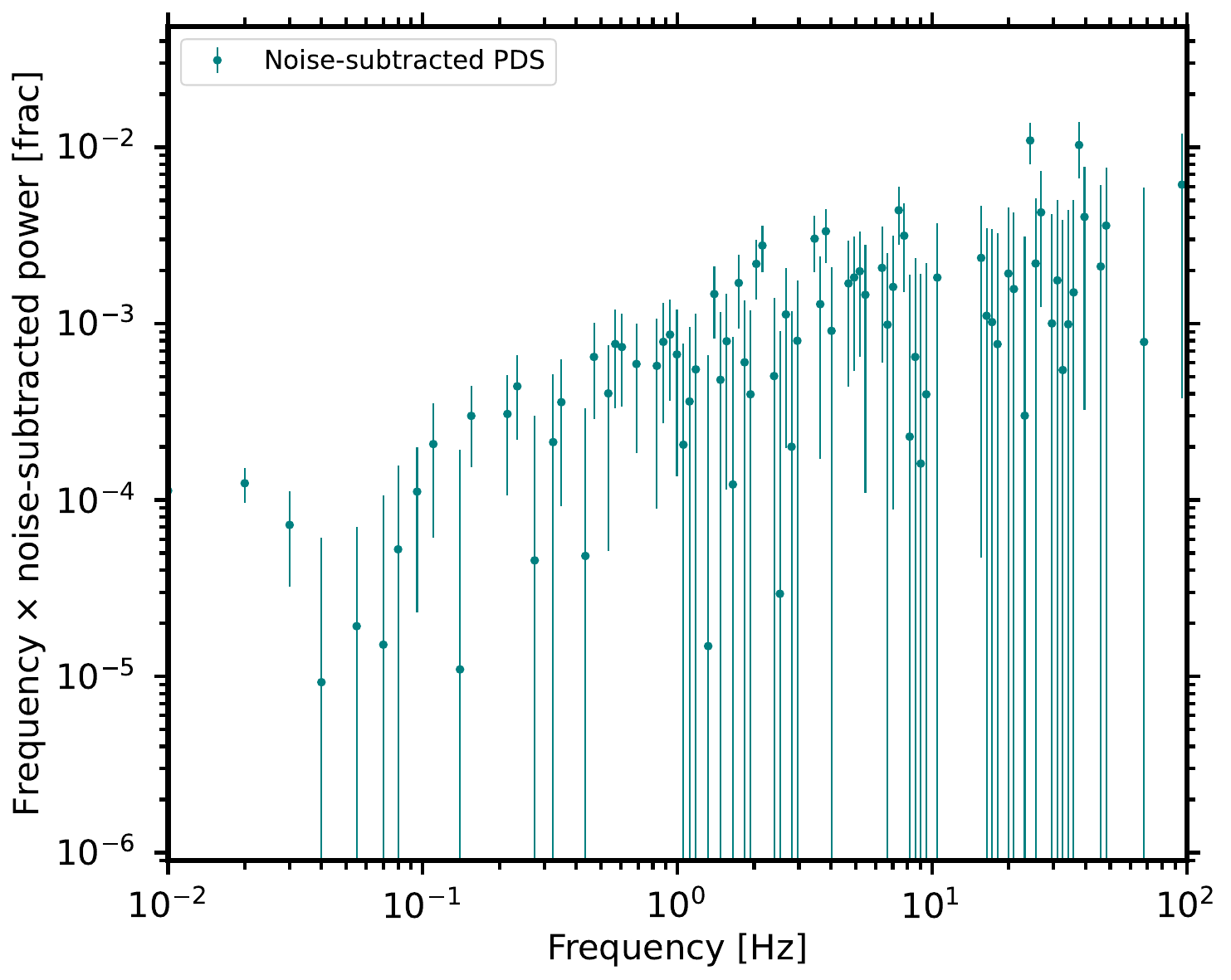}

    \vspace{-0.8em}

    \includegraphics[
        width=0.45\textwidth,
        trim=0 5 0 5,
        clip
    ]{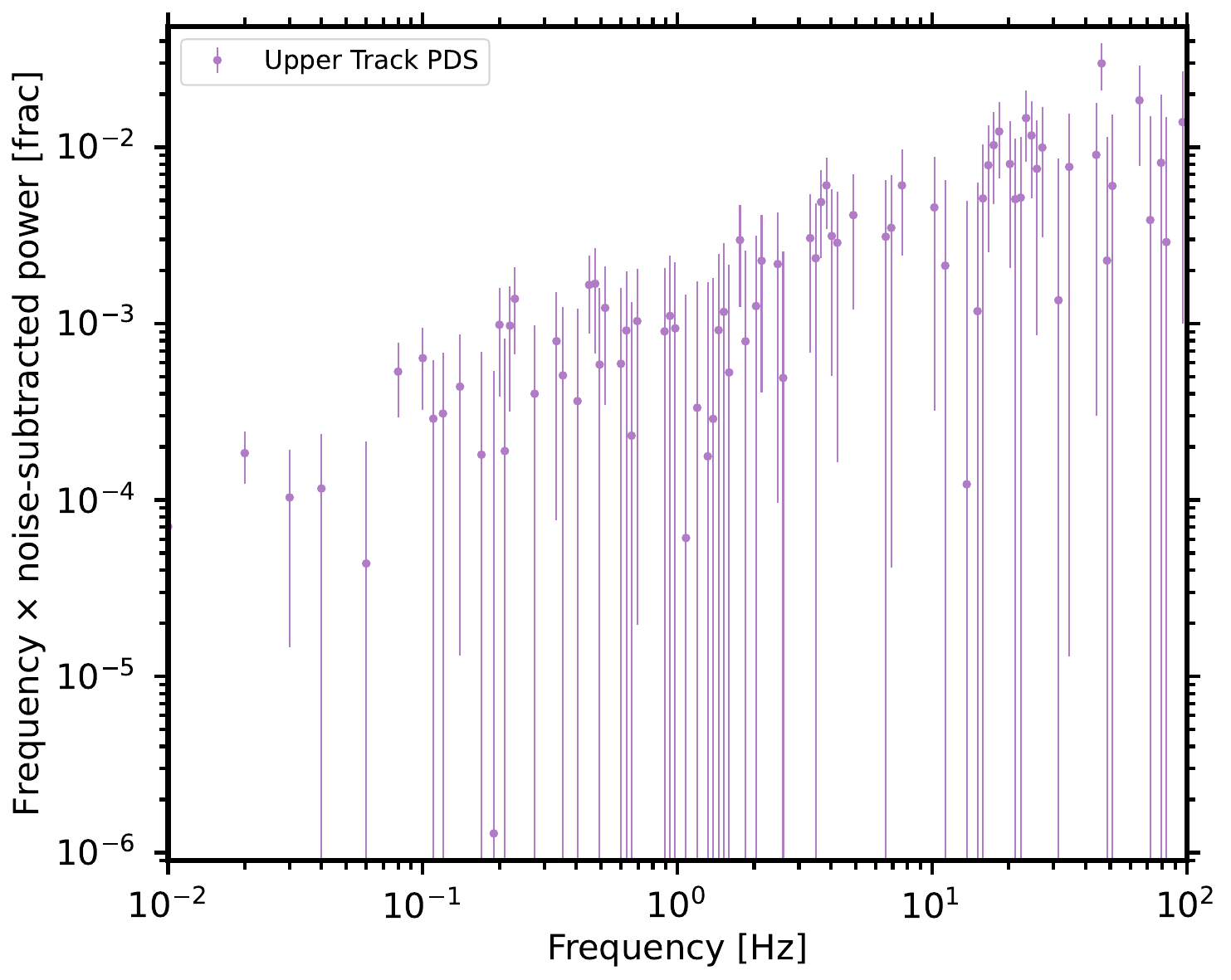}

    \vspace{-0.8em}

    \includegraphics[
        width=0.45\textwidth,
        trim=0 5 0 5,
        clip
    ]{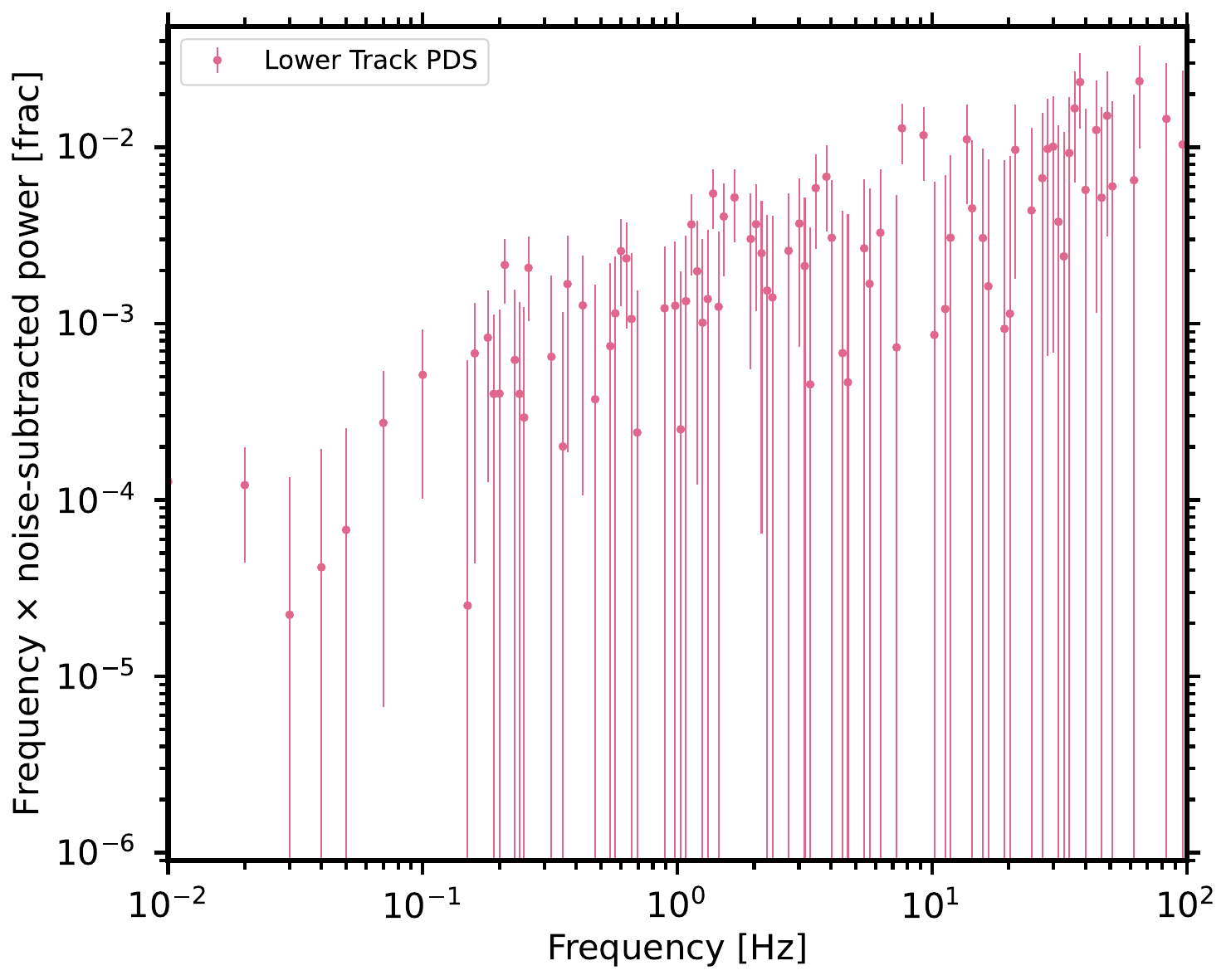}

    \vspace{-0.5em}

    \caption{Noise-subtracted PDS plotted as
    $\nu[P(\nu)-P_{\rm noise}]$ in units of fractional rms$^2$
    for the full GS~1354$-$64 \ixpe SPL observation \textbf{(top)},
    the upper track \textbf{(middle)}, and the lower track \textbf{(bottom)}.}
    \label{fig:1354_timing}
\end{figure}

The deadtime-corrected, logarithmically rebinned power density spectra (PDS) extracted from the full \ixpe SPL observation and separately from the upper and lower intensity tracks identified in the light curve (Fig.~\ref{fig:1354_lc}) are shown in Fig.~\ref{fig:1354_timing}. In contrast to the intermediate-state observation presented in \citealt{2026arXiv260303463R}, the SPL PDS exhibits characteristically weaker broadband variability and no evidence for QPOs. We examine both the full observation and the individual intensity tracks to track the variability across the SPL state.

We first fit the full SPL PDS in \texttt{XSPEC} using a constant continuum and a zero-centered Lorentzian to describe broadband variability, obtaining an acceptable fit with $\chi^2/\nu=171.01/154$.
%(null hypothesis probability $p=0.165$) 
Adding further Lorentzian components did not significantly improve the fit, and no additional broadband variability or QPO component was required by the SPL PDS.
% Allowing an additional Lorentzian with free centroid frequency and width improves the fit only modestly ($\chi^2/\nu=164.09/151$), insufficient to establish a significant QPO detection. %To specifically test for the persistence of the intermediate-state Type-C QPO, we fix the centroid frequency and full width at half maximum of the additional Lorentzian to the intermediate-state measurements of 4.69 and 0.89~Hz, respectively. The fitted normalization is consistent with zero, with no appreciable improvement over the continuum-only model ($\chi^2/\nu=171.01/153$). We obtain a 90\% confidence upper limit of $3.90\times10^{-4}$ on the QPO Lorentzian normalization ($\Delta\chi^2=2.7$), confirming 
We confirm that the previously detected oscillation is not significantly recovered in the full SPL observation. This nondetection, the collapsed broadband noise, and integrated 0.1--10~Hz rms power of \(3.5\%\)--\(5.3\%\), is consistent with softer states (e.g., \citealt{2005ApJ...629..403C,2010LNP...794...53B,2019NewAR..8501524I}).

We additionally fit the upper- and lower-track PDS using a power-law continuum and tested for additional broadband variability by adding a zero-centered Lorentzian. The measured noise-subtracted 0.1--10~Hz powers are \(1.25\times10^{-3}\) and \(2.82\times10^{-3}\ {\rm rms}^{2}\) for the upper and lower tracks, respectively, corresponding to fractional rms amplitudes of \(3.54\pm1.29\%\) and \(5.31\pm1.12\%\). The added Lorentzian therefore represents only weak broadband variability in both tracks, rather than evidence for a discrete QPO.

\begin{acknowledgments}
This work reports observations obtained with the Imaging X-ray Polarimetry Explorer (IXPE), a joint US (NASA) and Italian (ASI) mission, led by Marshall Space Flight Center (MSFC). The research uses data products provided by the IXPE Science Operations Center (MSFC), using algorithms developed by the IXPE Collaboration (MSFC, Istituto Nazionale di Astrofisica - INAF, Istituto Nazionale di Fisica Nucleare - INFN, ASI Space Science Data Center - SSDC), and distributed by the High-Energy Astrophysics Science Archive Research Center (HEASARC). This research has made use of data from the \nustar mission, a project led by the California Institute of Technology, managed by the Jet Propulsion Laboratory, and funded by the National Aeronautics and Space Administration. Data analysis was performed using the NuSTAR Data Analysis Software (NuSTARDAS), jointly developed by the ASI Science Data Center (SSDC, Italy) and the California Institute of Technology (USA).
This research has made use of MAXI data provided by RIKEN, JAXA and the MAXI team. We thank Andrzej Zdziarski for helpful comments on this source.

SR acknowledges support from NASA through the Imaging X-ray Polarimetry Explorer (\ixpe) General Observer Program under grant No.~80NSSC26K1006, supporting the project ``Energy-dependent Polarization Rotation: Testing General Relativistic Disk Models with \ixpe.''  JFS was supported by the Chandra X-ray Observatory, which is operated by the Smithsonian Astrophysical Observatory for and on behalf of NASA under contract NAS8-03060. AV acknowledges support from the Academy of Finland grants 355672 and 372881. AI is supported by the European Union (ERC, X-MAPS, 101169908). M.D., J.P., and  J.S. acknowledge support from the Czech Science Foundation grant 26-22614S.

\end{acknowledgments}

\facilities{\ixpe \citep{2022JATIS...8b6002W,2021AJ....162..208S}, \maxi \citep{2009PASJ...61..999M}, \nustar \citep{2013ApJ...770..103H}}

\software{astropy \citep{astropy:2013, astropy:2018, astropy:2022}, NumPy \citep{harris2020array}, SciPy \citep{2020SciPy-NMeth}, IPython \citep{PER-GRA:2007}, bilby \citep{2019ApJS..241...27A}, dynesty \citep{2020MNRAS.493.3132S}, ixpeobssim \citep{2022SoftX..1901194B}, QUEEN-BEE \citep{ravi2025queenbee, 2026ApJ...997...60R}, Stingray \citep{2019ApJ...881...39H, matteo_bachetti_2023_7970570}, SAOImage~DS9 \citep{2003ASPC..295..489J}, XSPEC \citep{1996ASPC..101...17A}}

\bibliography{1354}{}

@PHDTHESIS{1991PhDT........55E,
       author = {{Ebisawa}, K.},
        title = "{Spectral Study of Black Hole Candidates Observed with Ginga}",
       school = {JAXA, Institute of Space and Astronautical Science},
         year = 1991,
        month = feb,
       adsurl = {https://ui.adsabs.harvard.edu/abs/1991PhDT........55E}
}

@ARTICLE{1991ApJ...376...90L,
       author = {{Laor}, Ari},
        title = "{Line Profiles from a Disk around a Rotating Black Hole}",
      journal = {\apj},
         year = 1991,
        month = jul,
       volume = {376},
        pages = {90},
          doi = {10.1086/170257},
       adsurl = {https://ui.adsabs.harvard.edu/abs/1991ApJ...376...90L}
}

@ARTICLE{1996ApJ...465..487V,
       author = {{Verner}, D.~A. and {Ferland}, G.~J. and {Korista}, K.~T. and {Yakovlev}, D.~G.},
        title = "{Atomic Data for Astrophysics. II. New Analytic Fits for Photoionization Cross Sections of Atoms and Ions}",
      journal = {\apj},
         year = 1996,
        month = jul,
       volume = {465},
        pages = {487},
          doi = {10.1086/177435},
archivePrefix = {arXiv},
       eprint = {astro-ph/9601009},
 primaryClass = {astro-ph},
       adsurl = {https://ui.adsabs.harvard.edu/abs/1996ApJ...465..487V}
}

@ARTICLE{2016A&A...587A.151K,
       author = {{Kaastra}, J.~S. and {Bleeker}, J.~A.~M.},
        title = "{Optimal binning of X-ray spectra and response matrix design}",
      journal = {\aap},
         year = 2016,
        month = mar,
       volume = {587},
          eid = {A151},
        pages = {A151},
          doi = {10.1051/0004-6361/201527395},
archivePrefix = {arXiv},
       eprint = {1601.05309},
 primaryClass = {astro-ph.IM},
       adsurl = {https://ui.adsabs.harvard.edu/abs/2016A&A...587A.151K}
}

@INPROCEEDINGS{1996ASPC..101...17A,
       author = {{Arnaud}, K.~A.},
        title = "{XSPEC: The First Ten Years}",
    booktitle = {Astronomical Data Analysis Software and Systems V},
         year = 1996,
       editor = {{Jacoby}, George H. and {Barnes}, Jeannette},
       series = {Astronomical Society of the Pacific Conference Series},
       volume = {101},
        month = jan,
        pages = {17},
       adsurl = {https://ui.adsabs.harvard.edu/abs/1996ASPC..101...17A}
}

@ARTICLE{2021AJ....162..208S,
       author = {{Soffitta}, Paolo and {Baldini}, Luca and {Bellazzini}, Ronaldo and {Costa}, Enrico and {Latronico}, Luca and {Muleri}, Fabio and {Del Monte}, Ettore and {Fabiani}, Sergio and {Minuti}, Massimo and {Pinchera}, Michele and {Sgro'}, Carmelo and {Spandre}, Gloria and {Trois}, Alessio and {Amici}, Fabrizio and {Andersson}, Hans and {Attina'}, Primo and {Bachetti}, Matteo and {Barbanera}, Mattia and {Borotto}, Fabio and {Brez}, Alessandro and {Brienza}, Daniele and {Caporale}, Ciro and {Cardelli}, Claudia and {Carpentiero}, Rita and {Castellano}, Simone and {Castronuovo}, Marco and {Cavalli}, Luca and {Cavazzuti}, Elisabetta and {Ceccanti}, Marco and {Centrone}, Mauro and {Ciprini}, Stefano and {Citraro}, Saverio and {D'Amico}, Fabio and {D'Alba}, Elisa and {Di Cosimo}, Sergio and {Di Lalla}, Niccolo' and {Di Marco}, Alessandro and {Di Persio}, Giuseppe and {Donnarumma}, Immacolata and {Evangelista}, Yuri and {Ferrazzoli}, Riccardo and {Hayato}, Asami and {Kitaguchi}, Takao and {La Monaca}, Fabio and {Lefevre}, Carlo and {Loffredo}, Pasqualino and {Lorenzi}, Paolo and {Lucchesi}, Leonardo and {Magazzu}, Carlo and {Maldera}, Simone and {Manfreda}, Alberto and {Mangraviti}, Elio and {Marengo}, Marco and {Matt}, Giorgio and {Mereu}, Paolo and {Morbidini}, Alfredo and {Mosti}, Federico and {Nakano}, Toshio and {Nasimi}, Hikmat and {Negri}, Barbara and {Nenonen}, Seppo and {Nuti}, Alessio and {Orsini}, Leonardo and {Perri}, Matteo and {Pesce-Rollins}, Melissa and {Piazzolla}, Raffaele and {Pilia}, Maura and {Profeti}, Alessandro and {Puccetti}, Simonetta and {Rankin}, John and {Ratheesh}, Ajay and {Rubini}, Alda and {Santoli}, Francesco and {Sarra}, Paolo and {Scalise}, Emanuele and {Sciortino}, Andrea and {Tamagawa}, Toru and {Tardiola}, Marcello and {Tobia}, Antonino and {Vimercati}, Marco and {Xie}, Fei},
        title = "{The Instrument of the Imaging X-Ray Polarimetry Explorer}",
      journal = {\aj},
         year = 2021,
        month = nov,
       volume = {162},
       number = {5},
          eid = {208},
        pages = {208},
          doi = {10.3847/1538-3881/ac19b0},
archivePrefix = {arXiv},
       eprint = {2108.00284},
 primaryClass = {astro-ph.IM},
       adsurl = {https://ui.adsabs.harvard.edu/abs/2021AJ....162..208S}
}

@ARTICLE{2022JATIS...8b6002W,
       author = {{Weisskopf}, Martin C. and {Soffitta}, Paolo and {Baldini}, Luca and {Ramsey}, Brian D. and {O'Dell}, Stephen L. and {Romani}, Roger W. and {Matt}, Giorgio and {Deininger}, William D. and {Baumgartner}, Wayne H. and {Bellazzini}, Ronaldo and {Costa}, Enrico and {Kolodziejczak}, Jeffery J. and {Latronico}, Luca and {Marshall}, Herman L. and {Muleri}, Fabio and {Bongiorno}, Stephen D. and {Tennant}, Allyn and {Bucciantini}, Niccolo and {Dovciak}, Michal and {Marin}, Frederic and {Marscher}, Alan and {Poutanen}, Juri and {Slane}, Pat and {Turolla}, Roberto and {Kalinowski}, William and {Di Marco}, Alessandro and {Fabiani}, Sergio and {Minuti}, Massimo and {La Monaca}, Fabio and {Pinchera}, Michele and {Rankin}, John and {Sgro'}, Carmelo and {Trois}, Alessio and {Xie}, Fei and {Alexander}, Cheryl and {Allen}, D. Zachery and {Amici}, Fabrizio and {Andersen}, Jason and {Antonelli}, Angelo and {Antoniak}, Spencer and {Attin{\`a}}, Primo and {Barbanera}, Mattia and {Bachetti}, Matteo and {Baggett}, Randy M. and {Bladt}, Jeff and {Brez}, Alessandro and {Bonino}, Raffaella and {Boree}, Christopher and {Borotto}, Fabio and {Breeding}, Shawn and {Brienza}, Daniele and {Bygott}, H. Kyle and {Caporale}, Ciro and {Cardelli}, Claudia and {Carpentiero}, Rita and {Castellano}, Simone and {Castronuovo}, Marco and {Cavalli}, Luca and {Cavazzuti}, Elisabetta and {Ceccanti}, Marco and {Centrone}, Mauro and {Citraro}, Saverio and {D'Amico}, Fabio and {D'Alba}, Elisa and {Di Gesu}, Laura and {Del Monte}, Ettore and {Dietz}, Kurtis L. and {Di Lalla}, Niccolo' and {Persio}, Giuseppe Di and {Dolan}, David and {Donnarumma}, Immacolata and {Evangelista}, Yuri and {Ferrant}, Kevin and {Ferrazzoli}, Riccardo and {Ferrie}, MacKenzie and {Footdale}, Joseph and {Forsyth}, Brent and {Foster}, Michelle and {Garelick}, Benjamin and {Gunji}, Shuichi and {Gurnee}, Eli and {Head}, Michael and {Hibbard}, Grant and {Johnson}, Samantha and {Kelly}, Erik and {Kilaru}, Kiranmayee and {Lefevre}, Carlo and {Roy}, Shelley Le and {Loffredo}, Pasqualino and {Lorenzi}, Paolo and {Lucchesi}, Leonardo and {Maddox}, Tyler and {Magazzu}, Guido and {Maldera}, Simone and {Manfreda}, Alberto and {Mangraviti}, Elio and {Marengo}, Marco and {Marrocchesi}, Alessandra and {Massaro}, Francesco and {Mauger}, David and {McCracken}, Jeffrey and {McEachen}, Michael and {Mize}, Rondal and {Mereu}, Paolo and {Mitchell}, Scott and {Mitsuishi}, Ikuyuki and {Morbidini}, Alfredo and {Mosti}, Federico and {Nasimi}, Hikmat and {Negri}, Barbara and {Negro}, Michela and {Nguyen}, Toan and {Nitschke}, Isaac and {Nuti}, Alessio and {Onizuka}, Mitch and {Oppedisano}, Chiara and {Orsini}, Leonardo and {Osborne}, Darren and {Pacheco}, Richard and {Paggi}, Alessandro and {Painter}, Will and {Pavelitz}, Steven D. and {Pentz}, Christina and {Piazzolla}, Raffaele and {Perri}, Matteo and {Pesce-Rollins}, Melissa and {Peterson}, Colin and {Pilia}, Maura and {Profeti}, Alessandro and {Puccetti}, Simonetta and {Ranganathan}, Jaganathan and {Ratheesh}, Ajay and {Reedy}, Lee and {Root}, Noah and {Rubini}, Alda and {Ruswick}, Stephanie and {Sanchez}, Javier and {Sarra}, Paolo and {Santoli}, Francesco and {Scalise}, Emanuele and {Sciortino}, Andrea and {Schroeder}, Christopher and {Seek}, Tim and {Sosdian}, Kalie and {Spandre}, Gloria and {Speegle}, Chet O. and {Tamagawa}, Toru and {Tardiola}, Marcello and {Tobia}, Antonino and {Thomas}, Nicholas E. and {Valerie}, Robert and {Vimercati}, Marco and {Walden}, Amy L. and {Weddendorf}, Bruce and {Wedmore}, Jeffrey and {Welch}, David and {Zanetti}, Davide and {Zanetti}, Francesco},
        title = "{The Imaging X-Ray Polarimetry Explorer (IXPE): Pre-Launch}",
      journal = {Journal of Astronomical Telescopes, Instruments, and Systems},
         year = 2022,
        month = apr,
       volume = {8},
       number = {2},
          eid = {026002},
        pages = {026002},
          doi = {10.1117/1.JATIS.8.2.026002},
archivePrefix = {arXiv},
       eprint = {2112.01269},
 primaryClass = {astro-ph.IM},
       adsurl = {https://ui.adsabs.harvard.edu/abs/2022JATIS...8b6002W}
}

@ARTICLE{2023ApJ...958L..16V,
       author = {{Veledina}, Alexandra and {Muleri}, Fabio and {Dov{\v{c}}iak}, Michal and {Poutanen}, Juri and {Ratheesh}, Ajay and {Capitanio}, Fiamma and {Matt}, Giorgio and {Soffitta}, Paolo and {Tennant}, Allyn F. and {Negro}, Michela and {Kaaret}, Philip and {Costa}, Enrico and {Ingram}, Adam and {Svoboda}, Ji{\v{r}}{\'\i} and {Krawczynski}, Henric and {Bianchi}, Stefano and {Steiner}, James F. and {Garc{\'\i}a}, Javier A. and {Kravtsov}, Vadim and {Nitindala}, Anagha P. and {Ewing}, Melissa and {Mastroserio}, Guglielmo and {Marinucci}, Andrea and {Ursini}, Francesco and {Tombesi}, Francesco and {Tsygankov}, Sergey S. and {Yang}, Yi-Jung and {Weisskopf}, Martin C. and {Trushkin}, Sergei A. and {Egron}, Elise and {Iacolina}, Maria Noemi and {Pilia}, Maura and {Marra}, Lorenzo and {Miku{\v{s}}incov{\'a}}, Romana and {Nathan}, Edward and {Parra}, Maxime and {Petrucci}, Pierre-Olivier and {Podgorn{\'y}}, Jakub and {Tugliani}, Stefano and {Zane}, Silvia and {Zhang}, Wenda and {Agudo}, Iv{\'a}n and {Antonelli}, Lucio A. and {Bachetti}, Matteo and {Baldini}, Luca and {Baumgartner}, Wayne H. and {Bellazzini}, Ronaldo and {Bongiorno}, Stephen D. and {Bonino}, Raffaella and {Brez}, Alessandro and {Bucciantini}, Niccol{\`o} and {Castellano}, Simone and {Cavazzuti}, Elisabetta and {Chen}, Chien-Ting and {Ciprini}, Stefano and {De Rosa}, Alessandra and {Del Monte}, Ettore and {Di Gesu}, Laura and {Di Lalla}, Niccol{\`o} and {Di Marco}, Alessandro and {Donnarumma}, Immacolata and {Doroshenko}, Victor and {Ehlert}, Steven R. and {Enoto}, Teruaki and {Evangelista}, Yuri and {Fabiani}, Sergio and {Ferrazzoli}, Riccardo and {Gunji}, Shuichi and {Hayashida}, Kiyoshi and {Heyl}, Jeremy and {Iwakiri}, Wataru and {Jorstad}, Svetlana G. and {Karas}, Vladimir and {Kislat}, Fabian and {Kitaguchi}, Takao and {Kolodziejczak}, Jeffery J. and {La Monaca}, Fabio and {Latronico}, Luca and {Liodakis}, Ioannis and {Maldera}, Simone and {Manfreda}, Alberto and {Marin}, Fr{\'e}d{\'e}ric and {Marscher}, Alan P. and {Marshall}, Herman L. and {Massaro}, Francesco and {Mitsuishi}, Ikuyuki and {Mizuno}, Tsunefumi and {Ng}, Chi-Yung and {O'Dell}, Stephen L. and {Omodei}, Nicola and {Oppedisano}, Chiara and {Papitto}, Alessandro and {Pavlov}, George G. and {Peirson}, Abel L. and {Perri}, Matteo and {Pesce-Rollins}, Melissa and {Possenti}, Andrea and {Puccetti}, Simonetta and {Ramsey}, Brian D. and {Rankin}, John and {Roberts}, Oliver J. and {Romani}, Roger W. and {Sgr{\`o}}, Carmelo and {Slane}, Patrick and {Spandre}, Gloria and {Swartz}, Douglas A. and {Tamagawa}, Toru and {Tavecchio}, Fabrizio and {Taverna}, Roberto and {Tawara}, Yuzuru and {Thomas}, Nicholas E. and {Trois}, Alessio and {Turolla}, Roberto and {Vink}, Jacco and {Wu}, Kinwah and {Xie}, Fei},
        title = "{Discovery of X-Ray Polarization from the Black Hole Transient Swift J1727.8-1613}",
      journal = {\apjl},
         year = 2023,
        month = nov,
       volume = {958},
       number = {1},
          eid = {L16},
        pages = {L16},
          doi = {10.3847/2041-8213/ad0781},
archivePrefix = {arXiv},
       eprint = {2309.15928},
 primaryClass = {astro-ph.HE},
       adsurl = {https://ui.adsabs.harvard.edu/abs/2023ApJ...958L..16V}
}

@ARTICLE{2025A&A...701A.115K,
       author = {{Kravtsov}, Vadim and {Bocharova}, Anastasiia and {Veledina}, Alexandra and {Poutanen}, Juri and {Hughes}, Andrew K. and {Dov{\v{c}}iak}, Michal and {Egron}, Elise and {Muleri}, Fabio and {Podgorny}, Jakub and {Svoboda}, Ji{\v{r}}i and {Forsblom}, Sofia V. and {Berdyugin}, Andrei V. and {Blinov}, Dmitry and {Bright}, Joe S. and {Carotenuto}, Francesco and {Green}, David A. and {Ingram}, Adam and {Liodakis}, Ioannis and {Mandarakas}, Nikos and {Nitindala}, Anagha P. and {Rhodes}, Lauren and {Trushkin}, Sergei A. and {Tsygankov}, Sergey S. and {Brigitte}, Ma{\"\i}mouna and {Di Marco}, Alessandro and {Iacolina}, Noemi and {Krawczynski}, Henric and {La Monaca}, Fabio and {Loktev}, Vladislav and {Mastroserio}, Guglielmo and {Petrucci}, Pierre-Olivier and {Pilia}, Maura and {Tombesi}, Francesco and {Zdziarski}, Andrzej A.},
        title = "{Variability of X-ray polarization of Cyg X-1}",
      journal = {\aap},
         year = 2025,
        month = sep,
       volume = {701},
          eid = {A115},
        pages = {A115},
          doi = {10.1051/0004-6361/202555411},
archivePrefix = {arXiv},
       eprint = {2505.03942},
 primaryClass = {astro-ph.HE},
       adsurl = {https://ui.adsabs.harvard.edu/abs/2025A&A...701A.115K}
}

@ARTICLE{2022SoftX..1901194B,
       author = {{Baldini}, Luca and {Bucciantini}, Niccol{\`o} and {Lalla}, Niccol{\`o} Di and {Ehlert}, Steven and {Manfreda}, Alberto and {Negro}, Michela and {Omodei}, Nicola and {Pesce-Rollins}, Melissa and {Sgr{\`o}}, Carmelo and {Silvestri}, Stefano},
        title = "{ixpeobssim: A simulation and analysis framework for the imaging X-ray polarimetry explorer}",
      journal = {SoftwareX},
         year = 2022,
        month = jul,
       volume = {19},
          eid = {101194},
        pages = {101194},
          doi = {10.1016/j.softx.2022.101194},
archivePrefix = {arXiv},
       eprint = {2203.06384},
 primaryClass = {astro-ph.IM},
       adsurl = {https://ui.adsabs.harvard.edu/abs/2022SoftX..1901194B}
}

@ARTICLE{2015APh....68...45K,
       author = {{Kislat}, F. and {Clark}, B. and {Beilicke}, M. and {Krawczynski}, H.},
        title = "{Analyzing the data from X-ray polarimeters with Stokes parameters}",
      journal = {Astroparticle Physics},
         year = 2015,
        month = aug,
       volume = {68},
        pages = {45-51},
          doi = {10.1016/j.astropartphys.2015.02.007},
archivePrefix = {arXiv},
       eprint = {1409.6214},
 primaryClass = {astro-ph.IM},
       adsurl = {https://ui.adsabs.harvard.edu/abs/2015APh....68...45K}
}

@BOOK{Jeffreys1986,
    title = "{Theory of Probability}",
    author = {Harold Jeffreys},
    year = 1998,
    edition = {3rd},
    publisher = "{Oxford University Press}",
    location = "{England}"
}

@ARTICLE{2023AJ....165..143D,
       author = {{Di Marco}, Alessandro and {Soffitta}, Paolo and {Costa}, Enrico and {Ferrazzoli}, Riccardo and {La Monaca}, Fabio and {Rankin}, John and {Ratheesh}, Ajay and {Xie}, Fei and {Baldini}, Luca and {Del Monte}, Ettore and {Ehlert}, Steven R. and {Fabiani}, Sergio and {Kim}, Dawoon E. and {Muleri}, Fabio and {O'Dell}, Stephen L. and {Ramsey}, Brian D. and {Rubini}, Alda and {Sgr{\`o}}, Carmelo and {Silvestri}, Stefano and {Tennant}, Allyn F. and {Weisskopf}, Martin C.},
        title = "{Handling the Background in IXPE Polarimetric Data}",
      journal = {\aj},
         year = 2023,
        month = apr,
       volume = {165},
       number = {4},
          eid = {143},
        pages = {143},
          doi = {10.3847/1538-3881/acba0f},
archivePrefix = {arXiv},
       eprint = {2302.02927},
 primaryClass = {astro-ph.IM},
       adsurl = {https://ui.adsabs.harvard.edu/abs/2023AJ....165..143D}
}

@ARTICLE{2019ApJS..241...27A,
       author = {{Ashton}, Gregory and {H{\"u}bner}, Moritz and {Lasky}, Paul D. and {Talbot}, Colm and {Ackley}, Kendall and {Biscoveanu}, Sylvia and {Chu}, Qi and {Divakarla}, Atul and {Easter}, Paul J. and {Goncharov}, Boris and {Hernandez Vivanco}, Francisco and {Harms}, Jan and {Lower}, Marcus E. and {Meadors}, Grant D. and {Melchor}, Denyz and {Payne}, Ethan and {Pitkin}, Matthew D. and {Powell}, Jade and {Sarin}, Nikhil and {Smith}, Rory J.~E. and {Thrane}, Eric},
        title = "{BILBY: A User-friendly Bayesian Inference Library for Gravitational-wave Astronomy}",
      journal = {\apjs},
         year = 2019,
        month = apr,
       volume = {241},
       number = {2},
          eid = {27},
        pages = {27},
          doi = {10.3847/1538-4365/ab06fc},
archivePrefix = {arXiv},
       eprint = {1811.02042},
 primaryClass = {astro-ph.IM},
       adsurl = {https://ui.adsabs.harvard.edu/abs/2019ApJS..241...27A}
}

@ARTICLE{2020MNRAS.493.3132S,
       author = {{Speagle}, Joshua S.},
        title = "{DYNESTY: a dynamic nested sampling package for estimating Bayesian posteriors and evidences}",
      journal = {\mnras},
         year = 2020,
        month = apr,
       volume = {493},
       number = {3},
        pages = {3132-3158},
          doi = {10.1093/mnras/staa278},
archivePrefix = {arXiv},
       eprint = {1904.02180},
 primaryClass = {astro-ph.IM},
       adsurl = {https://ui.adsabs.harvard.edu/abs/2020MNRAS.493.3132S}
}

@ARTICLE{1984PASJ...36..741M,
       author = {{Mitsuda}, K. and {Inoue}, H. and {Koyama}, K. and {Makishima}, K. and {Matsuoka}, M. and {Ogawara}, Y. and {Shibazaki}, N. and {Suzuki}, K. and {Tanaka}, Y. and {Hirano}, T.},
        title = "{Energy spectra of low-mass binary X-ray sources observed from Tenma.}",
      journal = {\pasj},
         year = 1984,
        month = jan,
       volume = {36},
        pages = {741-759},
       adsurl = {https://ui.adsabs.harvard.edu/abs/1984PASJ...36..741M}
}

@ARTICLE{1996ApJ...470..249P,
       author = {{Poutanen}, Juri and {Svensson}, Roland},
        title = "{The Two-Phase Pair Corona Model for Active Galactic Nuclei and X-Ray Binaries: How to Obtain Exact Solutions}",
      journal = {\apj},
         year = 1996,
        month = oct,
       volume = {470},
        pages = {249},
          doi = {10.1086/177865},
archivePrefix = {arXiv},
       eprint = {astro-ph/9605073},
 primaryClass = {astro-ph},
       adsurl = {https://ui.adsabs.harvard.edu/abs/1996ApJ...470..249P}
}

@software{ravi2025queenbee,
  author       = {{Ravi}, Swati and {Ng}, Mason and {Marshall}, Herman L.},
  title        = {QUEEN-BEE},
  version      = {v3.2},
  year         = 2025,
  publisher    = {Zenodo},
  doi          = {10.5281/zenodo.17068831},
  url          = {https://doi.org/10.5281/zenodo.17068831}
}

@article{astropy:2013,
Adsurl = {http://adsabs.harvard.edu/abs/2013A%26A...558A..33A},
Archiveprefix = {arXiv},
Author = {{\!\!Astropy Collaboration} and {Robitaille}, T.~P. and {Tollerud}, E.~J. and {Greenfield}, P. and {Droettboom}, M. and {Bray}, E. and {Aldcroft}, T. and {Davis}, M. and {Ginsburg}, A. and {Price-Whelan}, A.~M. and {Kerzendorf}, W.~E. and {Conley}, A. and {Crighton}, N. and {Barbary}, K. and {Muna}, D. and {Ferguson}, H. and {Grollier}, F. and {Parikh}, M.~M. and {Nair}, P.~H. and {Unther}, H.~M. and {Deil}, C. and {Woillez}, J. and {Conseil}, S. and {Kramer}, R. and {Turner}, J.~E.~H. and {Singer}, L. and {Fox}, R. and {Weaver}, B.~A. and {Zabalza}, V. and {Edwards}, Z.~I. and {Azalee Bostroem}, K. and {Burke}, D.~J. and {Casey}, A.~R. and {Crawford}, S.~M. and {Dencheva}, N. and {Ely}, J. and {Jenness}, T. and {Labrie}, K. and {Lim}, P.~L. and {Pierfederici}, F. and {Pontzen}, A. and {Ptak}, A. and {Refsdal}, B. and {Servillat}, M. and {Streicher}, O.},
Doi = {10.1051/0004-6361/201322068},
Eid = {A33},
Eprint = {1307.6212},
Journal = {\aap},
Month = oct,
Pages = {A33},
Primaryclass = {astro-ph.IM},
Title = {{Astropy: A community Python package for astronomy}},
Volume = 558,
Year = 2013}

@ARTICLE{astropy:2018,
       author = {{Astropy Collaboration} and {Price-Whelan}, A.~M. and
         {Sip{\H{o}}cz}, B.~M. and {G{\"u}nther}, H.~M. and {Lim}, P.~L. and
         {Crawford}, S.~M. and {Conseil}, S. and {Shupe}, D.~L. and
         {Craig}, M.~W. and {Dencheva}, N. and {Ginsburg}, A. and {Vand
        erPlas}, J.~T. and {Bradley}, L.~D. and {P{\'e}rez-Su{\'a}rez}, D. and
         {de Val-Borro}, M. and {Aldcroft}, T.~L. and {Cruz}, K.~L. and
         {Robitaille}, T.~P. and {Tollerud}, E.~J. and {Ardelean}, C. and
         {Babej}, T. and {Bach}, Y.~P. and {Bachetti}, M. and {Bakanov}, A.~V. and
         {Bamford}, S.~P. and {Barentsen}, G. and {Barmby}, P. and
         {Baumbach}, A. and {Berry}, K.~L. and {Biscani}, F. and {Boquien}, M. and
         {Bostroem}, K.~A. and {Bouma}, L.~G. and {Brammer}, G.~B. and
         {Bray}, E.~M. and {Breytenbach}, H. and {Buddelmeijer}, H. and
         {Burke}, D.~J. and {Calderone}, G. and {Cano Rodr{\'\i}guez}, J.~L. and
         {Cara}, M. and {Cardoso}, J.~V.~M. and {Cheedella}, S. and {Copin}, Y. and
         {Corrales}, L. and {Crichton}, D. and {D'Avella}, D. and {Deil}, C. and
         {Depagne}, {\'E}. and {Dietrich}, J.~P. and {Donath}, A. and
         {Droettboom}, M. and {Earl}, N. and {Erben}, T. and {Fabbro}, S. and
         {Ferreira}, L.~A. and {Finethy}, T. and {Fox}, R.~T. and
         {Garrison}, L.~H. and {Gibbons}, S.~L.~J. and {Goldstein}, D.~A. and
         {Gommers}, R. and {Greco}, J.~P. and {Greenfield}, P. and
         {Groener}, A.~M. and {Grollier}, F. and {Hagen}, A. and {Hirst}, P. and
         {Homeier}, D. and {Horton}, A.~J. and {Hosseinzadeh}, G. and {Hu}, L. and
         {Hunkeler}, J.~S. and {Ivezi{\'c}}, {\v{Z}}. and {Jain}, A. and
         {Jenness}, T. and {Kanarek}, G. and {Kendrew}, S. and {Kern}, N.~S. and
         {Kerzendorf}, W.~E. and {Khvalko}, A. and {King}, J. and {Kirkby}, D. and
         {Kulkarni}, A.~M. and {Kumar}, A. and {Lee}, A. and {Lenz}, D. and
         {Littlefair}, S.~P. and {Ma}, Z. and {Macleod}, D.~M. and
         {Mastropietro}, M. and {McCully}, C. and {Montagnac}, S. and
         {Morris}, B.~M. and {Mueller}, M. and {Mumford}, S.~J. and {Muna}, D. and
         {Murphy}, N.~A. and {Nelson}, S. and {Nguyen}, G.~H. and
         {Ninan}, J.~P. and {N{\"o}the}, M. and {Ogaz}, S. and {Oh}, S. and
         {Parejko}, J.~K. and {Parley}, N. and {Pascual}, S. and {Patil}, R. and
         {Patil}, A.~A. and {Plunkett}, A.~L. and {Prochaska}, J.~X. and
         {Rastogi}, T. and {Reddy Janga}, V. and {Sabater}, J. and
         {Sakurikar}, P. and {Seifert}, M. and {Sherbert}, L.~E. and
         {Sherwood-Taylor}, H. and {Shih}, A.~Y. and {Sick}, J. and
         {Silbiger}, M.~T. and {Singanamalla}, S. and {Singer}, L.~P. and
         {Sladen}, P.~H. and {Sooley}, K.~A. and {Sornarajah}, S. and
         {Streicher}, O. and {Teuben}, P. and {Thomas}, S.~W. and
         {Tremblay}, G.~R. and {Turner}, J.~E.~H. and {Terr{\'o}n}, V. and
         {van Kerkwijk}, M.~H. and {de la Vega}, A. and {Watkins}, L.~L. and
         {Weaver}, B.~A. and {Whitmore}, J.~B. and {Woillez}, J. and
         {Zabalza}, V. and {Astropy Contributors}},
        title = "{The Astropy Project: Building an Open-science Project and Status of the v2.0 Core Package}",
      journal = {\aj},
         year = 2018,
        month = sep,
       volume = {156},
       number = {3},
          eid = {123},
        pages = {123},
          doi = {10.3847/1538-3881/aabc4f},
archivePrefix = {arXiv},
       eprint = {1801.02634},
 primaryClass = {astro-ph.IM},
       adsurl = {https://ui.adsabs.harvard.edu/abs/2018AJ....156..123A}
}

@ARTICLE{astropy:2022,
       author = {{Astropy Collaboration} and {Price-Whelan}, Adrian M. and {Lim}, Pey Lian and {Earl}, Nicholas and {Starkman}, Nathaniel and {Bradley}, Larry and {Shupe}, David L. and {Patil}, Aarya A. and {Corrales}, Lia and {Brasseur}, C.~E. and {N{"o}the}, Maximilian and {Donath}, Axel and {Tollerud}, Erik and {Morris}, Brett M. and {Ginsburg}, Adam and {Vaher}, Eero and {Weaver}, Benjamin A. and {Tocknell}, James and {Jamieson}, William and {van Kerkwijk}, Marten H. and {Robitaille}, Thomas P. and {Merry}, Bruce and {Bachetti}, Matteo and {G{"u}nther}, H. Moritz and {Aldcroft}, Thomas L. and {Alvarado-Montes}, Jaime A. and {Archibald}, Anne M. and {B{'o}di}, Attila and {Bapat}, Shreyas and {Barentsen}, Geert and {Baz{'a}n}, Juanjo and {Biswas}, Manish and {Boquien}, M{'e}d{'e}ric and {Burke}, D.~J. and {Cara}, Daria and {Cara}, Mihai and {Conroy}, Kyle E. and {Conseil}, Simon and {Craig}, Matthew W. and {Cross}, Robert M. and {Cruz}, Kelle L. and {D'Eugenio}, Francesco and {Dencheva}, Nadia and {Devillepoix}, Hadrien A.~R. and {Dietrich}, J{"o}rg P. and {Eigenbrot}, Arthur Davis and {Erben}, Thomas and {Ferreira}, Leonardo and {Foreman-Mackey}, Daniel and {Fox}, Ryan and {Freij}, Nabil and {Garg}, Suyog and {Geda}, Robel and {Glattly}, Lauren and {Gondhalekar}, Yash and {Gordon}, Karl D. and {Grant}, David and {Greenfield}, Perry and {Groener}, Austen M. and {Guest}, Steve and {Gurovich}, Sebastian and {Handberg}, Rasmus and {Hart}, Akeem and {Hatfield-Dodds}, Zac and {Homeier}, Derek and {Hosseinzadeh}, Griffin and {Jenness}, Tim and {Jones}, Craig K. and {Joseph}, Prajwel and {Kalmbach}, J. Bryce and {Karamehmetoglu}, Emir and {Ka{l}uszy{'n}ski}, Miko{l}aj and {Kelley}, Michael S.~P. and {Kern}, Nicholas and {Kerzendorf}, Wolfgang E. and {Koch}, Eric W. and {Kulumani}, Shankar and {Lee}, Antony and {Ly}, Chun and {Ma}, Zhiyuan and {MacBride}, Conor and {Maljaars}, Jakob M. and {Muna}, Demitri and {Murphy}, N.~A. and {Norman}, Henrik and {O'Steen}, Richard and {Oman}, Kyle A. and {Pacifici}, Camilla and {Pascual}, Sergio and {Pascual-Granado}, J. and {Patil}, Rohit R. and {Perren}, Gabriel I. and {Pickering}, Timothy E. and {Rastogi}, Tanuj and {Roulston}, Benjamin R. and {Ryan}, Daniel F. and {Rykoff}, Eli S. and {Sabater}, Jose and {Sakurikar}, Parikshit and {Salgado}, Jes{'u}s and {Sanghi}, Aniket and {Saunders}, Nicholas and {Savchenko}, Volodymyr and {Schwardt}, Ludwig and {Seifert-Eckert}, Michael and {Shih}, Albert Y. and {Jain}, Anany Shrey and {Shukla}, Gyanendra and {Sick}, Jonathan and {Simpson}, Chris and {Singanamalla}, Sudheesh and {Singer}, Leo P. and {Singhal}, Jaladh and {Sinha}, Manodeep and {Sip{H{o}}cz}, Brigitta M. and {Spitler}, Lee R. and {Stansby}, David and {Streicher}, Ole and {{{S}}umak}, Jani and {Swinbank}, John D. and {Taranu}, Dan S. and {Tewary}, Nikita and {Tremblay}, Grant R. and {Val-Borro}, Miguel de and {Van Kooten}, Samuel J. and {Vasovi{'c}}, Zlatan and {Verma}, Shresth and {de Miranda Cardoso}, Jos{'e} Vin{'i}cius and {Williams}, Peter K.~G. and {Wilson}, Tom J. and {Winkel}, Benjamin and {Wood-Vasey}, W.~M. and {Xue}, Rui and {Yoachim}, Peter and {Zhang}, Chen and {Zonca}, Andrea and {Astropy Project Contributors}},
        title = "{The Astropy Project: Sustaining and Growing a Community-oriented Open-source Project and the Latest Major Release (v5.0) of the Core Package}",
      journal = {\apj},
         year = 2022,
        month = aug,
       volume = {935},
       number = {2},
          eid = {167},
        pages = {167},
          doi = {10.3847/1538-4357/ac7c74},
archivePrefix = {arXiv},
       eprint = {2206.14220},
 primaryClass = {astro-ph.IM},
       adsurl = {https://ui.adsabs.harvard.edu/abs/2022ApJ...935..167A}
}

@Article{         harris2020array,
 title         = {Array programming with {NumPy}},
 author        = {Charles R. Harris and K. Jarrod Millman and St{\'{e}}fan J.
                 van der Walt and Ralf Gommers and Pauli Virtanen and David
                 Cournapeau and Eric Wieser and Julian Taylor and Sebastian
                 Berg and Nathaniel J. Smith and Robert Kern and Matti Picus
                 and Stephan Hoyer and Marten H. van Kerkwijk and Matthew
                 Brett and Allan Haldane and Jaime Fern{\'{a}}ndez del
                 R{\'{i}}o and Mark Wiebe and Pearu Peterson and Pierre
                 G{\'{e}}rard-Marchant and Kevin Sheppard and Tyler Reddy and
                 Warren Weckesser and Hameer Abbasi and Christoph Gohlke and
                 Travis E. Oliphant},
 year          = {2020},
 month         = sep,
 journal       = {Nature},
 volume        = {585},
 number        = {7825},
 pages         = {357--362},
 doi           = {10.1038/s41586-020-2649-2},
 publisher     = {Springer Science and Business Media {LLC}},
 url           = {https://doi.org/10.1038/s41586-020-2649-2}
}

@ARTICLE{2020SciPy-NMeth,
  author  = {Virtanen, Pauli and Gommers, Ralf and Oliphant, Travis E. and
            Haberland, Matt and Reddy, Tyler and Cournapeau, David and
            Burovski, Evgeni and Peterson, Pearu and Weckesser, Warren and
            Bright, Jonathan and {van der Walt}, St{\'e}fan J. and
            Brett, Matthew and Wilson, Joshua and Millman, K. Jarrod and
            Mayorov, Nikolay and Nelson, Andrew R. J. and Jones, Eric and
            Kern, Robert and Larson, Eric and Carey, C J and
            Polat, {\.I}lhan and Feng, Yu and Moore, Eric W. and
            {VanderPlas}, Jake and Laxalde, Denis and Perktold, Josef and
            Cimrman, Robert and Henriksen, Ian and Quintero, E. A. and
            Harris, Charles R. and Archibald, Anne M. and
            Ribeiro, Ant{\^o}nio H. and Pedregosa, Fabian and
            {van Mulbregt}, Paul and {SciPy 1.0 Contributors}},
  title   = {{{SciPy} 1.0: Fundamental Algorithms for Scientific
            Computing in Python}},
  journal = {Nature Methods},
  year    = {2020},
  volume  = {17},
  pages   = {261--272},
  adsurl  = {https://rdcu.be/b08Wh},
  doi     = {10.1038/s41592-019-0686-2},
}

@Article{PER-GRA:2007,
  Author    = {P\'erez, Fernando and Granger, Brian E.},
  Title     = {{IP}ython: a System for Interactive Scientific Computing},
  Journal   = {Computing in Science and Engineering},
  Volume    = {9},
  Number    = {3},
  Pages     = {21--29},
  month     = may,
  year      = 2007,
  url       = "https://ipython.org",
  ISSN      = "1521-9615",
  doi       = {10.1109/MCSE.2007.53},
  publisher = {IEEE Computer Society},
}

@ARTICLE{2013ApJ...770..103H,
       author = {{Harrison}, Fiona A. and {Craig}, William W. and {Christensen}, Finn E. and {Hailey}, Charles J. and {Zhang}, William W. and {Boggs}, Steven E. and {Stern}, Daniel and {Cook}, W. Rick and {Forster}, Karl and {Giommi}, Paolo and {Grefenstette}, Brian W. and {Kim}, Yunjin and {Kitaguchi}, Takao and {Koglin}, Jason E. and {Madsen}, Kristin K. and {Mao}, Peter H. and {Miyasaka}, Hiromasa and {Mori}, Kaya and {Perri}, Matteo and {Pivovaroff}, Michael J. and {Puccetti}, Simonetta and {Rana}, Vikram R. and {Westergaard}, Niels J. and {Willis}, Jason and {Zoglauer}, Andreas and {An}, Hongjun and {Bachetti}, Matteo and {Barri{\`e}re}, Nicolas M. and {Bellm}, Eric C. and {Bhalerao}, Varun and {Brejnholt}, Nicolai F. and {Fuerst}, Felix and {Liebe}, Carl C. and {Markwardt}, Craig B. and {Nynka}, Melania and {Vogel}, Julia K. and {Walton}, Dominic J. and {Wik}, Daniel R. and {Alexander}, David M. and {Cominsky}, Lynn R. and {Hornschemeier}, Ann E. and {Hornstrup}, Allan and {Kaspi}, Victoria M. and {Madejski}, Greg M. and {Matt}, Giorgio and {Molendi}, Silvano and {Smith}, David M. and {Tomsick}, John A. and {Ajello}, Marco and {Ballantyne}, David R. and {Balokovi{\'c}}, Mislav and {Barret}, Didier and {Bauer}, Franz E. and {Blandford}, Roger D. and {Brandt}, W. Niel and {Brenneman}, Laura W. and {Chiang}, James and {Chakrabarty}, Deepto and {Chenevez}, Jerome and {Comastri}, Andrea and {Dufour}, Francois and {Elvis}, Martin and {Fabian}, Andrew C. and {Farrah}, Duncan and {Fryer}, Chris L. and {Gotthelf}, Eric V. and {Grindlay}, Jonathan E. and {Helfand}, David J. and {Krivonos}, Roman and {Meier}, David L. and {Miller}, Jon M. and {Natalucci}, Lorenzo and {Ogle}, Patrick and {Ofek}, Eran O. and {Ptak}, Andrew and {Reynolds}, Stephen P. and {Rigby}, Jane R. and {Tagliaferri}, Gianpiero and {Thorsett}, Stephen E. and {Treister}, Ezequiel and {Urry}, C. Megan},
        title = "{The Nuclear Spectroscopic Telescope Array (NuSTAR) High-energy X-Ray Mission}",
      journal = {\apj},
         year = 2013,
        month = jun,
       volume = {770},
       number = {2},
          eid = {103},
        pages = {103},
          doi = {10.1088/0004-637X/770/2/103},
archivePrefix = {arXiv},
       eprint = {1301.7307},
 primaryClass = {astro-ph.IM},
       adsurl = {https://ui.adsabs.harvard.edu/abs/2013ApJ...770..103H}
}

@ARTICLE{2014SSRv..183..295M,
       author = {{McClintock}, Jeffrey E. and {Narayan}, Ramesh and {Steiner}, James F.},
        title = "{Black Hole Spin via Continuum Fitting and the Role of Spin in Powering Transient Jets}",
      journal = {\ssr},
         year = 2014,
        month = sep,
       volume = {183},
       number = {1-4},
        pages = {295-322},
          doi = {10.1007/s11214-013-0003-9},
archivePrefix = {arXiv},
       eprint = {1303.1583},
 primaryClass = {astro-ph.HE},
       adsurl = {https://ui.adsabs.harvard.edu/abs/2014SSRv..183..295M}
}

@ARTICLE{2006ARA&A..44...49R,
       author = {{Remillard}, Ronald A. and {McClintock}, Jeffrey E.},
        title = "{X-Ray Properties of Black-Hole Binaries}",
      journal = {\araa},
         year = 2006,
        month = sep,
       volume = {44},
       number = {1},
        pages = {49-92},
          doi = {10.1146/annurev.astro.44.051905.092532},
archivePrefix = {arXiv},
       eprint = {astro-ph/0606352},
 primaryClass = {astro-ph},
       adsurl = {https://ui.adsabs.harvard.edu/abs/2006ARA&A..44...49R}
}

@ARTICLE{2007A&ARv..15....1D,
       author = {{Done}, Chris and {Gierli{\'n}ski}, Marek and {Kubota}, Aya},
        title = "{Modelling the behaviour of accretion flows in X-ray binaries. Everything you always wanted to know about accretion but were afraid to ask}",
      journal = {\aapr},
         year = 2007,
        month = dec,
       volume = {15},
       number = {1},
        pages = {1-66},
          doi = {10.1007/s00159-007-0006-1},
archivePrefix = {arXiv},
       eprint = {0708.0148},
 primaryClass = {astro-ph},
       adsurl = {https://ui.adsabs.harvard.edu/abs/2007A&ARv..15....1D}
}

@ARTICLE{2009ApJ...701.1175S,
       author = {{Schnittman}, Jeremy D. and {Krolik}, Julian H.},
        title = "{X-ray Polarization from Accreting Black Holes: The Thermal State}",
      journal = {\apj},
         year = 2009,
        month = aug,
       volume = {701},
       number = {2},
        pages = {1175-1187},
          doi = {10.1088/0004-637X/701/2/1175},
archivePrefix = {arXiv},
       eprint = {0902.3982},
 primaryClass = {astro-ph.HE},
       adsurl = {https://ui.adsabs.harvard.edu/abs/2009ApJ...701.1175S}
}

@ARTICLE{2008MNRAS.391...32D,
       author = {{Dov{\v{c}}iak}, M. and {Muleri}, F. and {Goosmann}, R.~W. and {Karas}, V. and {Matt}, G.},
        title = "{Thermal disc emission from a rotating black hole: X-ray polarization signatures}",
      journal = {\mnras},
         year = 2008,
        month = nov,
       volume = {391},
       number = {1},
        pages = {32-38},
          doi = {10.1111/j.1365-2966.2008.13872.x},
archivePrefix = {arXiv},
       eprint = {0809.0418},
 primaryClass = {astro-ph},
       adsurl = {https://ui.adsabs.harvard.edu/abs/2008MNRAS.391...32D}
}

@ARTICLE{2023MNRAS.519.6138M,
       author = {{Mikusincova}, Romana and {Dovciak}, Michal and {Bursa}, Michal and {Lalla}, Niccolo Di and {Matt}, Giorgio and {Svoboda}, Jiri and {Taverna}, Roberto and {Zhang}, Wenda},
        title = "{X-ray polarimetry as a tool to measure the black hole spin in microquasars: simulations of IXPE capabilities}",
      journal = {\mnras},
         year = 2023,
        month = mar,
       volume = {519},
       number = {4},
        pages = {6138-6148},
          doi = {10.1093/mnras/stad077},
archivePrefix = {arXiv},
       eprint = {2301.04002},
 primaryClass = {astro-ph.HE},
       adsurl = {https://ui.adsabs.harvard.edu/abs/2023MNRAS.519.6138M}
}

@BOOK{1960ratr.book.....C,
       author = {{Chandrasekhar}, Subrahmanyan},
        title = "{Radiative transfer}",
         year = 1960,
       adsurl = {https://ui.adsabs.harvard.edu/abs/1960ratr.book.....C}
}

@ARTICLE{1977Natur.269..128C,
       author = {{Connors}, P.~A. and {Stark}, R.~F.},
        title = "{Observable gravitational effects on polarised radiation coming from near a black hole}",
      journal = {\nat},
         year = 1977,
        month = sep,
       volume = {269},
       number = {5624},
        pages = {128-129},
          doi = {10.1038/269128a0},
       adsurl = {https://ui.adsabs.harvard.edu/abs/1977Natur.269..128C}
}

@ARTICLE{1977Natur.266..429S,
       author = {{Stark}, R.~F. and {Connors}, P.~A.},
        title = "{Observational test for the existence of a rotating black hole in Cyg X-1}",
      journal = {\nat},
         year = 1977,
        month = mar,
       volume = {266},
       number = {5601},
        pages = {429-430},
          doi = {10.1038/266429a0},
       adsurl = {https://ui.adsabs.harvard.edu/abs/1977Natur.266..429S}
}

@ARTICLE{2025ApJ...978L..19M,
       author = {{Mastroserio}, G. and {De Marco}, B. and {Baglio}, M.~C. and {Carotenuto}, F. and {Fabiani}, S. and {Russell}, T.~D. and {Capitanio}, F. and {Cavecchi}, Y. and {Motta}, S. and {Russell}, D.~M. and {Dov{\v{c}}iak}, M. and {Del Santo}, M. and {Alabarta}, K. and {Ambrifi}, A. and {Campana}, S. and {Casella}, P. and {Covino}, S. and {Illiano}, G. and {Kara}, E. and {Lai}, E.~V. and {Lodato}, G. and {Manca}, A. and {Mariani}, I. and {Marino}, A. and {Miceli}, C. and {Saikia}, P. and {Shaw}, A.~W. and {Svoboda}, J. and {Vincentelli}, F.~M. and {Wang}, J.},
        title = "{X-Ray and Optical Polarization Aligned with the Radio Jet Ejecta in GX 339─4}",
      journal = {\apjl},
         year = 2025,
        month = jan,
       volume = {978},
       number = {2},
          eid = {L19},
        pages = {L19},
          doi = {10.3847/2041-8213/ad9913},
archivePrefix = {arXiv},
       eprint = {2408.06856},
 primaryClass = {astro-ph.HE},
       adsurl = {https://ui.adsabs.harvard.edu/abs/2025ApJ...978L..19M}
}

@ARTICLE{2024ApJ...969L..30S,
       author = {{Steiner}, James F. and {Nathan}, Edward and {Hu}, Kun and {Krawczynski}, Henric and {Dov{\v{c}}iak}, Michal and {Veledina}, Alexandra and {Muleri}, Fabio and {Svoboda}, Jiri and {Alabarta}, Kevin and {Parra}, Maxime and {Bhargava}, Yash and {Matt}, Giorgio and {Poutanen}, Juri and {Petrucci}, Pierre-Olivier and {Tennant}, Allyn F. and {Baglio}, M. Cristina and {Baldini}, Luca and {Barnier}, Samuel and {Bhattacharyya}, Sudip and {Bianchi}, Stefano and {Brigitte}, Maimouna and {Cabezas}, Mauricio and {Cangemi}, Floriane and {Capitanio}, Fiamma and {Casey}, Jacob and {Rodriguez Cavero}, Nicole and {Castellano}, Simone and {Cavazzuti}, Elisabetta and {Chun}, Sohee and {Churazov}, Eugene and {Costa}, Enrico and {Di Lalla}, Niccol{\`o} and {Di Marco}, Alessandro and {Egron}, Elise and {Ewing}, Melissa and {Fabiani}, Sergio and {Garc{\'\i}a}, Javier A. and {Green}, David A. and {Grinberg}, Victoria and {Hadrava}, Petr and {Ingram}, Adam and {Kaaret}, Philip and {Kislat}, Fabian and {Kitaguchi}, Takao and {Kravtsov}, Vadim and {Kub{\'a}tov{\'a}}, Brankica and {La Monaca}, Fabio and {Latronico}, Luca and {Loktev}, Vladislav and {Malacaria}, Christian and {Marin}, Fr{\'e}d{\'e}ric and {Marinucci}, Andrea and {Maryeva}, Olga and {Mastroserio}, Guglielmo and {Mizuno}, Tsunefumi and {Negro}, Michela and {Omodei}, Nicola and {Podgorn{\'y}}, Jakub and {Rankin}, John and {Ratheesh}, Ajay and {Rhodes}, Lauren and {Russell}, David M. and {{\v{S}}lechta}, Miroslav and {Soffitta}, Paolo and {Spooner}, Sean and {Suleimanov}, Valery and {Tombesi}, Francesco and {Trushkin}, Sergei A. and {Weisskopf}, Martin C. and {Zane}, Silvia and {Zdziarski}, Andrzej A. and {Zhang}, Sixuan and {Zhang}, Wenda and {Zhou}, Menglei and {Agudo}, Iv{\'a}n and {Antonelli}, Lucio A. and {Bachetti}, Matteo and {Baumgartner}, Wayne H. and {Bellazzini}, Ronaldo and {Bongiorno}, Stephen D. and {Bonino}, Raffaella and {Brez}, Alessandro and {Bucciantini}, Niccol{\`o} and {Chen}, Chien-Ting and {Ciprini}, Stefano and {De Rosa}, Alessandra and {Del Monte}, Ettore and {Di Gesu}, Laura and {Donnarumma}, Immacolata and {Doroshenko}, Victor and {Ehlert}, Steven R. and {Enoto}, Teruaki and {Evangelista}, Yuri and {Ferrazzoli}, Riccardo and {Gunji}, Shuichi and {Hayashida}, Kiyoshi and {Heyl}, Jeremy and {Iwakiri}, Wataru and {Jorstad}, Svetlana G. and {Karas}, Vladimir and {Kolodziejczak}, Jeffery J. and {Liodakis}, Ioannis and {Maldera}, Simone and {Manfreda}, Alberto and {Marscher}, Alan P. and {Marshall}, Herman L. and {Massaro}, Francesco and {Mitsuishi}, Ikuyuki and {Ng}, Chi-Yung and {O'Dell}, Stephen L. and {Oppedisano}, Chiara and {Papitto}, Alessandro and {Pavlov}, George G. and {Peirson}, Abel L. and {Perri}, Matteo and {Pesce-Rollins}, Melissa and {Pilia}, Maura and {Possenti}, Andrea and {Puccetti}, Simonetta and {Ramsey}, Brian D. and {Roberts}, Oliver J. and {Romani}, Roger W. and {Sgr{\`o}}, Carmelo and {Slane}, Patrick and {Spandre}, Gloria and {Swartz}, Douglas A. and {Tamagawa}, Toru and {Tavecchio}, Fabrizio and {Taverna}, Roberto and {Tawara}, Yuzuru and {Thomas}, Nicholas E. and {Trois}, Alessio and {Tsygankov}, Sergey S. and {Turolla}, Roberto and {Vink}, Jacco and {Wu}, Kinwah and {Xie}, Fei},
        title = "{An IXPE-led X-Ray Spectropolarimetric Campaign on the Soft State of Cygnus X-1: X-Ray Polarimetric Evidence for Strong Gravitational Lensing}",
      journal = {\apjl},
         year = 2024,
        month = jul,
       volume = {969},
       number = {2},
          eid = {L30},
        pages = {L30},
          doi = {10.3847/2041-8213/ad58e4},
archivePrefix = {arXiv},
       eprint = {2406.12014},
 primaryClass = {astro-ph.HE},
       adsurl = {https://ui.adsabs.harvard.edu/abs/2024ApJ...969L..30S}
}

@ARTICLE{2024A&A...684A..95M,
       author = {{Marra}, L. and {Brigitte}, M. and {Rodriguez Cavero}, N. and {Chun}, S. and {Steiner}, J.~F. and {Dov{\v{c}}iak}, M. and {Nowak}, M. and {Bianchi}, S. and {Capitanio}, F. and {Ingram}, A. and {Matt}, G. and {Muleri}, F. and {Podgorn{\'y}}, J. and {Poutanen}, J. and {Svoboda}, J. and {Taverna}, R. and {Ursini}, F. and {Veledina}, A. and {De Rosa}, A. and {Garc{\'\i}a}, J.~A. and {Lutovinov}, A.~A. and {Mereminskiy}, I.~A. and {Farinelli}, R. and {Gunji}, S. and {Kaaret}, P. and {Kallman}, T. and {Krawczynski}, H. and {Kan}, Y. and {Hu}, K. and {Marinucci}, A. and {Mastroserio}, G. and {Mikus̆incov{\'a}}, R. and {Parra}, M. and {Petrucci}, P.-O. and {Ratheesh}, A. and {Soffitta}, P. and {Tombesi}, F. and {Zane}, S. and {Agudo}, I. and {Antonelli}, L.~A. and {Bachetti}, M. and {Baldini}, L. and {Baumgartner}, W.~H. and {Bellazzini}, R. and {Bongiorno}, S.~D. and {Bonino}, R. and {Brez}, A. and {Bucciantini}, N. and {Castellano}, S. and {Cavazzuti}, E. and {Chen}, C. and {Ciprini}, S. and {Costa}, E. and {Del Monte}, E. and {Di Gesu}, L. and {Di Lalla}, N. and {Di Marco}, A. and {Donnarumma}, I. and {Doroshenko}, V. and {Ehlert}, S.~R. and {Enoto}, T. and {Evangelista}, Y. and {Fabiani}, S. and {Ferrazzoli}, R. and {Hayashida}, K. and {Heyl}, J. and {Iwakiri}, W. and {Jorstad}, S.~G. and {Karas}, V. and {Kislat}, F. and {Kitaguchi}, T. and {Kolodziejczak}, J.~J. and {La Monaca}, F. and {Latronico}, L. and {Liodakis}, I. and {Maldera}, S. and {Manfreda}, A. and {Marin}, F. and {Marscher}, A.~P. and {Marshall}, H.~L. and {Massaro}, F. and {Mitsuishi}, I. and {Mizuno}, T. and {Negro}, M. and {Ng}, C.~Y. and {O'Dell}, S.~L. and {Omodei}, N. and {Oppedisano}, C. and {Papitto}, A. and {Pavlov}, G.~G. and {Peirson}, A.~L. and {Perri}, M. and {Pesce-Rollins}, M. and {Pilia}, M. and {Possenti}, A. and {Puccetti}, S. and {Ramsey}, B.~D. and {Rankin}, J. and {Roberts}, O.~J. and {Romani}, R.~W. and {Sgr{\`o}}, C. and {Slane}, P. and {Spandre}, G. and {Swartz}, D.~A. and {Tamagawa}, T. and {Tavecchio}, F. and {Tawara}, Y. and {Tennant}, A.~F. and {Thomas}, N.~E. and {Trois}, A. and {Tsygankov}, S.~S. and {Turolla}, R. and {Vink}, J. and {Weisskopf}, M.~C. and {Wu}, K. and {Xie}, F.},
        title = "{IXPE observation confirms a high spin in the accreting black hole 4U 1957+115}",
      journal = {\aap},
         year = 2024,
        month = apr,
       volume = {684},
          eid = {A95},
        pages = {A95},
          doi = {10.1051/0004-6361/202348277},
archivePrefix = {arXiv},
       eprint = {2310.11125},
 primaryClass = {astro-ph.HE},
       adsurl = {https://ui.adsabs.harvard.edu/abs/2024A&A...684A..95M}
}

@ARTICLE{2026ApJ...997...60R,
       author = {{Ravi}, Swati and {Ng}, Mason and {Marshall}, Herman L. and {Gnarini}, Andrea},
        title = "{What's the Buzz about GX 13+1? Constraining Coronal Geometry with QUEEN-BEE: A Bayesian Nested Sampling Framework for X-Ray Polarization Rotation Analysis}",
      journal = {\apj},
         year = 2026,
        month = jan,
       volume = {997},
       number = {1},
          eid = {60},
        pages = {60},
          doi = {10.3847/1538-4357/ae21bd},
archivePrefix = {arXiv},
       eprint = {2509.07059},
 primaryClass = {astro-ph.HE},
       adsurl = {https://ui.adsabs.harvard.edu/abs/2026ApJ...997...60R}
}

@ARTICLE{2009PASJ...61..999M,
       author = {{Matsuoka}, Masaru and {Kawasaki}, Kazuyoshi and {Ueno}, Shiro and {Tomida}, Hiroshi and {Kohama}, Mitsuhiro and {Suzuki}, Motoko and {Adachi}, Yasuki and {Ishikawa}, Masaki and {Mihara}, Tatehiro and {Sugizaki}, Mutsumi and {Isobe}, Naoki and {Nakagawa}, Yujin and {Tsunemi}, Hiroshi and {Miyata}, Emi and {Kawai}, Nobuyuki and {Kataoka}, Jun and {Morii}, Mikio and {Yoshida}, Atsumasa and {Negoro}, Hitoshi and {Nakajima}, Motoki and {Ueda}, Yoshihiro and {Chujo}, Hirotaka and {Yamaoka}, Kazutaka and {Yamazaki}, Osamu and {Nakahira}, Satoshi and {You}, Tetsuya and {Ishiwata}, Ryoji and {Miyoshi}, Sho and {Eguchi}, Satoshi and {Hiroi}, Kazuo and {Katayama}, Haruyoshi and {Ebisawa}, Ken},
        title = "{The MAXI Mission on the ISS: Science and Instruments for Monitoring All-Sky X-Ray Images}",
      journal = {\pasj},
         year = 2009,
        month = oct,
       volume = {61},
        pages = {999},
          doi = {10.1093/pasj/61.5.999},
archivePrefix = {arXiv},
       eprint = {0906.0631},
 primaryClass = {astro-ph.IM},
       adsurl = {https://ui.adsabs.harvard.edu/abs/2009PASJ...61..999M}
}

@ARTICLE{2005Ap&SS.300..107H,
       author = {{Homan}, Jeroen and {Belloni}, Tomaso},
        title = "{The Evolution of Black Hole States}",
      journal = {\apss},
         year = 2005,
        month = nov,
       volume = {300},
       number = {1-3},
        pages = {107-117},
          doi = {10.1007/s10509-005-1197-4},
archivePrefix = {arXiv},
       eprint = {astro-ph/0412597},
 primaryClass = {astro-ph},
       adsurl = {https://ui.adsabs.harvard.edu/abs/2005Ap&SS.300..107H}
}

@ARTICLE{2004MNRAS.355.1105F,
       author = {{Fender}, R.~P. and {Belloni}, T.~M. and {Gallo}, E.},
        title = "{Towards a unified model for black hole X-ray binary jets}",
      journal = {\mnras},
         year = 2004,
        month = dec,
       volume = {355},
       number = {4},
        pages = {1105-1118},
          doi = {10.1111/j.1365-2966.2004.08384.x},
archivePrefix = {arXiv},
       eprint = {astro-ph/0409360},
 primaryClass = {astro-ph},
       adsurl = {https://ui.adsabs.harvard.edu/abs/2004MNRAS.355.1105F}
}

@ARTICLE{2022Sci...378..650K,
       author = {{Krawczynski}, Henric and {Muleri}, Fabio and {Dov{\v{c}}iak}, Michal and {Veledina}, Alexandra and {Rodriguez Cavero}, Nicole and {Svoboda}, Jiri and {Ingram}, Adam and {Matt}, Giorgio and {Garcia}, Javier A. and {Loktev}, Vladislav and {Negro}, Michela and {Poutanen}, Juri and {Kitaguchi}, Takao and {Podgorn{\'y}}, Jakub and {Rankin}, John and {Zhang}, Wenda and {Berdyugin}, Andrei and {Berdyugina}, Svetlana V. and {Bianchi}, Stefano and {Blinov}, Dmitry and {Capitanio}, Fiamma and {Di Lalla}, Niccol{\`o} and {Draghis}, Paul and {Fabiani}, Sergio and {Kagitani}, Masato and {Kravtsov}, Vadim and {Kiehlmann}, Sebastian and {Latronico}, Luca and {Lutovinov}, Alexander A. and {Mandarakas}, Nikos and {Marin}, Fr{\'e}d{\'e}ric and {Marinucci}, Andrea and {Miller}, Jon M. and {Mizuno}, Tsunefumi and {Molkov}, Sergey V. and {Omodei}, Nicola and {Petrucci}, Pierre-Olivier and {Ratheesh}, Ajay and {Sakanoi}, Takeshi and {Semena}, Andrei N. and {Skalidis}, Raphael and {Soffitta}, Paolo and {Tennant}, Allyn F. and {Thalhammer}, Phillipp and {Tombesi}, Francesco and {Weisskopf}, Martin C. and {Wilms}, Joern and {Zhang}, Sixuan and {Agudo}, Iv{\'a}n and {Antonelli}, Lucio A. and {Bachetti}, Matteo and {Baldini}, Luca and {Baumgartner}, Wayne H. and {Bellazzini}, Ronaldo and {Bongiorno}, Stephen D. and {Bonino}, Raffaella and {Brez}, Alessandro and {Bucciantini}, Niccol{\`o} and {Castellano}, Simone and {Cavazzuti}, Elisabetta and {Ciprini}, Stefano and {Costa}, Enrico and {De Rosa}, Alessandra and {Del Monte}, Ettore and {Di Gesu}, Laura and {Di Marco}, Alessandro and {Donnarumma}, Immacolata and {Doroshenko}, Victor and {Ehlert}, Steven R. and {Enoto}, Teruaki and {Evangelista}, Yuri and {Ferrazzoli}, Riccardo and {Gunji}, Shuichi and {Hayashida}, Kiyoshi and {Heyl}, Jeremy and {Iwakiri}, Wataru and {Jorstad}, Svetlana G. and {Karas}, Vladimir and {Kolodziejczak}, Jeffery J. and {La Monaca}, Fabio and {Liodakis}, Ioannis and {Maldera}, Simone and {Manfreda}, Alberto and {Marscher}, Alan P. and {Marshall}, Herman L. and {Mitsuishi}, Ikuyuki and {Ng}, Chi-Yung and {O{\textquoteright}Dell}, Stephen L. and {Oppedisano}, Chiara and {Papitto}, Alessandro and {Pavlov}, George G. and {Peirson}, Abel L. and {Perri}, Matteo and {Pesce-Rollins}, Melissa and {Pilia}, Maura and {Possenti}, Andrea and {Puccetti}, Simonetta and {Ramsey}, Brian D. and {Romani}, Roger W. and {Sgr{\`o}}, Carmelo and {Slane}, Patrick and {Spandre}, Gloria and {Tamagawa}, Toru and {Tavecchio}, Fabrizio and {Taverna}, Roberto and {Tawara}, Yuzuru and {Thomas}, Nicholas E. and {Trois}, Alessio and {Tsygankov}, Sergey and {Turolla}, Roberto and {Vink}, Jacco and {Wu}, Kinwah and {Xie}, Fei and {Zane}, Silvia},
        title = "{Polarized x-rays constrain the disk-jet geometry in the black hole x-ray binary Cygnus X-1}",
      journal = {Science},
         year = 2022,
        month = nov,
       volume = {378},
       number = {6620},
        pages = {650-654},
          doi = {10.1126/science.add5399},
archivePrefix = {arXiv},
       eprint = {2206.09972},
 primaryClass = {astro-ph.HE},
       adsurl = {https://ui.adsabs.harvard.edu/abs/2022Sci...378..650K}
}

@ARTICLE{2024ApJ...966L..35S,
       author = {{Svoboda}, Ji{\v{r}}{\'\i} and {Dov{\v{c}}iak}, Michal and {Steiner}, James F. and {Kaaret}, Philip and {Podgorn{\'y}}, Jakub and {Poutanen}, Juri and {Veledina}, Alexandra and {Muleri}, Fabio and {Taverna}, Roberto and {Krawczynski}, Henric and {Brigitte}, Ma{\"\i}mouna and {Datta}, Sudeb Ranjan and {Bianchi}, Stefano and {Mu{\~n}oz-Darias}, Teo and {Negro}, Michela and {Rodriguez Cavero}, Nicole and {Castro Segura}, Noel and {Bollemeijer}, Niek and {Garc{\'\i}a}, Javier A. and {Ingram}, Adam and {Matt}, Giorgio and {Nathan}, Edward and {Weisskopf}, Martin C. and {Altamirano}, Diego and {Baldini}, Luca and {Capitanio}, Fiamma and {Egron}, Elise and {Emami}, Razieh and {Hu}, Kun and {Marra}, Lorenzo and {Mastroserio}, Guglielmo and {Petrucci}, Pierre-Olivier and {Ratheesh}, Ajay and {Soffitta}, Paolo and {Tombesi}, Francesco and {Yang}, Yi-Jung and {Zhang}, Yuexin},
        title = "{Dramatic Drop in the X-Ray Polarization of Swift J1727.8{\textendash}1613 in the Soft Spectral State}",
      journal = {\apjl},
         year = 2024,
        month = may,
       volume = {966},
       number = {2},
          eid = {L35},
        pages = {L35},
          doi = {10.3847/2041-8213/ad402e},
archivePrefix = {arXiv},
       eprint = {2403.04689},
 primaryClass = {astro-ph.HE},
       adsurl = {https://ui.adsabs.harvard.edu/abs/2024ApJ...966L..35S}
}

@ARTICLE{2024ApJ...968...76I,
       author = {{Ingram}, Adam and {Bollemeijer}, Niek and {Veledina}, Alexandra and {Dov{\v{c}}iak}, Michal and {Poutanen}, Juri and {Egron}, Elise and {Russell}, Thomas D. and {Trushkin}, Sergei A. and {Negro}, Michela and {Ratheesh}, Ajay and {Capitanio}, Fiamma and {Connors}, Riley and {Neilsen}, Joseph and {Kraus}, Alexander and {Iacolina}, Maria Noemi and {Pellizzoni}, Alberto and {Pilia}, Maura and {Carotenuto}, Francesco and {Matt}, Giorgio and {Mastroserio}, Guglielmo and {Kaaret}, Philip and {Bianchi}, Stefano and {Garc{\'\i}a}, Javier A. and {Bachetti}, Matteo and {Wu}, Kinwah and {Costa}, Enrico and {Ewing}, Melissa and {Kravtsov}, Vadim and {Krawczynski}, Henric and {Loktev}, Vladislav and {Marinucci}, Andrea and {Marra}, Lorenzo and {Miku{\v{s}}incov{\'a}}, Romana and {Nathan}, Edward and {Parra}, Maxime and {Petrucci}, Pierre-Olivier and {Righini}, Simona and {Soffitta}, Paolo and {Steiner}, James F. and {Svoboda}, Ji{\v{r}}{\'\i} and {Tombesi}, Francesco and {Tugliani}, Stefano and {Ursini}, Francesco and {Yang}, Yi-Jung and {Zane}, Silvia and {Zhang}, Wenda and {Agudo}, Iv{\'a}n and {Antonelli}, Lucio A. and {Baldini}, Luca and {Baumgartner}, Wayne H. and {Bellazzini}, Ronaldo and {Bongiorno}, Stephen D. and {Bonino}, Raffaella and {Brez}, Alessandro and {Bucciantini}, Niccol{\`o} and {Castellano}, Simone and {Cavazzuti}, Elisabetta and {Chen}, Chien-Ting and {Ciprini}, Stefano and {De Rosa}, Alessandra and {Del Monte}, Ettore and {Di Gesu}, Laura and {Di Lalla}, Niccol{\`o} and {Di Marco}, Alessandro and {Donnarumma}, Immacolata and {Doroshenko}, Victor and {Ehlert}, Steven R. and {Enoto}, Teruaki and {Evangelista}, Yuri and {Fabiani}, Sergio and {Ferrazzoli}, Riccardo and {Gunji}, Shuichi and {Hayashida}, Kiyoshi and {Heyl}, Jeremy and {Iwakiri}, Wataru and {Jorstad}, Svetlana G. and {Karas}, Vladimir and {Kislat}, Fabian and {Kitaguchi}, Takao and {Kolodziejczak}, Jeffery J. and {La Monaca}, Fabio and {Latronico}, Luca and {Liodakis}, Ioannis and {Maldera}, Simone and {Manfreda}, Alberto and {Marin}, Fr{\'e}d{\'e}ric and {Marscher}, Alan P. and {Marshall}, Herman L. and {Massaro}, Francesco and {Mitsuishi}, Ikuyuki and {Mizuno}, Tsunefumi and {Muleri}, Fabio and {Ng}, Chi-Yung and {O'Dell}, Stephen L. and {Omodei}, Nicola and {Oppedisano}, Chiara and {Papitto}, Alessandro and {Pavlov}, George G. and {Peirson}, Abel L. and {Perri}, Matteo and {Pesce-Rollins}, Melissa and {Possenti}, Andrea and {Puccetti}, Simonetta and {Ramsey}, Brian D. and {Rankin}, John and {Roberts}, Oliver J. and {Romani}, Roger W. and {Sgr{\`o}}, Carmelo and {Slane}, Patrick and {Spandre}, Gloria and {Swartz}, Douglas A. and {Tamagawa}, Toru and {Tavecchio}, Fabrizio and {Taverna}, Roberto and {Tawara}, Yuzuru and {Tennant}, Allyn F. and {Thomas}, Nicholas E. and {Trois}, Alessio and {Tsygankov}, Sergey S. and {Turolla}, Roberto and {Vink}, Jacco and {Weisskopf}, Martin C. and {Xie}, Fei and {IXPE Collaboration}},
        title = "{Tracking the X-Ray Polarization of the Black Hole Transient Swift J1727.8{\textendash}1613 during a State Transition}",
      journal = {\apj},
         year = 2024,
        month = jun,
       volume = {968},
       number = {2},
          eid = {76},
        pages = {76},
          doi = {10.3847/1538-4357/ad3faf},
archivePrefix = {arXiv},
       eprint = {2311.05497},
 primaryClass = {astro-ph.HE},
       adsurl = {https://ui.adsabs.harvard.edu/abs/2024ApJ...968...76I}
}

@ARTICLE{2024A&A...686L..12P,
       author = {{Podgorn{\'y}}, J. and {Svoboda}, J. and {Dov{\v{c}}iak}, M. and {Veledina}, A. and {Poutanen}, J. and {Kaaret}, P. and {Bianchi}, S. and {Ingram}, A. and {Capitanio}, F. and {Datta}, S.~R. and {Egron}, E. and {Krawczynski}, H. and {Matt}, G. and {Muleri}, F. and {Petrucci}, P.-O. and {Russell}, T.~D. and {Steiner}, J.~F. and {Bollemeijer}, N. and {Brigitte}, M. and {Castro Segura}, N. and {Emami}, R. and {Garc{\'\i}a}, J.~A. and {Hu}, K. and {Iacolina}, M.~N. and {Kravtsov}, V. and {Marra}, L. and {Mastroserio}, G. and {Mu{\~n}oz-Darias}, T. and {Nathan}, E. and {Negro}, M. and {Ratheesh}, A. and {Rodriguez Cavero}, N. and {Taverna}, R. and {Tombesi}, F. and {Yang}, Y.~J. and {Zhang}, W. and {Zhang}, Y.},
        title = "{Recovery of the X-ray polarisation of Swift J1727.8‒1613 after the soft-to-hard spectral transition}",
      journal = {\aap},
         year = 2024,
        month = jun,
       volume = {686},
          eid = {L12},
        pages = {L12},
          doi = {10.1051/0004-6361/202450566},
archivePrefix = {arXiv},
       eprint = {2404.19601},
 primaryClass = {astro-ph.HE},
       adsurl = {https://ui.adsabs.harvard.edu/abs/2024A&A...686L..12P}
}

@ARTICLE{1990ApJ...361..590K,
       author = {{Kitamoto}, Shunji and {Tsunemi}, Hiroshi and {Pedersen}, Holger and {Ilovaisky}, Sergio A. and {van der Klis}, Michiel},
        title = "{Outburst, Identification, and X-Ray Light Curve of GS 1354-64 (=MX 1353-64?, Centaurus X-2?)}",
      journal = {\apj},
         year = 1990,
        month = oct,
       volume = {361},
        pages = {590},
          doi = {10.1086/169222},
       adsurl = {https://ui.adsabs.harvard.edu/abs/1990ApJ...361..590K}
}

@ARTICLE{2001MNRAS.323..517B,
       author = {{Brocksopp}, C. and {Jonker}, P.~G. and {Fender}, R.~P. and {Groot}, P.~J. and {van der Klis}, M. and {Tingay}, S.~J.},
        title = "{The 1997 hard-state outburst of the X-ray transient GS 1354-64/BW Cir}",
      journal = {\mnras},
         year = 2001,
        month = may,
       volume = {323},
       number = {2},
        pages = {517-528},
          doi = {10.1046/j.1365-8711.2001.04193.x},
archivePrefix = {arXiv},
       eprint = {astro-ph/0011145},
 primaryClass = {astro-ph},
       adsurl = {https://ui.adsabs.harvard.edu/abs/2001MNRAS.323..517B}
}

@ARTICLE{2016MNRAS.460..942K,
       author = {{Koljonen}, K.~I.~I. and {Russell}, D.~M. and {Corral-Santana}, J.~M. and {Armas Padilla}, M. and {Mu{\~n}oz-Darias}, T. and {Lewis}, F. and {Coriat}, M. and {Bauer}, F.~E.},
        title = "{A `high-hard' outburst of the black hole X-ray binary GS 1354-64}",
      journal = {\mnras},
         year = 2016,
        month = jul,
       volume = {460},
       number = {1},
        pages = {942-955},
          doi = {10.1093/mnras/stw1007},
archivePrefix = {arXiv},
       eprint = {1602.06414},
 primaryClass = {astro-ph.HE},
       adsurl = {https://ui.adsabs.harvard.edu/abs/2016MNRAS.460..942K}
}

@ARTICLE{2017MNRAS.469..193P,
       author = {{Pahari}, Mayukh and {Gandhi}, Poshak and {Charles}, Philip A. and {Kotze}, Marissa M. and {Altamirano}, Diego and {Misra}, Ranjeev},
        title = "{Simultaneous optical/X-ray study of GS 1354-64 (=BW Cir) during hard outburst: evidence for optical cyclo-synchrotron emission from the hot accretion flow}",
      journal = {\mnras},
         year = 2017,
        month = jul,
       volume = {469},
       number = {1},
        pages = {193-205},
          doi = {10.1093/mnras/stx840},
archivePrefix = {arXiv},
       eprint = {1704.00474},
 primaryClass = {astro-ph.HE},
       adsurl = {https://ui.adsabs.harvard.edu/abs/2017MNRAS.469..193P}
}

@ARTICLE{2016MNRAS.459.4038S,
       author = {{Stiele}, H. and {Kong}, A.~K.~H.},
        title = "{The 2015 hard-state only outburst of GS 1354-64}",
      journal = {\mnras},
         year = 2016,
        month = jul,
       volume = {459},
       number = {4},
        pages = {4038-4045},
          doi = {10.1093/mnras/stw903},
archivePrefix = {arXiv},
       eprint = {1604.05727},
 primaryClass = {astro-ph.HE},
       adsurl = {https://ui.adsabs.harvard.edu/abs/2016MNRAS.459.4038S}
}

@ARTICLE{2025ATel17563....1N,
       author = {{Negoro}, H. and {Fujiwara}, K. and {Nakajima}, M. and {Takagi}, K. and {Nishio}, H. and {Mihara}, T. and {Tamagawa}, T. and {Kawai}, N. and {Matsuoka}, M. and {Sakamoto}, T. and {Serino}, M. and {Sugita}, S. and {Kawakubo}, Y. and {Hiramatsu}, H. and {Kondo}, Y. and {Yoshida}, A. and {Tsuboi}, Y. and {Sugai}, H. and {Nagashima}, N. and {Ishihara}, Y. and {Shidatsu}, M. and {Kang}, C. and {Nakamoto}, T. and {Uenishi}, M. and {Usuki}, T. and {Yatsuzuka}, S. and {Takahashi}, I. and {Yatsu}, Y. and {Nakahira}, S. and {Ueno}, S. and {Tomida}, H. and {Ogawa}, S. and {Kurihara}, M. and {Ueda}, Y. and {Kobayashi}, S. and {Yamauchi}, M. and {Nishio}, M. and {Hiraizumi}, C. and {Yamaoka}, K. and {Sugizaki}, M. and {Iwakiri}, W. and {Kawamuro}, T. and {Yamada}, S.},
        title = "{MAXI/GSC detection of renewed activity of the black hole X-ray binary GS 1354-64}",
      journal = {The Astronomer's Telegram},
         year = 2025,
        month = dec,
       volume = {17563},
        pages = {1},
       adsurl = {https://ui.adsabs.harvard.edu/abs/2025ATel17563....1N}
}

@ARTICLE{2009ApJS..181..238C,
       author = {{Casares}, J. and {Orosz}, J.~A. and {Zurita}, C. and {Shahbaz}, T. and {Corral-Santana}, J.~M. and {McClintock}, J.~E. and {Garcia}, M.~R. and {Mart{\'\i}nez-Pais}, I.~G. and {Charles}, P.~A. and {Fender}, R.~P. and {Remillard}, R.~A.},
        title = "{Refined Orbital Solution and Quiescent Variability in the Black Hole Transient GS 1354-64 (= BW Cir)}",
      journal = {\apjs},
         year = 2009,
        month = mar,
       volume = {181},
       number = {1},
        pages = {238-243},
          doi = {10.1088/0067-0049/181/1/238},
       adsurl = {https://ui.adsabs.harvard.edu/abs/2009ApJS..181..238C}
}

@ARTICLE{2004ApJ...613L.133C,
       author = {{Casares}, J. and {Zurita}, C. and {Shahbaz}, T. and {Charles}, P.~A. and {Fender}, R.~P.},
        title = "{Evidence of a Black Hole in the X-Ray Transient GS 1354-64 (=BW Circini)}",
      journal = {\apjl},
         year = 2004,
        month = oct,
       volume = {613},
       number = {2},
        pages = {L133-L136},
          doi = {10.1086/425145},
archivePrefix = {arXiv},
       eprint = {astro-ph/0408331},
 primaryClass = {astro-ph},
       adsurl = {https://ui.adsabs.harvard.edu/abs/2004ApJ...613L.133C}
}

@ARTICLE{2026ATel17618....1N,
       author = {{Negoro}, H. and {Nakajima}, M. and {Takagi}, K. and {Nishio}, H. and {Mihara}, T. and {Tamagawa}, T. and {Kawai}, N. and {Matsuoka}, M. and {Sakamoto}, T. and {Serino}, M. and {Sugita}, S. and {Kawakubo}, Y. and {Hiramatsu}, H. and {Kondo}, Y. and {Yoshida}, A. and {Tsuboi}, Y. and {Sugai}, H. and {Nagashima}, N. and {Ishihara}, Y. and {Shidatsu}, M. and {Kang}, C. and {Nakamoto}, T. and {Uenishi}, M. and {Usuki}, T. and {Yatsuzuka}, S. and {Takahashi}, I. and {Yatsu}, Y. and {Nakahira}, S. and {Ueno}, S. and {Tomida}, H. and {Ogawa}, S. and {Kurihara}, M. and {Ueda}, Y. and {Fujiwara}, K. and {Kobayashi}, S. and {Yamauchi}, M. and {Nishio}, M. and {Hiraizumi}, C. and {Yamaoka}, K. and {Sugizaki}, M. and {Iwakiri}, W. and {Kawamuro}, T. and {Yamada}, S.},
        title = "{GS 1354-64 is likely undergoing a hard-to-soft transition}",
      journal = {The Astronomer's Telegram},
         year = 2026,
        month = jan,
       volume = {17618},
        pages = {1},
       adsurl = {https://ui.adsabs.harvard.edu/abs/2026ATel17618....1N}
}

@ARTICLE{2010ApJ...712..908S,
       author = {{Schnittman}, Jeremy D. and {Krolik}, Julian H.},
        title = "{X-ray Polarization from Accreting Black Holes: Coronal Emission}",
      journal = {\apj},
         year = 2010,
        month = apr,
       volume = {712},
       number = {2},
        pages = {908-924},
          doi = {10.1088/0004-637X/712/2/908},
archivePrefix = {arXiv},
       eprint = {0912.0907},
 primaryClass = {astro-ph.HE},
       adsurl = {https://ui.adsabs.harvard.edu/abs/2010ApJ...712..908S}
}

@ARTICLE{2019ApJ...881...39H,
          author = {{Huppenkothen}, Daniela and {Bachetti}, Matteo and
                    {Stevens}, Abigail L. and {Migliari}, Simone and {Balm}, Paul and
                    {Hammad}, Omar and {Khan}, Usman Mahmood and {Mishra}, Himanshu and
                    {Rashid}, Haroon and {Sharma}, Swapnil and {Martinez Ribeiro}, Evandro and
                    {Valles Blanco}, Ricardo},
          title = "{Stingray: A Modern Python Library for Spectral Timing}",
          journal = {apj},
          year = 2019,
          month = aug,
          volume = {881},
          number = {1},
          eid = {39},
          pages = {39},
          doi = {10.3847/1538-4357/ab258d},
          archivePrefix = {arXiv},
          eprint = {1901.07681},
          primaryClass = {astro-ph.IM},
          adsurl = {https://ui.adsabs.harvard.edu/abs/2019ApJ...881...39H}
        }

@software{matteo_bachetti_2023_7970570,
  author       = {Matteo Bachetti and
                  Daniela Huppenkothen and
                  Usman Khan and
                  Himanshu Mishra and
                  Abbie Stevens and
                  Swapnil Sharma and
                  John Swinbank and
                  Amogh Desai and
                  Haroon Rashid and
                  Evandro Martinez Ribeiro and
                  Mihir Tripathi and
                  Brigitta SipÅcz and
                  Dhruv Vats and
                  tappina and
                  Guglielmo Mastroserio and
                  omargamal8 and
                  Meg Davis and
                  Achilles Rasquinha and
                  Paul Balm and
                  Stuart Mumford and
                  Devansh Shukla and
                  Riccardo Campana and
                  parkma99 and
                  Nitish Garg and
                  Akash Tandon and
                  Anurag Hota and
                  Arnav Anand and
                  Nick and
                  Rashmi Raj and
                  Sashank Mishra},
  title        = {StingraySoftware/stingray: Version 1.1.2},
  month        = may,
  year         = 2023,
  publisher    = {Zenodo},
  version      = {v1.1.2},
  doi          = {10.5281/zenodo.7970570},
  url          = {https://doi.org/10.5281/zenodo.7970570},
}

@ARTICLE{2019MNRAS.485.2642G,
       author = {{Gandhi}, Poshak and {Rao}, Anjali and {Johnson}, Michael A.~C. and {Paice}, John A. and {Maccarone}, Thomas J.},
        title = "{Gaia Data Release 2 distances and peculiar velocities for Galactic black hole transients}",
      journal = {\mnras},
         year = 2019,
        month = may,
       volume = {485},
       number = {2},
        pages = {2642-2655},
          doi = {10.1093/mnras/stz438},
archivePrefix = {arXiv},
       eprint = {1804.11349},
 primaryClass = {astro-ph.HE},
       adsurl = {https://ui.adsabs.harvard.edu/abs/2019MNRAS.485.2642G}
}

@ARTICLE{2024ApJ...969...40D,
       author = {{Draghis}, Paul A. and {Miller}, Jon M. and {Costantini}, Elisa and {Gallo}, Luigi C. and {Reynolds}, Mark and {Tomsick}, John A. and {Zoghbi}, Abderahmen},
        title = "{Systematically Revisiting All NuSTAR Spins of Black Holes in X-Ray Binaries}",
      journal = {\apj},
         year = 2024,
        month = jul,
       volume = {969},
       number = {1},
          eid = {40},
        pages = {40},
          doi = {10.3847/1538-4357/ad43ea},
archivePrefix = {arXiv},
       eprint = {2311.16225},
 primaryClass = {astro-ph.HE},
       adsurl = {https://ui.adsabs.harvard.edu/abs/2024ApJ...969...40D}
}

@ARTICLE{2023ApJ...951..145L,
       author = {{Liu}, Honghui and {Jiang}, Jiachen and {Zhang}, Zuobin and {Bambi}, Cosimo and {Fabian}, Andrew C. and {Garc{\'\i}a}, Javier A. and {Ingram}, Adam and {Kara}, Erin and {Steiner}, James F. and {Tomsick}, John A. and {Walton}, Dominic J. and {Young}, Andrew J.},
        title = "{High-density Reflection Spectroscopy of Black Hole X-Ray Binaries in the Hard State}",
      journal = {\apj},
         year = 2023,
        month = jul,
       volume = {951},
       number = {2},
          eid = {145},
        pages = {145},
          doi = {10.3847/1538-4357/acd8b9},
archivePrefix = {arXiv},
       eprint = {2303.10593},
 primaryClass = {astro-ph.HE},
       adsurl = {https://ui.adsabs.harvard.edu/abs/2023ApJ...951..145L}
}

@INPROCEEDINGS{2003ASPC..295..489J,
       author = {{Joye}, W.~A. and {Mandel}, E.},
        title = "{New Features of SAOImage DS9}",
    booktitle = {Astronomical Data Analysis Software and Systems XII},
         year = 2003,
       editor = {{Payne}, H.~E. and {Jedrzejewski}, R.~I. and {Hook}, R.~N.},
       series = {Astronomical Society of the Pacific Conference Series},
       volume = {295},
        month = jan,
        pages = {489},
       adsurl = {https://ui.adsabs.harvard.edu/abs/2003ASPC..295..489J}
}

@ARTICLE{2018ApJ...853L..21B,
       author = {{Bachetti}, Matteo and {Huppenkothen}, Daniela},
        title = "{No Time for Dead Time: Use the Fourier Amplitude Differences to Normalize Dead-time-affected Periodograms}",
      journal = {\apjl},
         year = 2018,
        month = feb,
       volume = {853},
       number = {2},
          eid = {L21},
        pages = {L21},
          doi = {10.3847/2041-8213/aaa83b},
archivePrefix = {arXiv},
       eprint = {1709.09700},
 primaryClass = {astro-ph.HE},
       adsurl = {https://ui.adsabs.harvard.edu/abs/2018ApJ...853L..21B}
}

@ARTICLE{2005ApJ...629..403C,
       author = {{Casella}, P. and {Belloni}, T. and {Stella}, L.},
        title = "{The ABC of Low-Frequency Quasi-periodic Oscillations in Black Hole Candidates: Analogies with Z Sources}",
      journal = {\apj},
         year = 2005,
        month = aug,
       volume = {629},
       number = {1},
        pages = {403-407},
          doi = {10.1086/431174},
archivePrefix = {arXiv},
       eprint = {astro-ph/0504318},
 primaryClass = {astro-ph},
       adsurl = {https://ui.adsabs.harvard.edu/abs/2005ApJ...629..403C}
}

@INCOLLECTION{2010LNP...794...53B,
       author = {{Belloni}, T.~M.},
        title = "{States and Transitions in Black Hole Binaries}",
    booktitle = {Lecture Notes in Physics, Berlin Springer Verlag},
         year = 2010,
       editor = {{Belloni}, Tomaso},
       volume = {794},
        pages = {53},
          doi = {10.1007/978-3-540-76937-8_3},
       adsurl = {https://ui.adsabs.harvard.edu/abs/2010LNP...794...53B}
}

@ARTICLE{2019NewAR..8501524I,
       author = {{Ingram}, Adam R. and {Motta}, Sara E.},
        title = "{A review of quasi-periodic oscillations from black hole X-ray binaries: Observation and theory}",
      journal = {\nar},
         year = 2019,
        month = sep,
       volume = {85},
          eid = {101524},
        pages = {101524},
          doi = {10.1016/j.newar.2020.101524},
archivePrefix = {arXiv},
       eprint = {2001.08758},
 primaryClass = {astro-ph.HE},
       adsurl = {https://ui.adsabs.harvard.edu/abs/2019NewAR..8501524I}
}

@ARTICLE{2000ApJ...542..914W,
       author = {{Wilms}, J. and {Allen}, A. and {McCray}, R.},
        title = "{On the Absorption of X-Rays in the Interstellar Medium}",
      journal = {\apj},
         year = 2000,
        month = oct,
       volume = {542},
       number = {2},
        pages = {914-924},
          doi = {10.1086/317016},
archivePrefix = {arXiv},
       eprint = {astro-ph/0008425},
 primaryClass = {astro-ph},
       adsurl = {https://ui.adsabs.harvard.edu/abs/2000ApJ...542..914W}
}

@ARTICLE{2025MNRAS.541.1774E,
       author = {{Ewing}, Melissa and {Parra}, Maxime and {Mastroserio}, Guglielmo and {Veledina}, Alexandra and {Ingram}, Adam and {Dov{\v{c}}iak}, Michal and {Garc{\'\i}a}, Javier A. and {Russell}, Thomas D. and {Baglio}, Maria C. and {Poutanen}, Juri and {Adegoke}, Oluwashina and {Bianchi}, Stefano and {Capitanio}, Fiamma and {Connors}, Riley and {Del Santo}, Melania and {De Marco}, Barbara and {Trigo}, Mar{\'\i}a D{\'\i}az and {Gandhi}, Poshak and {Gupta}, Maitrayee and {Kang}, Chulsoo and {Kammoun}, Elias and {Loktev}, Vladislav and {Marra}, Lorenzo and {Matt}, Giorgio and {Nathan}, Edward and {Petrucci}, Pierre-Olivier and {Shidatsu}, Megumi and {Steiner}, James F. and {Tombesi}, Francesco and {Vincentelli}, Federico M.},
        title = "{The very high X-ray polarization of accreting black hole IGR J17091‑3624 in the hard state}",
      journal = {\mnras},
         year = 2025,
        month = aug,
       volume = {541},
       number = {2},
        pages = {1774-1781},
          doi = {10.1093/mnras/staf859},
archivePrefix = {arXiv},
       eprint = {2503.22665},
 primaryClass = {astro-ph.HE},
       adsurl = {https://ui.adsabs.harvard.edu/abs/2025MNRAS.541.1774E}
}

@ARTICLE{2024ApJ...964...77R,
       author = {{Ratheesh}, Ajay and {Dov{\v{c}}iak}, Michal and {Krawczynski}, Henric and {Podgorn{\'y}}, Jakub and {Marra}, Lorenzo and {Veledina}, Alexandra and {Suleimanov}, Valery F. and {Rodriguez Cavero}, Nicole and {Steiner}, James F. and {Svoboda}, Ji{\v{r}}{\'\i} and {Marinucci}, Andrea and {Bianchi}, Stefano and {Negro}, Michela and {Matt}, Giorgio and {Tombesi}, Francesco and {Poutanen}, Juri and {Ingram}, Adam and {Taverna}, Roberto and {West}, Andrew and {Karas}, Vladimir and {Ursini}, Francesco and {Soffitta}, Paolo and {Capitanio}, Fiamma and {Viscolo}, Domenico and {Manfreda}, Alberto and {Muleri}, Fabio and {Parra}, Maxime and {Beheshtipour}, Banafsheh and {Chun}, Sohee and {Cibrario}, Nicol{\`o} and {Di Lalla}, Niccol{\`o} and {Fabiani}, Sergio and {Hu}, Kun and {Kaaret}, Philip and {Loktev}, Vladislav and {Miku{\v{s}}incov{\'a}}, Romana and {Mizuno}, Tsunefumi and {Omodei}, Nicola and {Petrucci}, Pierre-Olivier and {Puccetti}, Simonetta and {Rankin}, John and {Zane}, Silvia and {Zhang}, Sixuan and {Agudo}, Iv{\'a}n and {Antonelli}, Lucio A. and {Bachetti}, Matteo and {Baldini}, Luca and {Baumgartner}, Wayne H. and {Bellazzini}, Ronaldo and {Bongiorno}, Stephen D. and {Bonino}, Raffaella and {Brez}, Alessandro and {Bucciantini}, Niccol{\`o} and {Castellano}, Simone and {Cavazzuti}, Elisabetta and {Chen}, Chien-Ting and {Ciprini}, Stefano and {Costa}, Enrico and {De Rosa}, Alessandra and {Del Monte}, Ettore and {Di Gesu}, Laura and {Di Marco}, Alessandro and {Donnarumma}, Immacolata and {Doroshenko}, Victor and {Ehlert}, Steven R. and {Enoto}, Teruaki and {Evangelista}, Yuri and {Ferrazzoli}, Riccardo and {Garcia}, Javier A. and {Gunji}, Shuichi and {Hayashida}, Kiyoshi and {Heyl}, Jeremy and {Iwakiri}, Wataru and {Jorstad}, Svetlana G. and {Kislat}, Fabian and {Kitaguchi}, Takao and {Kolodziejczak}, Jeffery J. and {La Monaca}, Fabio and {Latronico}, Luca and {Liodakis}, Ioannis and {Maldera}, Simone and {Marin}, Fr{\'e}d{\'e}ric and {Marscher}, Alan P. and {Marshall}, Herman L. and {Massaro}, Francesco and {Mitsuishi}, Ikuyuki and {Ng}, Stephen C.-Y. and {O'Dell}, Stephen L. and {Oppedisano}, Chiara and {Papitto}, Alessandro and {Pavlov}, George G. and {Peirson}, Abel L. and {Perri}, Matteo and {Pesce-Rollins}, Melissa and {Pilia}, Maura and {Possenti}, Andrea and {Ramsey}, Brian D. and {Roberts}, Oliver J. and {Romani}, Roger W. and {Sgr{\`o}}, Carmelo and {Slane}, Patrick and {Spandre}, Gloria and {Swartz}, Douglas A. and {Tamagawa}, Toru and {Tavecchio}, Fabrizio and {Tawara}, Yuzuru and {Tennant}, Allyn F. and {Thomas}, Nicholas E. and {Trois}, Alessio and {Tsygankov}, Sergey S. and {Turolla}, Roberto and {Vink}, Jacco and {Weisskopf}, Martin C. and {Wu}, Kinwah and {Xie}, Fei},
        title = "{X-Ray Polarization of the Black Hole X-Ray Binary 4U 1630─47 Challenges the Standard Thin Accretion Disk Scenario}",
      journal = {\apj},
         year = 2024,
        month = mar,
       volume = {964},
       number = {1},
          eid = {77},
        pages = {77},
          doi = {10.3847/1538-4357/ad226e},
archivePrefix = {arXiv},
       eprint = {2304.12752},
 primaryClass = {astro-ph.HE},
       adsurl = {https://ui.adsabs.harvard.edu/abs/2024ApJ...964...77R}
}

@ARTICLE{2024ApJ...960....3S,
       author = {{Svoboda}, Ji{\v{r}}{\'\i} and {Dov{\v{c}}iak}, Michal and {Steiner}, James F. and {Muleri}, Fabio and {Ingram}, Adam and {Yilmaz}, Anastasiya and {Rodriguez Cavero}, Nicole and {Marra}, Lorenzo and {Poutanen}, Juri and {Veledina}, Alexandra and {Mojaver}, Mehrnoosh Rahbardar and {Bianchi}, Stefano and {Garc{\'\i}a}, Javier A. and {Kaaret}, Philip and {Krawczynski}, Henric and {Matt}, Giorgio and {Podgorn{\'y}}, Jakub and {Weisskopf}, Martin C. and {Kislat}, Fabian and {Petrucci}, Pierre-Olivier and {Brigitte}, Maimouna and {Bursa}, Michal and {Fabiani}, Sergio and {Hu}, Kun and {Chun}, Sohee and {Mastroserio}, Guglielmo and {Mikus̆incov{\'a}}, Romana and {Ratheesh}, Ajay and {Romani}, Roger W. and {Soffitta}, Paolo and {Ursini}, Francesco and {Zane}, Silvia and {Agudo}, Iv{\'a}n and {Antonelli}, Lucio A. and {Bachetti}, Matteo and {Baldini}, Luca and {Baumgartner}, Wayne H. and {Bellazzini}, Ronaldo and {Bongiorno}, Stephen D. and {Bonino}, Raffaella and {Brez}, Alessandro and {Bucciantini}, Niccol{\`o} and {Capitanio}, Fiamma and {Castellano}, Simone and {Cavazzuti}, Elisabetta and {Chen}, Chien-Ting and {Ciprini}, Stefano and {Costa}, Enrico and {De Rosa}, Alessandra and {Del Monte}, Ettore and {Di Gesu}, Laura and {Di Lalla}, Niccol{\`o} and {Di Marco}, Alessandro and {Donnarumma}, Immacolata and {Doroshenko}, Victor and {Ehlert}, Steven R. and {Enoto}, Teruaki and {Evangelista}, Yuri and {Ferrazzoli}, Riccardo and {Gunji}, Shuichi and {Hayashida}, Kiyoshi and {Heyl}, Jeremy and {Iwakiri}, Wataru and {Jorstad}, Svetlana G. and {Karas}, Vladim{\'\i}r and {Kitaguchi}, Takao and {Kolodziejczak}, Jeffery J. and {La Monaca}, Fabio and {Latronico}, Luca and {Liodakis}, Ioannis and {Maldera}, Simone and {Manfreda}, Alberto and {Marin}, Fr{\'e}d{\'e}ric and {Marinucci}, Andrea and {Marscher}, Alan P. and {Marshall}, Herman L. and {Massaro}, Francesco and {Mitsuishi}, Ikuyuki and {Mizuno}, Tsunefumi and {Negro}, Michela and {Ng}, Chi-Yung and {O'Dell}, Stephen L. and {Omodei}, Nicola and {Oppedisano}, Chiara and {Papitto}, Alessandro and {Pavlov}, George G. and {Peirson}, Abel L. and {Perri}, Matteo and {Pesce-Rollins}, Melissa and {Pilia}, Maura and {Possenti}, Andrea and {Puccetti}, Simonetta and {Ramsey}, Brian D. and {Rankin}, John and {Roberts}, Oliver J. and {Sgr{\`o}}, Carmelo and {Slane}, Patrick and {Spandre}, Gloria and {Swartz}, Douglas A. and {Tamagawa}, Toru and {Tavecchio}, Fabrizio and {Taverna}, Roberto and {Tawara}, Yuzuru and {Tennant}, Allyn F. and {Thomas}, Nicholas E. and {Tombesi}, Francesco and {Trois}, Alessio and {Tsygankov}, Sergey S. and {Turolla}, Roberto and {Vink}, Jacco and {Wu}, Kinwah and {Xie}, Fei},
        title = "{First X-Ray Polarization Measurement Confirms the Low Black Hole Spin in LMC X-3}",
      journal = {\apj},
         year = 2024,
        month = jan,
       volume = {960},
       number = {1},
          eid = {3},
        pages = {3},
          doi = {10.3847/1538-4357/ad0842},
archivePrefix = {arXiv},
       eprint = {2309.10813},
 primaryClass = {astro-ph.HE},
       adsurl = {https://ui.adsabs.harvard.edu/abs/2024ApJ...960....3S}
}

@ARTICLE{2016ApJ...826L..12E,
       author = {{El-Batal}, A.~M. and {Miller}, J.~M. and {Reynolds}, M.~T. and {Boggs}, S.~E. and {Chistensen}, F.~E. and {Craig}, W.~W. and {Fuerst}, F. and {Hailey}, C.~J. and {Harrison}, F.~A. and {Stern}, D.~K. and {Tomsick}, J. and {Walton}, D.~J. and {Zhang}, W.~W.},
        title = "{NuSTAR Observations of the Black Hole GS 1354-645: Evidence of Rapid Black Hole Spin}",
      journal = {\apjl},
         year = 2016,
        month = jul,
       volume = {826},
       number = {1},
          eid = {L12},
        pages = {L12},
          doi = {10.3847/2041-8205/826/1/L12},
archivePrefix = {arXiv},
       eprint = {1607.00343},
 primaryClass = {astro-ph.HE},
       adsurl = {https://ui.adsabs.harvard.edu/abs/2016ApJ...826L..12E}
}

@ARTICLE{2026arXiv260306883L,
       author = {{Liu}, Honghui and {Kong}, Lingda and {Adegoke}, Oluwashina K. and {Jiang}, Jiachen and {Bambi}, Cosimo and {Fabian}, Andrew C. and {Ingram}, Adam and {Ravi}, Swati and {Steiner}, James F. and {Shui}, Qingcang and {Walton}, Dominic J. and {Xu}, Yerong and {Young}, Andrew J. and {Zhang}, Yuexin and {Zhang}, Zuobin and {Santangelo}, Andrea},
        title = "{A clean broad iron line in GS 1354--64 as seen by XRISM}",
      journal = {arXiv e-prints},
         year = 2026,
        month = mar,
          eid = {arXiv:2603.06883},
        pages = {arXiv:2603.06883},
          doi = {10.48550/arXiv.2603.06883},
archivePrefix = {arXiv},
       eprint = {2603.06883},
 primaryClass = {astro-ph.HE},
       adsurl = {https://ui.adsabs.harvard.edu/abs/2026arXiv260306883L}
}

@ARTICLE{2026arXiv260303463R,
       author = {{Ravi}, Swati and {Marra}, Lorenzo and {Steiner}, James F. and {Mastroserio}, Guglielmo and {Ng}, Mason and {Neilsen}, Joey and {Marshall}, Herman L. and {Capitanio}, Fiamma and {Ranjan Datta}, Sudeb and {Egron}, Elise and {Garcia}, Javier A. and {Ingram}, Adam and {Kaaret}, Philip and {Koenig}, Ole and {Liu}, Honghui and {Mikusincova}, Romana and {Nathan}, Edward J.~R. and {Petrucci}, P.-O. and {Podgorny}, Jakub and {Salvaggio}, Chiara and {Svoboda}, Jiri and {Veledina}, Alexandra and {Zhang}, Yuexin},
        title = "{Discovery of Strong Energy-Dependent X-ray Polarization in the Intermediate State of GS 1354-64}",
      journal = {arXiv e-prints},
         year = 2026,
        month = mar,
          eid = {arXiv:2603.03463},
        pages = {arXiv:2603.03463},
          doi = {10.48550/arXiv.2603.03463},
archivePrefix = {arXiv},
       eprint = {2603.03463},
 primaryClass = {astro-ph.HE},
       adsurl = {https://ui.adsabs.harvard.edu/abs/2026arXiv260303463R}
}

@ARTICLE{2021ARA&A..59..117R,
       author = {{Reynolds}, Christopher S.},
        title = "{Observational Constraints on Black Hole Spin}",
      journal = {\araa},
         year = 2021,
        month = sep,
       volume = {59},
        pages = {117-154},
          doi = {10.1146/annurev-astro-112420-035022},
archivePrefix = {arXiv},
       eprint = {2011.08948},
 primaryClass = {astro-ph.HE},
       adsurl = {https://ui.adsabs.harvard.edu/abs/2021ARA&A..59..117R}
}

@ARTICLE{2026arXiv260726144L,
       author = {{Ling}, Yu-shan and {Xie}, Fei and {Di Marco}, Alessandro and {La Monaca}, Fabio and {Ge}, Ming-Yu},
        title = "{Unchanged X-Ray Polarization During Accretion Dips in the Low Hard State of Cygnus X-1}",
      journal = {arXiv e-prints},
         year = 2026,
        month = jul,
          eid = {arXiv:2607.26144},
        pages = {arXiv:2607.26144},
          doi = {10.48550/arXiv.2607.26144},
archivePrefix = {arXiv},
       eprint = {2607.26144},
 primaryClass = {astro-ph.HE},
       adsurl = {https://ui.adsabs.harvard.edu/abs/2026arXiv260726144L}
}

@ARTICLE{2024A&A...683A.131V,
       author = {{Veronese}, S. and {Vignali}, C. and {Severgnini}, P. and {Matzeu}, G.~A. and {Cignoni}, M.},
        title = "{Interpreting the long-term variability of the changing-look AGN Mrk 1018}",
      journal = {\aap},
         year = 2024,
        month = mar,
       volume = {683},
          eid = {A131},
        pages = {A131},
          doi = {10.1051/0004-6361/202348098},
archivePrefix = {arXiv},
       eprint = {2312.07663},
 primaryClass = {astro-ph.HE},
       adsurl = {https://ui.adsabs.harvard.edu/abs/2024A&A...683A.131V}
}

@ARTICLE{2007ApJ...659L.111R,
       author = {{Risaliti}, G. and {Elvis}, M. and {Fabbiano}, G. and {Baldi}, A. and {Zezas}, A. and {Salvati}, M.},
        title = "{Occultation Measurement of the Size of the X-Ray-emitting Region in the Active Galactic Nucleus of NGC 1365}",
      journal = {\apjl},
         year = 2007,
        month = apr,
       volume = {659},
       number = {2},
        pages = {L111-L114},
          doi = {10.1086/517884},
archivePrefix = {arXiv},
       eprint = {astro-ph/0703173},
 primaryClass = {astro-ph},
       adsurl = {https://ui.adsabs.harvard.edu/abs/2007ApJ...659L.111R}
}

@ARTICLE{2023MNRAS.519.3903L,
       author = {{Lawther}, D. and {Vestergaard}, M. and {Raimundo}, S. and {Koay}, J.~Y. and {Peterson}, B.~M. and {Fan}, X. and {Grupe}, D. and {Mathur}, S.},
        title = "{Flares in the changing look AGN Mrk 590 - I. The UV response to X-ray outbursts suggests a more complex reprocessing geometry than a standard disc}",
      journal = {\mnras},
         year = 2023,
        month = mar,
       volume = {519},
       number = {3},
        pages = {3903-3922},
          doi = {10.1093/mnras/stac3515},
archivePrefix = {arXiv},
       eprint = {2211.14371},
 primaryClass = {astro-ph.HE},
       adsurl = {https://ui.adsabs.harvard.edu/abs/2023MNRAS.519.3903L}
}

@ARTICLE{2021MNRAS.503.3784P,
       author = {{Plotkin}, R.~M. and {Bahramian}, A. and {Miller-Jones}, J.~C.~A. and {Reynolds}, M.~T. and {Atri}, P. and {Maccarone}, T.~J. and {Shaw}, A.~W. and {Gandhi}, P.},
        title = "{Towards a larger sample of radio jets from quiescent black hole X-ray binaries}",
      journal = {\mnras},
         year = 2021,
        month = may,
       volume = {503},
       number = {3},
        pages = {3784-3795},
          doi = {10.1093/mnras/stab644},
archivePrefix = {arXiv},
       eprint = {2103.02178},
 primaryClass = {astro-ph.HE},
       adsurl = {https://ui.adsabs.harvard.edu/abs/2021MNRAS.503.3784P}
}

@ARTICLE{2003A&A...409..697M,
       author = {{Maccarone}, T.~J.},
        title = "{Do X-ray binary spectral state transition luminosities vary?}",
      journal = {\aap},
         year = 2003,
        month = oct,
       volume = {409},
        pages = {697-706},
          doi = {10.1051/0004-6361:20031146},
archivePrefix = {arXiv},
       eprint = {astro-ph/0308036},
 primaryClass = {astro-ph},
       adsurl = {https://ui.adsabs.harvard.edu/abs/2003A&A...409..697M}
}

@ARTICLE{2010MNRAS.403...61D,
       author = {{Dunn}, R.~J.~H. and {Fender}, R.~P. and {K{\"o}rding}, E.~G. and {Belloni}, T. and {Cabanac}, C.},
        title = "{A global spectral study of black hole X-ray binaries}",
      journal = {\mnras},
         year = 2010,
        month = mar,
       volume = {403},
       number = {1},
        pages = {61-82},
          doi = {10.1111/j.1365-2966.2010.16114.x},
archivePrefix = {arXiv},
       eprint = {0912.0142},
 primaryClass = {astro-ph.HE},
       adsurl = {https://ui.adsabs.harvard.edu/abs/2010MNRAS.403...61D}
}

@ARTICLE{2009PASP..121.1279S,
       author = {{Steiner}, James F. and {Narayan}, Ramesh and {McClintock}, Jeffrey E. and {Ebisawa}, Ken},
        title = "{A Simple Comptonization Model}",
      journal = {\pasp},
         year = 2009,
        month = nov,
       volume = {121},
       number = {885},
        pages = {1279},
          doi = {10.1086/648535},
archivePrefix = {arXiv},
       eprint = {0810.1758},
 primaryClass = {astro-ph},
       adsurl = {https://ui.adsabs.harvard.edu/abs/2009PASP..121.1279S}
}

@ARTICLE{2026PASA...43...44M,
       author = {{Maccarone}, Thomas and {Runnoe}, Jessie C and {Marcel}, Gregoire and {J{\"a}rvel{\"a}}, Emilia and {Buisson}, Douglas and {Kashyap}, Unnati and {Vincentelli}, Federico},
        title = "{Flip-flop states in X-ray binaries and changing-state AGN}",
      journal = {\pasa},
         year = 2026,
        month = mar,
       volume = {43},
          eid = {e044},
        pages = {e044},
          doi = {10.1017/pasa.2026.10176},
archivePrefix = {arXiv},
       eprint = {2603.07317},
 primaryClass = {astro-ph.HE},
       adsurl = {https://ui.adsabs.harvard.edu/abs/2026PASA...43...44M}
}

@INCOLLECTION{2006csxs.book..157M,
       author = {{McClintock}, Jeffrey E. and {Remillard}, Ronald A.},
        title = "{Black hole binaries}",
    booktitle = {Compact stellar X-ray sources},
         year = 2006,
       editor = {{Lewin}, Walter H.~G. and {van der Klis}, Michiel},
       volume = {39},
        pages = {157-213},
          doi = {10.48550/arXiv.astro-ph/0306213},
       adsurl = {https://ui.adsabs.harvard.edu/abs/2006csxs.book..157M}
}

@ARTICLE{2023ApJ...958L...8R,
       author = {{Rodriguez Cavero}, Nicole and {Marra}, Lorenzo and {Krawczynski}, Henric and {Dov{\v{c}}iak}, Michal and {Bianchi}, Stefano and {Steiner}, James F. and {Svoboda}, Jiri and {Capitanio}, Fiamma and {Matt}, Giorgio and {Negro}, Michela and {Ingram}, Adam and {Veledina}, Alexandra and {Taverna}, Roberto and {Karas}, Vladimir and {Ursini}, Francesco and {Podgorn{\'y}}, Jakub and {Ratheesh}, Ajay and {Suleimanov}, Valery and {Miku{\v{s}}incov{\'a}}, Romana and {Zane}, Silvia and {Kaaret}, Philip and {Muleri}, Fabio and {Poutanen}, Juri and {Malacaria}, Christian and {Petrucci}, Pierre-Olivier and {Gau}, Ephraim and {Hu}, Kun and {Chun}, Sohee and {Agudo}, Iv{\'a}n and {Antonelli}, Lucio A. and {Bachetti}, Matteo and {Baldini}, Luca and {Baumgartner}, Wayne H. and {Bellazzini}, Ronaldo and {Bongiorno}, Stephen D. and {Bonino}, Raffaella and {Brez}, Alessandro and {Bucciantini}, Niccol{\`o} and {Castellano}, Simone and {Cavazzuti}, Elisabetta and {Chen}, Chien-Ting and {Ciprini}, Stefano and {Costa}, Enrico and {De Rosa}, Alessandra and {Del Monte}, Ettore and {Di Gesu}, Laura and {Di Lalla}, Niccol{\`o} and {Di Marco}, Alessandro and {Donnarumma}, Immacolata and {Doroshenko}, Victor and {Ehlert}, Steven R. and {Enoto}, Teruaki and {Evangelista}, Yuri and {Fabiani}, Sergio and {Ferrazzoli}, Riccardo and {Garc{\'\i}a}, Javier A. and {Gunji}, Shuichi and {Hayashida}, Kiyoshi and {Heyl}, Jeremy and {Iwakiri}, Wataru and {Jorstad}, Svetlana G. and {Kislat}, Fabian and {Kitaguchi}, Takao and {Kolodziejczak}, Jeffery J. and {La Monaca}, Fabio and {Latronico}, Luca and {Liodakis}, Ioannis and {Maldera}, Simone and {Manfreda}, Alberto and {Marin}, Fr{\'e}d{\'e}ric and {Marinucci}, Andrea and {Marscher}, Alan P. and {Marshall}, Herman L. and {Massaro}, Francesco and {Mitsuishi}, Ikuyuki and {Mizuno}, Tsunefumi and {Ng}, Chi-Yung and {O'Dell}, Stephen L. and {Omodei}, Nicola and {Oppedisano}, Chiara and {Papitto}, Alessandro and {Pavlov}, George G. and {Peirson}, Abel L. and {Perri}, Matteo and {Pesce-Rollins}, Melissa and {Pilia}, Maura and {Possenti}, Andrea and {Puccetti}, Simonetta and {Ramsey}, Brian D. and {Rankin}, John and {Roberts}, Oliver J. and {Romani}, Roger W. and {Sgr{\`o}}, Carmelo and {Slane}, Patrick and {Spandre}, Gloria and {Soffitta}, Paolo and {Swartz}, Douglas A. and {Tamagawa}, Toru and {Tavecchio}, Fabrizio and {Tawara}, Yuzuru and {Tennant}, Allyn F. and {Thomas}, Nicholas E. and {Tombesi}, Francesco and {Trois}, Alessio and {Tsygankov}, Sergey S. and {Turolla}, Roberto and {Vink}, Jacco and {Weisskopf}, Martin C. and {Wu}, Kinwah and {Xie}, Fei},
        title = "{The First X-Ray Polarization Observation of the Black Hole X-Ray Binary 4U 1630-47 in the Steep Power-law State}",
      journal = {\apjl},
         year = 2023,
        month = nov,
       volume = {958},
       number = {1},
          eid = {L8},
        pages = {L8},
          doi = {10.3847/2041-8213/acfd2c},
archivePrefix = {arXiv},
       eprint = {2305.10630},
 primaryClass = {astro-ph.HE},
       adsurl = {https://ui.adsabs.harvard.edu/abs/2023ApJ...958L...8R}
}

@ARTICLE{LightmanShapiro1975,
       author = {{Lightman}, A.~P. and {Shapiro}, S.~L.},
        title = "{Spectrum and polarization of X-rays from accretion disks around black holes.}",
      journal = {\apjl},
         year = 1975,
        month = jun,
       volume = {198},
        pages = {L73-L75},
          doi = {10.1086/181815},
       adsurl = {https://ui.adsabs.harvard.edu/abs/1975ApJ...198L..73L}
}

@ARTICLE{Rees1975,
       author = {{Rees}, M.~J.},
        title = "{Expected polarization properties of binary X-ray sources.}",
      journal = {\mnras},
         year = 1975,
        month = jun,
       volume = {171},
        pages = {457-465},
          doi = {10.1093/mnras/171.3.457},
       adsurl = {https://ui.adsabs.harvard.edu/abs/1975MNRAS.171..457R}
}

@BOOK{Sobolev1963,
       author = {{Sobolev}, Viktor Viktorovich},
        title = "{A treatise on radiative transfer.}",
         year = 1963,
       adsurl = {https://ui.adsabs.harvard.edu/abs/1963trt..book.....S}
}

@ARTICLE{Taverna2021,
       author = {{Taverna}, R. and {Marra}, L. and {Bianchi}, S. and {Dov{\v{c}}iak}, M. and {Goosmann}, R. and {Marin}, F. and {Matt}, G. and {Zhang}, W.},
        title = "{Spectral and polarization properties of black hole accretion disc emission: including absorption effects}",
      journal = {\mnras},
         year = 2021,
        month = mar,
       volume = {501},
       number = {3},
        pages = {3393-3405},
          doi = {10.1093/mnras/staa3859},
archivePrefix = {arXiv},
       eprint = {2012.06504},
 primaryClass = {astro-ph.HE},
       adsurl = {https://ui.adsabs.harvard.edu/abs/2021MNRAS.501.3393T}
}

@ARTICLE{1999MNRAS.309..561Z,
       author = {{{\.Z}ycki}, Piotr T. and {Done}, Chris and {Smith}, David A.},
        title = "{The 1989 May outburst of the soft X-ray transient GS 2023+338 (V404 Cyg)}",
      journal = {\mnras},
         year = 1999,
        month = nov,
       volume = {309},
       number = {3},
        pages = {561-575},
          doi = {10.1046/j.1365-8711.1999.02885.x},
archivePrefix = {arXiv},
       eprint = {astro-ph/9904304},
 primaryClass = {astro-ph},
       adsurl = {https://ui.adsabs.harvard.edu/abs/1999MNRAS.309..561Z}
}

@ARTICLE{2019MNRAS.485.2942N,
       author = {{Nied{\'z}wiecki}, Andrzej and {Szanecki}, Micha{\l} and {Zdziarski}, Andrzej A.},
        title = "{Improved spectral models for relativistic reflection}",
      journal = {\mnras},
         year = 2019,
        month = may,
       volume = {485},
       number = {2},
        pages = {2942-2955},
          doi = {10.1093/mnras/stz487},
archivePrefix = {arXiv},
       eprint = {1805.06065},
 primaryClass = {astro-ph.HE},
       adsurl = {https://ui.adsabs.harvard.edu/abs/2019MNRAS.485.2942N}
}

@ARTICLE{2009ApJ...690L..97P,
       author = {{Poutanen}, Juri and {Vurm}, Indrek},
        title = "{On the Origin of Spectral States in Accreting Black Holes}",
      journal = {\apjl},
         year = 2009,
        month = jan,
       volume = {690},
       number = {2},
        pages = {L97-L100},
          doi = {10.1088/0004-637X/690/2/L97},
archivePrefix = {arXiv},
       eprint = {0807.3073},
 primaryClass = {astro-ph},
       adsurl = {https://ui.adsabs.harvard.edu/abs/2009ApJ...690L..97P}
}

@ARTICLE{2025arXiv250208718B,
       author = {{Buisson}, D.~J.~K. and {Marcel}, G. and {L{\'o}pez-Barquero}, V. and {Motta}, S.~E. and {Turner}, S.~G.~D. and {Vincentelli}, F.~M.},
        title = "{Flip-flop QPO changes during state transitions: a case study of GX339-4 and theoretical discussion}",
      journal = {arXiv e-prints},
         year = 2025,
        month = feb,
          eid = {arXiv:2502.08718},
        pages = {arXiv:2502.08718},
          doi = {10.48550/arXiv.2502.08718},
archivePrefix = {arXiv},
       eprint = {2502.08718},
 primaryClass = {astro-ph.HE},
       adsurl = {https://ui.adsabs.harvard.edu/abs/2025arXiv250208718B}
}

@ARTICLE{2023NatAs...7.1282R,
       author = {{Ricci}, Claudio and {Trakhtenbrot}, Benny},
        title = "{Changing-look active galactic nuclei}",
      journal = {Nature Astronomy},
         year = 2023,
        month = nov,
       volume = {7},
        pages = {1282-1294},
          doi = {10.1038/s41550-023-02108-4},
archivePrefix = {arXiv},
       eprint = {2211.05132},
 primaryClass = {astro-ph.GA},
       adsurl = {https://ui.adsabs.harvard.edu/abs/2023NatAs...7.1282R}
}

@ARTICLE{2018MNRAS.480.3898N,
       author = {{Noda}, Hirofumi and {Done}, Chris},
        title = "{Explaining changing-look AGN with state transition triggered by rapid mass accretion rate drop}",
      journal = {\mnras},
         year = 2018,
        month = nov,
       volume = {480},
       number = {3},
        pages = {3898-3906},
          doi = {10.1093/mnras/sty2032},
archivePrefix = {arXiv},
       eprint = {1805.07873},
 primaryClass = {astro-ph.GA},
       adsurl = {https://ui.adsabs.harvard.edu/abs/2018MNRAS.480.3898N}
}

@ARTICLE{2019ApJ...883...94T,
       author = {{Trakhtenbrot}, Benny and {Arcavi}, Iair and {MacLeod}, Chelsea L. and {Ricci}, Claudio and {Kara}, Erin and {Graham}, Melissa L. and {Stern}, Daniel and {Harrison}, Fiona A. and {Burke}, Jamison and {Hiramatsu}, Daichi and {Hosseinzadeh}, Griffin and {Howell}, D. Andrew and {Smartt}, Stephen J. and {Rest}, Armin and {Prieto}, Jose L. and {Shappee}, Benjamin J. and {Holoien}, Thomas W.-S. and {Bersier}, David and {Filippenko}, Alexei V. and {Brink}, Thomas G. and {Zheng}, WeiKang and {Li}, Ruancun and {Remillard}, Ronald A. and {Loewenstein}, Michael},
        title = "{1ES 1927+654: An AGN Caught Changing Look on a Timescale of Months}",
      journal = {\apj},
         year = 2019,
        month = sep,
       volume = {883},
       number = {1},
          eid = {94},
        pages = {94},
          doi = {10.3847/1538-4357/ab39e4},
archivePrefix = {arXiv},
       eprint = {1903.11084},
 primaryClass = {astro-ph.GA},
       adsurl = {https://ui.adsabs.harvard.edu/abs/2019ApJ...883...94T}
}

@ARTICLE{1979Afz....15..241L,
       author = {{Loskutov}, V.~M. and {Sobolev}, V.~V.},
        title = "{Polarization of scattered light in atmospheres with embedded sources.}",
      journal = {Astrofizika},
         year = 1979,
        month = apr,
       volume = {15},
        pages = {241-252},
       adsurl = {https://ui.adsabs.harvard.edu/abs/1979Afz....15..241L}
}

@ARTICLE{2022ApJ...939L..16Z,
       author = {{Zeltyn}, Grisha and {Trakhtenbrot}, Benny and {Eracleous}, Michael and {Runnoe}, Jessie and {Trump}, Jonathan R. and {Stern}, Jonathan and {Shen}, Yue and {Hern{\'a}ndez-Garc{\'\i}a}, Lorena and {Bauer}, Franz E. and {Yang}, Qian and {Dwelly}, Tom and {Ricci}, Claudio and {Green}, Paul and {Anderson}, Scott F. and {Assef}, Roberto J. and {Guolo}, Muryel and {MacLeod}, Chelsea and {Davis}, Megan C. and {Fries}, Logan and {Gezari}, Suvi and {Grogin}, Norman A. and {Homan}, David and {Koekemoer}, Anton M. and {Krumpe}, Mirko and {LaMassa}, Stephanie and {Liu}, Xin and {Merloni}, Andrea and {Mart{\'\i}nez-Aldama}, Mary Loli and {Schneider}, Donald P. and {Temple}, Matthew J. and {Brownstein}, Joel R. and {Ibarra-Medel}, Hector and {Burke}, Jamison and {Pellegrino}, Craig and {Kollmeier}, Juna A.},
        title = "{A Transient ``Changing-look'' Active Galactic Nucleus Resolved on Month Timescales from First-year Sloan Digital Sky Survey V Data}",
      journal = {\apjl},
         year = 2022,
        month = nov,
       volume = {939},
       number = {1},
          eid = {L16},
        pages = {L16},
          doi = {10.3847/2041-8213/ac9a47},
archivePrefix = {arXiv},
       eprint = {2210.07258},
 primaryClass = {astro-ph.GA},
       adsurl = {https://ui.adsabs.harvard.edu/abs/2022ApJ...939L..16Z}
}
\bibliographystyle{aasjournalv7}

\end{document}